\documentclass[journal,onecolumn]{IEEEtran}
\ifCLASSINFOpdf
\else
\fi
\usepackage[T1]{fontenc}
\usepackage[linesnumbered,ruled]{algorithm2e}
\usepackage{stfloats,color}
\usepackage{amsfonts}
\usepackage{amssymb}
\usepackage[cmex10]{amsmath}
\usepackage{graphicx}
\usepackage{setspace}
\usepackage{cite}
\usepackage{array}
\usepackage{subcaption}
\usepackage{times}
\usepackage{epsfig}
\usepackage{latexsym}
\usepackage{epstopdf}
\usepackage{verbatim}
\usepackage{units}
\usepackage{amsthm}
\usepackage{placeins}
\usepackage{afterpage}
\usepackage{dsfont}
\usepackage{soul}
\usepackage{multicol}
\usepackage{multirow}
\usepackage{mathtools}
\usepackage[cmintegrals]{newtxmath}
\newcolumntype{P}[1]{>{\centering\arraybackslash}p{#1}}
\newcolumntype{M}[1]{>{\centering\arraybackslash}m{#1}}

\newcommand{\defeq}{\ensuremath{\triangleq}}

\newtheorem{remark}{Remark}

\begin{document}
%
\title{Straggler-aware Distributed Learning: Communication Computation Latency Trade-off}
%
%
%

\author{Emre~Ozfatura,
        Sennur~Ulukus,
        and~Deniz~G{\"u}nd{\"u}z
\thanks{This paper was presented in part at the 2019  IEEE International Symposium on Information Theory (ISIT) in Paris, France, and at the 2019 IEEE Data Science Workshop in Minneapolis, USA.}
\thanks{Emre Ozfatura and Deniz G{\"u}nd{\"u}z are with Information Processing and Communications Lab, Department of Electrical and Electronic Engineering,
Imperial College London Email: \{m.ozfatura, d.gunduz\} @imperial.ac.uk.}
\thanks{Sennur Ulukus is with Department of Electrical and Computer Engineering, 
University of Maryland.} 
\thanks{This work was supported in part by the Marie Sklodowska-Curie Action SCAVENGE (grant agreement no. 675891), and by the European Research Council (ERC) Starting Grant BEACON (grant agreement no. 677854).}}

\maketitle

\begin{abstract}
 When gradient descent (GD) is scaled to many parallel \textit{workers} for large scale machine learning problems, its per-iteration computation time is limited by the \textit{straggling} workers. Straggling workers can be tolerated by assigning redundant computations and coding across data and computations, but in most existing schemes, each non-straggling worker transmits one message per iteration to the parameter server (PS) after completing all its computations. Imposing such a limitation results in two  main drawbacks; {\em over-computation} due to inaccurate prediction of the straggling behaviour, and
{\em under-utilization} due to treating workers as straggler/non-straggler and discarding partial computations carried out by stragglers. In this paper, to overcome these drawbacks, we consider multi-message communication (MMC) by allowing multiple computations to be conveyed from each worker per iteration, and design straggler avoidance techniques accordingly. Then, we analyze how the proposed designs can be employed efficiently to seek a balance between the computation and communication latency to minimize the overall latency. Furthermore, through extensive simulations, both model-based and real implementation on Amazon EC2 servers, we identify the advantages and disadvantages of these designs in different settings, and demonstrate that MMC can help improve upon existing straggler avoidance schemes. 
\end{abstract} 
\begin{IEEEkeywords}
Coded computation, distributed computation, gradient descent, gradient coding, machine learning, parallel computing, polynomial codes.
\end{IEEEkeywords}

\section{Introduction} \label{sec:introduction}
Machine learning techniques have become highly popular thanks to their success in a wide variety of classification and regression tasks. This success can be partially attributed to the availability of high-quality large training datasets. Unfortunately, as the size of the datasets increases, memory storage, management and maintenance become unmanageable within the resources of a single machine. An efficient way to deal with such colossal computing tasks within a reasonable training time is to exploit
computation and memory resources of multiple machines in parallel.\\
\indent In many supervised machine learning problems, the objective is to minimize the following {\em parameterized empirical loss function} for given training dataset $\mathcal{D}$ of $(x,y)$ pairs, where $x$ denotes the input sample while $y$ is the output (label for classification problems):
\begin{equation}
L(\boldsymbol{\theta}) \triangleq \frac{1}{\vert\mathcal{D}\vert}\sum_{(x, y)\in \mathcal{D}}l\left((x, y),\boldsymbol{\theta} \right),
\end{equation}
where $\boldsymbol{\theta}\in \mathbb{R}^{d}$ is the parameter vector and $l$ is an application specific loss function. This optimization problem can be solved by gradient descent (GD), where, at each iteration $t$, the parameter vector $\boldsymbol{\theta}_{t}\in\mathbb{R}^{d}$ is updated along the GD direction:
\begin{equation}\label{update}
\boldsymbol{\theta}_{t+1} = \boldsymbol{\theta}_{t} - \eta_{t} \nabla_{\boldsymbol{\theta}_{t}} L(\boldsymbol{\theta}_{t}), ~
\end{equation}
where $\eta_{t}$ is the learning rate at iteration $t$,
and the gradient with respect to current parameter vector is given by
\begin{equation}
\nabla_{\boldsymbol{\theta}_{t}}L(\boldsymbol{\theta}_{t})=\frac{1}{\vert\mathcal{D}\vert}\sum_{(x, y)\in \mathcal{D}}\nabla_{\boldsymbol{\theta}_{t}}l\left((x,y),\boldsymbol{\theta}_{t}\right).
\end{equation}
\indent When the dataset $\mathcal{D}$ is large,
distributed GD (DGD) \cite{PSGD1,PSGD2,PSGD3,PSGD4} can be utilized to reduce the computation time, and hence, the overall training time. In the  {\em naive parameter server (PS)} type implementation with $K$ workers, denoted by $w_{1},\ldots w_{K}$, first the dataset $\mathcal{D}$ is divided into $K$ non-overlapping equal-size subsets $D_{1},\ldots,D_{K}$. Then, at the beginning of each iteration $t$, the PS broadcasts the current parameter model $\boldsymbol{\theta}_{t}$ to all the workers. Each worker $w_{k}$ computes the partial gradient $g^{(t)}_{k}$ with respect to $\boldsymbol{\theta}_{t}$, based on the assigned dataset, i.e.,
\begin{equation}
g^{(t)}_{k}=\frac{1}{\vert \mathcal{D}_{k} \vert} \sum_{(x, y)\in \mathcal{D}_{k}} \nabla_{\boldsymbol{\theta}_{t}} l((x, y),\boldsymbol{\theta}_{t}).
\end{equation}
Finally, the PS waits to receive the partial gradients, $g^{(t)}_{k}$, from all the workers, and aggregates them to obtain the full gradient $\nabla_{\boldsymbol{\theta}_{t}}L(\boldsymbol{\theta}_{t})=\frac{1}{K}\sum^{K}_{k=1}g^{(t)}_{k}$. The PS updates the parameter vector according to (\ref{update}). While distributed computation is essential to handle large datasets, due to {\em synchronised updates} the completion time of each iteration is constrained by the slowest, so-called \textit{straggling worker(s)}\footnote{Although the straggling behaviour is often used to described the processing delay of the machine, we use this term for a wide range of delays including connection failures or queuing delays of assigned tasks.}, which can be detrimental for the convergence of the algorithm.\\
\indent A wealth of straggler avoidance techniques have been proposed in recent years for DGD as well as other distributed computation tasks \cite{UCUT.2,UCUT.3,UCUT.4,UCUT.5,UCUT.6,UCUT.7,UCUT.8,UCCT.1,UCCT.2,UCCT.3,UCCT.4,UCCT.5,CC.1,CC.2,CC.3,CC.4,CC.5,CC.6,CC.7,CC.8,CC.9,CC.10,CC.11,CC.12,CC.13,CC.14,CC.15,CC.16,CC.17,CC.18,CC.19,CC.20,CC.21,CC.22,CC.23,CC.30,CC.31,CC.32,CC.33,CMR1,CMR2,CMR3,CMR4,CMR5}. The common design notion behind all these schemes is the assignment of redundant computations/tasks to workers, such that faster workers can compensate for the stragglers. The main challenge is that the computation and communication latency of workers vary over time, and these values are not known in advance. This randomness can be treated as packet erasures in a communication channel \cite{CC.1}, and erasure coding techniques can be used to efficiently combat stragglers rather than simple task replication \cite{UCCT.1, UCCT.2, UCCT.3, UCCT.4, CC.3, CC.4, CC.5, CC.14}. However, most of the existing techniques, such as gradient coding (GC) \cite{UCCT.1}, Langrange coded computation (LCC) \cite{CC.4}, and their variations, suffer from two drawbacks: {\em over-computation} and {\em under-utilization}. By assigning redundant computations to workers, each iteration can be terminated with results from only a subset of the workers, and the minimum number of workers that must complete the assigned computation is called the \textit{non-straggling threshold}. The non-straggling threshold can be reduced by increasing the redundancy; however, a smaller threshold does not necessarily imply a lower completion time. Workers may be assigned more tasks than required due to an inaccurate prediction of the straggling behaviour, which we refer to as {\em over-computation}. Besides, in those schemes straggling behaviour is treated as `all or nothing' (straggler/non-straggler), and the computations carried out by stragglers are discarded as long as they cannot complete all their assigned computations. However, in practice, \textit{non-persistent} straggling servers are capable of completing a certain (sometimes significant) portion of their assigned tasks. This leads to {\em under-utilization} of the computational resources.
Therefore, our main objective in this paper is to introduce straggling avoidance techniques to mitigate under-utilization and over-computation. This will be achieved by allowing each worker to send multiple messages to the PS at each iteration, which we refer to as {\em multi-message communication (MMC)}. However, MMC may introduce additional delays due to the communication overhead. Hence, in this paper we also address the communication-computation latency trade-off, and provide flexible designs that can balance the two.\\
\indent Our contributions can be summarized as follows. First, we propose new straggler avoidance techniques specially designed to benefit from MMC. Second, in order to account for the additional communication load that may be introduced due to MMC, we provide designs that can provide a balance between the communication and computation latencies. Third, through extensive numerical simulations we illustrate the main advantages/disadvantages of the proposed designs compared to existing schemes. Finally, based on real experiments on Amazon EC2 servers, we show that the proposed schemes can improve upon existing straggler avoidance techniques.
\begin{table*}
\begin{center}
    \begin{tabular}{ | M{4cm} | M{4cm} |M{4cm} |}
   \hline
   Uncoded computation & Coded transmission & Coded computation\\ \hline
   \cite{UCUT.1,UCUT.2,UCUT.3,UCUT.4,UCUT.5,UCUT.6,UCUT.7,UCUT.8} & \cite{UCCT.1,UCCT.2,UCCT.3,UCCT.4,UCCT.5} & \cite{CC.1,CC.2,CC.3,CC.4,CC.5,CC.6,CC.7,CC.8,CC.9,CC.10,CC.11,CC.12,CC.13,CC.14,CC.15,CC.16,CC.17,CC.18}\\ \hline
     \end{tabular}
 \end{center}
 \caption{Classification of the DGD algorithms in the literature according to the straggler avoidance approach used.}\label{table:coded_uncoded}
\end{table*}
\section{An Overview of Existing Straggler Avoidance Techniques}
 There is already a rich literature on straggler avoidance methods in distributed learning/computation, many of them employing some form of coding. To provide a better understanding, we classify those schemes under three groups based on whether coding is employed or not, and if so, at which stage; namely, 1) coded computation,  2) coded communication, and finally, 3) uncoded computation. Before explaining these schemes, we first introduce two design parameters: {\em computation load} and {\em communication load}. Computation load, denoted by $r$, measures the redundancy of computations assigned to each worker compared to naive distributed  computation, where each computation task is assigned to a single worker. Communication load characterizes the total number of messages conveyed from the workers to the PS per iteration, where the size of each message is equal to the size of the parameter vector, $d$.
\subsection{Coded computation schemes}
In some problems, the gradient can be expressed as an explicit function of the dataset and the parameter vector, and more efficient straggler mitigation techniques can be introduced exploiting this particular relation. For example, for the least squares linear regression problem, the loss function can be explicitly written as
\begin{equation}
L(\boldsymbol{\theta}) = \frac{1}{2N}\sum_{(\mathbf{x},y)\in \mathcal{D}}(y-\mathbf{x}^{T}\boldsymbol{\theta})^{2} ~,
\end{equation}
where $\mathbf{x} \in \mathbb{R}^{d}$ is the input vector, $y\in \mathbb{R}$ the corresponding output, and $N$ is the size of the dataset. For this particular loss function, the gradient is given by 
\begin{eqnarray}\label{coded_grad}
\nabla_{\boldsymbol{\theta}} L(\boldsymbol{\theta}) = \mathbf{X}^{T} \mathbf{X} \boldsymbol{\theta}-\mathbf{X}^{T}\mathbf{y},
\end{eqnarray}
where $\mathbf{X}=[\mathbf{x}_{1},\ldots,\mathbf{x}_{N}]^{T}$ and  $\mathbf{y}=[y_{1},\ldots,y_{N}]^{T}$ are concatenation of all input vectors and output values, respectively. Since the second term, $\mathbf{X}^{T}\mathbf{y}$, does not include the term $\boldsymbol{\theta}$, it remains the same throughout  the iterations. Therefore, the main computation task is to compute $\mathbf{X}^{T}\mathbf{X}\boldsymbol{\theta}_{t}$ at each iteration. In this particular case the problem can be reduced to distributed matrix-matrix multiplication, or matrix-vector multiplication if $\mathbf{X}^{T}\mathbf{X}$ is computed beforehand, and this simplified form allows exploiting novel ideas from coding theory.\\
\indent In the naive distributed computation scenario, $\mathbf{X}$ can be divided into $K$ submatrices (assume, for simplicity, that $K$ divides $N$), $\mathbf{X}_{1},\ldots,\mathbf{X}_{K}$, each of size $N/K \times d$, such that $k$th worker computes $\mathbf{X}_{k}^{T}\mathbf{X}_{k}\boldsymbol{\theta}_{t}$ at iteration $t$. Since the following equality holds
\begin{equation}
\mathbf{X}^{T}\mathbf{X}\boldsymbol{\theta}=\sum_{k=1}^{K}\mathbf{X}_{k}^{T}\mathbf{X}_{k}\boldsymbol{\theta},
\end{equation}
PS can obtain the full gradient receiving the computation results from all the workers. In contrast to the naive approach, coded computation schemes for distributed matrix multiplication \cite{CC.6,CC.7,CC.16,CC.18}
first encode the submatrices, and then assign them to the workers to achieve a certain tolerance against slow/straggling workers.\\
\indent We note that $\mathbf{W}\defeq\mathbf{X}^{T}\mathbf{X}$ in (\ref{coded_grad}) also remains unchanged throughout GD iterations. Hence, if $\mathbf{W}$ can be computed at the beginning, the main computational task reduces to linear operations at each iteration, which allows employing various linear coding structures, e.g.,  maximum distance separable (MDS) codes, or rateless codes, to encode rows of $\mathbf{W}$ to achieve robustness against stragglers \cite{CC.1,CC.2,CC.3,CC.9,CC.12,CC.13}.\\
\indent We want to reemphasize that coded computation schemes are mostly designed for the full recovery of the main task, such as the recovery of the full gradient in DGD. However, in DGD implementations approximate/partial gradient can also be used instead of the full gradient to seek a balance between the computation time and accuracy, and to eventually reduce the convergence time. Approximate GC and partial gradient recovery schemes have also been studied in \cite{CC.AG1,CC.AG2,CC.AG3} and \cite{CCP.1,CCP.2,CC.13}, respectively. 
In the scope of this paper, we limit our focus to full gradient recovery and leave the MMC variation of partial gradient recovery \cite{CC.13} as a future work.
\subsection{Coded transmission schemes}
Let $\mathcal{G}=\{g_{1},\ldots,g_{K}\}$ be the set of partial gradients corresponding to datasets $\mathcal{D}_{1},\ldots,\mathcal{D}_{K}$. In the GC scheme with computation load $r$,  $r$ partial gradient computations, denoted by $\mathcal{G}_{k}$, are assigned to worker $k$ \cite{UCCT.1}. After computing these $r$ partial gradients, each worker sends a linear combination of the results,
\begin{equation}
c^{(t)}_{k}\defeq\mathcal{L}_{k}(g^{(t)}_{i}:g_{i}\in\mathcal{G}_{k}).
\end{equation}
We refer to these linear combinations $c_{1},\ldots,c_{K}$ as {\em coded  partial gradients}. The PS waits until it receives sufficiently many coded partial gradients to recover the full gradient. It is shown in \cite{UCCT.1} that, for any set of non-straggler workers $\mathcal{W}\subseteq[K]$ with  $ \lvert \mathcal{W}   \rvert = K-r+1 $, there exists a set of coefficients $\mathcal{A}_{\mathcal{W}}=\left \{a_{k}:k\in\mathcal{W}\right\}$ such that
\begin{equation}
\sum_{k\in\mathcal{W}} a_{k}c^{(t)}_{k}=\frac{1}{K}\sum^{K}_{k=1}g^{(t)}_{k}.
\end{equation}
Hence, GC can tolerate up to $r-1$ persistent stragglers at each iteration.
GC can also be interpreted as a polynomial interpolation problem \cite{UCCT.3}. In this model, the gradient assignment matrix is called a {\em mask matrix}, and the {\em support} of the $k$th row $\mathbf{M}_{(k \text{ } , \text{ }:)}$, denoted by $supp(\mathbf{M}_{(k \text{ } , \text{ }:)})$, gives the index of the partial gradients assigned to $w_{k}$, $\mathcal{G}_{k}$, and for given redundancy $r$, $\vert\vert \mathbf{M}_{(k \text{ } , \text{ }:)} \vert\vert_{1}=r$. For a given mask matrix $\mathbf{M}$, GC is equivalent to interpolating a polynomial with {\em degree} $h$, where $h=K-\min_{k}{\vert\vert \mathbf{M}_{(: \text{ } , \text{ }k)}\vert\vert_{1}}$; in other words, $h$ is equal to the number of zeros in the most sparse column\footnote{If $\mathbf{M}$ is a $K \times K$ matrix then $\vert\vert \mathbf{M}_{(: \text{ } , \text{ }k)}\vert\vert_{1}=\vert\vert \mathbf{M}_{(k \text{ } , \text{ }:)}\vert\vert_{1}=r$, $\forall k\in[K]$.} of $\mathbf{M}$.\\
\indent In a broad sense, each partial gradient $g_{k}$ is embedded into a polynomial $f_{k}$, and each worker evaluates the  polynomials $f_{1},\ldots,f_{K}$ at preassigned points, and sends their sum to the PS. Let polynomial $f_{k}$ be constructed as
\begin{equation}
f_{k}(x)= \prod_{i:g_{k}\notin \mathcal{G}_i}(x-\alpha_{i})
\end{equation}
for some distinct $\alpha_{1},\ldots,\alpha_{K}$. We define another polynomial:
\begin{equation}
h(x)=\sum^{K}_{k=1} g_{k}f_{k}(x).
\end{equation}
At each iteration, each worker $w_{i}$ sends $h(\alpha_{i})$ to the PS. The key design trick here is that, worker $w_{i}$ does not need to compute $g_{k}$, if $f_{k}(x)$ has a root at $\alpha_{i}$, and can compute $h(\alpha_{i})$ only with the knowledge of $g_{k}$s in the set $\mathcal{G}_{i}$. \\
\indent To explain the decoding stage, consider the following mask matrix:  
\begin{equation}
\mathbf{M}=
  \begin{bmatrix}
    1 & 1 & 1 & 0 & 0 & 0  \\
    0 & 1 & 1 & 1 & 0 & 0 \\
    0 & 0 & 1 & 1 & 1 & 0 \\
    0 & 0 & 0 & 1 & 1 & 1 \\
    1 & 0 & 0 & 0 & 1 & 1\\
    1 & 1 & 0 & 0 & 0 & 1 \\
  \end{bmatrix}.\label{mask1}
\end{equation}
There will be six partial gradients $g_{1},\ldots,g_{6}$ and six  corresponding polynomials $f_{1}(x),\ldots,f_{6}(x)$ to embed their values. Observe that the number of zeros in each column, $K-r$, is equivalent to the number of roots of the corresponding polynomial $f$, which is three for all polynomials, in our example. Then, the leading coefficient of $h(x)$ is equal to $g=\sum^{6}_{k=1}g_{k}$, and it has degree three since the degree of each polynomial $f_{k}$ is three. Therefore, at each iteration, $h(x)$ can be interpolated using its value at any $K-r+1$, 4 for the given example, different points. Accordingly, to recover $g$, any $K-r+1$ results are sufficient, which implies a non-straggling threshold of $K_{th}=K-r+1$. We sketch the general design strategy and the corresponding non-straggling threshold, however implementation of the encoding and decoding procedures, and their complexity (see \cite{UCCT.3} for further details), also affect the completion time; nevertheless, in the scope of this paper we omit the complexity analysis and focus on computation and communication latency. \\
\indent In \cite{UCCT.2}, the GC scheme is extended to seek a trade-off between the communication latency and the non-straggler threshold, and it is shown that the length of the coded partial gradient $c_{k}$ can be reduced with an increase in the non-straggling threshold. The trade-off between the communication latency and straggler tolerance is also studied in \cite{UCCT.4}, and it is shown that the PS can recover the full gradient faster when each worker is allowed to send more than one coded partial gradient. We classify these schemes as  coded transmission since computations are carried out using uncoded data, but the computations are transmitted to the PS in a coded manner.
\subsection{Uncoded computation schemes}
This class includes schemes that do not employ any coding. In the naive distributed approach the computation task is divided into disjoint sub-tasks to be executed in parallel. To mitigate the stragglers each worker may perform some backup computations \cite{UCUT.2,UCUT.3,UCUT.4,CC.5}, certain unfinished subtasks (slow workers) can be relaunched  at the fast workers \cite{UCUT.7,UCUT.8}, or some additional backup workers can be employed \cite{UCUT.6}. Alternatively, PS can terminate an iteration after receiving results from a subset of workers \cite{UCUT.1,UCUT.9}.


Existing  straggler tolerant DGD schemes focus on minimizing the non-straggling threshold, which does not necessarily capture the average completion time statistics for one iteration of the GD algorithm. Indeed, in certain regimes of computation load $r$, the average completion time may be increasing as the non-straggling threshold decreases. Accordingly, in  this paper, we consider the average completion time as the main performance metric, and allow workers to send  multiple messages at each iteration to reduce the per-iteration completion time.\\
\indent MMC can be easily applied in uncoded computation by assigning each computation task to multiple workers \cite{UCUT.4,CC.5}. Workers can then return each of their computations as soon as it is completed, and the iteration is completed when each computation task is completed by at least one worker. Multi-message coded computation is also studied in \cite{CC.2, CC.9}. However, these schemes are limited to matrix-vector multiplication. Furthermore, they ignore the communication overhead due to MMC and its impact on the communication latency, and focus only on the computation time.

\section{Coded Computation with MMC}\label{sec:codedcomp}
For the coded computation we employ the LCC method introduced in \cite{CC.6,CC.4}, which utilizes polynomial interpolation for the code design. In this section, we first explain the structure of the Lagrange polynomial, then explain how it is utilized for coded computation, and finally discuss how it can be modified to benefit from MMC.

\subsection{Lagrange Coded Computation (LCC)}
First, $\mathbf{X}$ is divided into $K$ submatrices (assume, for simplicity, that $K$ divides $N$), $\mathbf{X}_{1}, \ldots, \mathbf{X}_{K}$, each of size $N/K \times d$. For given $r$, assuming $K$ is divisible by $r$, these $K$ submatrices are further divided into $r$  disjoint groups, each containing $K/r$ submatrices. Let $\mathbf{x}_{q,j}$ denote the $j$th submatrix in the $q$th group, and $\mathbf{X}_{q}$ denote all the submatrices in the $q$th group; that is, $\mathbf{X}_{q}$ is an $N/r \times d$ submatrix of $\mathbf{X}$. Then, for distinct real numbers $\alpha_1, \ldots, \alpha_{K/r}$, we form the following $r$ structurally identical polynomials of degree $K/r-1$, taking the submatrices of $\mathbf{X}_{q}$ as their coefficients:
\begin{equation}
f_{q}(z)=\sum_{i=1}^{K/r}\mathbf{x}_{q,i}\prod_{j=1,j \neq i}^{K/r} \frac{z-\alpha_{j}}{\alpha_{i}-\alpha_{j}},\text{ } q\in[r],
\end{equation}
which satisfy $f_{q}(\alpha_{i})= x_{k,i},$ $\forall k,i$. Then, we define
\begin{equation}
 H(z)\defeq\sum_{q=1}^{r}f_{q}(z)^{T}f_{q}(z)\boldsymbol{\theta}_{t}.
\end{equation}
Coded submatrices $\tilde{\mathbf{x}}_{k}^{(q)}$, $q\in[r]$, for worker $w_{k}$, $k\in[K]$ are obtained by evaluating
$f_{q}(z)$ polynomials  at distinct values, $\beta_{k} \in \mathbb{R}$, i.e.,
$\tilde{\mathbf{x}}_{k}^{(q)} = f_{q}(\beta_{k})$. 
At each iteration $w_{k}$ returns the value of 
 \begin{equation}
 H(\beta_{k})=\sum_{q=1}^{r}(\tilde{\mathbf{x}}_{k}^{(q)})^{T}\tilde{\mathbf{x}}_{k}^{(q)}\boldsymbol{\theta}_{t}.
 \end{equation}
The degree of polynomial $H(z)$ is $2K/r-2$; and thus, the non-straggling threshold for LCC is given by $K_{LCC}(r)=2K/r-1$; that is, having received the value of $H(z)$ at $K_{LCC}(r)$ distinct points, the PS can extrapolate $ H(z)$ and compute
 \begin{equation}
 \sum_{i=1}^{K/r}H(\alpha_{i})=\mathbf{X}^{T}\mathbf{X}\boldsymbol{\theta}_{t},
 \end{equation}
\indent When $N$ is not divisible by $r$, zero-valued data points can be added to $\mathbf{X}$ to make it divisible by $r$. Hence, in general the non-straggling threshold is given by $K_{LCC}(r)=2\lceil N/r \rceil-1$. 
\subsection{LCC with MMC}
 Here, we introduce  LCC with MMC by using a single polynomial $f(z)$ of degree $K-1$, instead of using $r$ different polynomials each of degree $K/r-1$. We define
\begin{equation}
f(z)\defeq\sum_{i=1}^{K}\mathbf{x}_{i}\prod_{j=1,j\neq i}^{K} \frac{z-\alpha_{j}}{\alpha_{i}-\alpha_{j}},
\end{equation}
where $\alpha_{1},\ldots,\alpha_{K}$ are  $K$ distinct real numbers, and we construct
\begin{equation}
h(z)\defeq f(z)^{T}f(z)\boldsymbol{\theta}_{t},
\end{equation}
such that $h(\alpha_{i})=\mathbf{x}_{i}^{T}\mathbf{x}_{i}\boldsymbol{\theta}_{t}$. Consequently, if the polynomial $h(z)$ is known at the PS, then the full gradient $\sum_{i=1}^{K}h(\alpha_{i})=\sum_{i=1}^{N}\mathbf{x}_{i}^{T}\mathbf{x}_{i}\boldsymbol{\theta}_{t}$ can be obtained. To this end, $Kr$ coded submatrices $\tilde{\mathbf{x}}_{k}^{(q)}, k\in [K], \text{ }q\in [r]$, are constructed by evaluating $f(z)$ at $Kr$ different points, $\beta_{k}^{(q)}$, i.e.,
\begin{equation}
\tilde{\mathbf{x}}_{k}^{(q)}=f(\beta_{k}^{(q)}), \text{ }k\in[K], q\in[r],
\end{equation}
 and $\tilde{\mathbf{x}}_{k}^{(1)},\tilde{\mathbf{x}}_{k}^{(r)}$ are assigned to $w_{k}$, $k\in [K]$. $w_{k}$ computes $(\tilde{\mathbf{x}}_{k}^{(1)})^{T}\tilde{\mathbf{x}}_{k}^{(1)}\boldsymbol{\theta}_{t},\ldots,(\tilde{\mathbf{x}}_{k}^{(r)})^{T}\tilde{\mathbf{x}}_{k}^{(r)}\boldsymbol{\theta}_{t}$ sequentially, and transmits each of these results to the PS as soon as it is computed. Coded computation corresponding to coded data point $\tilde{\mathbf{x}}_{k}^{(q)}$ at $w_{k}$ provides the value of polynomial $h(z)$ at point $\beta_{k}^{(q)}$. The degrees of  polynomials $f(z)$ and $h(z)$ are $K-1$ and $2(K-1)$, respectively, which implies that  $h(z)$ can be interpolated from its values at any $2K-1$ distinct points. Hence, any $2K-1$ computations received from any subset of the workers are sufficient to obtain the full gradient.\\
\indent We note that, in the original LCC scheme coded data points are constructed evaluating $r$ different polynomials at the same data point, whereas in the multi-message LCC scheme, coded data points are constructed evaluating a single polynomial at $r$ distinct points. Per iteration completion time can be reduced with MMC since the partial computations of the non-persistent stragglers are also utilized; however, at the expense of an increase in the communication load. Nevertheless, it is possible to set the number of polynomials to a different value to seek a balance between the communication load and the per iteration completion time. This will be explored in Section \ref{s:numerical_results}.

\section{GC with MMC} \label{sec:codedcomp}
In the original GC scheme of \cite{UCCT.1}, the number of messages transmitted to the PS per-iteration per-worker is limited to one. Due to the synchronized model update, the workers that complete their computations stay idle until they receive the updated parameter vector to start the next iteration. To prevent under-utilization of the computation resources, we will allow each worker to send coded partial gradients to the PS; that is, at each iteration each worker sends multiple coded partial gradients instead of sending a single coded computation result. In the scope of this paper, we will present two different approaches to design coded partial gradients, namely {\em correlated code design} and {\em uncorrelated code design}, which are explained next.

\subsection{Correlated code design}
In GC, the number of partial gradients linearly combined to form the transmitted message from a worker is equal to the computation load $r$. In MMC, we allow each worker to compute and transmit multiple coded partial gradients, each of which will be generated by combining $m \leq r$ gradient computations. We will refer to $m$ as the {\em order} of the corresponding partial gradient. In particular, each worker will be able to send up to $l= r-m+1$ different messages, each of order $m$; that is, each of the coded partial gradients will be a linear combination of the $m$ most recently computed partial gradients. 

As an example, let $r=3$ and $m=2$, and consider the worker with assigned partial gradients $g_{1}, g_{2}$ and $g_{3}$ in this order\footnote{In the rest of the paper, to simplify the notation we drop the time index from the gradients when we focus on a single iteration of the algorithm.}. After computing $g_{1}$ and $g_{2}$ the worker will send a linear combination of these two partial gradients as a coded message to the PS, and after computing $g_{3}$, it will send a linear combination of partial gradients $g_{2}$ and $g_{3}$.

In general, the proposed scheme consists of two steps:  coded message construction and message assignment. In the coded message construction step, structure of the coded messages are designed according to the order $m$ by simply setting $r=m$ in the original GC scheme. Then, in the coded message assignment step, constructed messages are assigned to the workers based on the assigned partial gradients. We present the following example to clarify these steps.

\textbf{Example 1:} Let $K=6$, $r=3$, $m=2$, and consider the assignment matrix $\mathbf{M}$, whose $i$th row indicates the mini-batches assigned to the $i$th worker; that is  $\mathbf{M}(i,j)=1$ means that partial gradient $g_{j}$ will be computed by the $i$th worker. In GC with $K=6$ and $r=3$, we have the following assignment matrix.
\begin{equation}
\mathbf{M}=
  \begin{bmatrix}
    1 & 1 & {\color{red}1} & 0 & 0 & 0  \\
    0 & 1 & 1 & {\color{red}1} & 0 & 0 \\
    0 & 0 & 1 & 1 & {\color{red}1} & 0 \\
    0 & 0 & 0 & 1 & 1 & {\color{red}1} \\
    {\color{red}1} & 0 & 0 & 0 & 1 & 1\\
    1 & {\color{red}1} & 0 & 0 & 0 & 1 \\
  \end{bmatrix}
\end{equation}
When  $m=2$, coded gradients are obtained according to the assignment matrix $\tilde{M}$, which is obtained by removing the last $r-m$ of the 1s in each row (shown in {\color{red}red} above). When the assignment matrix $\tilde{M}$ is used to design GC, a total of $K=6$ coded partial gradients, each of order two, are constructed; and the full gradient can be obtained from any $K-m+1=5$ coded partial gradients. Let ${c_{1},\ldots,c_{6}}$ denote the corresponding coded partial gradients. We remark that $c_{1}$ is a  linear combination of $g_{1}$ and $g_{2}$, while $c_{2}$ is a linear combination of $g_{2}$ and $g_{3}$. Since $g_{1},g_{2}$ and $g_{3}$ are assigned to the first worker, it can send both coded messages $c_{1}$ and $c_{2}$. 
To illustrate the assignment of coded partial gradients, we use the assignment matrix $\mathbf{C}$, where the $i$th column shows the assigned coded gradients to the $i$th worker in the order of computation. In Example 1, we have
\begin{equation}
\mathbf{C}=
  \begin{bmatrix}
  c_{1} & c_{2} & c_{3} & c_{4} & c_{5} & c_{6} \\
   c_{2} & c_{3} & c_{4} & c_{5} & c_{6} & c_{1} \\
  \end{bmatrix}.
\end{equation}
We call this approach {\em correlated code design}, since the same coded partial gradient can be computed and sent by more than one worker, e.g., in Example 1, $c_{2}$ can be sent by both $w_{1}$ and $w_{2}$. In Example 1, the original GC algorithm needs to receive computations from at least four workers in order to complete an iteration; whereas the proposed scheme can complete an iteration with results from only three workers. For instance, when workers 1, 2 and 4 each send two coded partial gradients, the PS will obtain $c_{1},c_{2},c_{3},c_{4},c_{5}$, and recover the full gradient. In the next section, we will analyze the uncorrelated coded design approach, where each coded gradient is assigned to exactly one worker. 
\subsection{Uncorrelated code design}
Here, we present another code construction to extend GC to the multi-message scenario. Consider the partial gradient assignment to 6 workers governed by the mask matrix in (\ref{mask1}). Assume that each worker sends a coded partial gradient after computing the first two assigned partial gradients, and then sends a second coded partial gradient after computing all its assigned partial gradients. Now, consider the scenario with 12 workers and the following mask matrix:
\begin{equation}
\tilde{\mathbf{M}}=
  \begin{bmatrix}
    {\color{red} 1} & {\color{red}1} & {\color{red}1} & {\color{red}0} & {\color{red}0} & {\color{red}0}  \\
    {\color{blue} 1} & {\color{blue}1} & {\color{blue}0} & {\color{blue}0} & {\color{blue}0} & {\color{blue}0}  \\
    {\color{red}0} & {\color{red}1} & {\color{red}1} & {\color{red}1} & {\color{red}0} & {\color{red}0} \\
        {\color{blue} 0} & {\color{blue}1} & {\color{blue}1} & {\color{blue}0} & {\color{blue}0} & {\color{blue}0}  \\
    {\color{red}0} & {\color{red}0} & {\color{red}1} & {\color{red}1} & {\color{red}1} & {\color{red}0} \\
        {\color{blue} 0} & {\color{blue}0} & {\color{blue}1} & {\color{blue}1} & {\color{blue}0} & {\color{blue}0}  \\
    {\color{red}0} & {\color{red}0} & {\color{red}0} & {\color{red}1} & {\color{red}1} & {\color{red}1} \\
        {\color{blue} 0} & {\color{blue}0} & {\color{blue}0} & {\color{blue}1} & {\color{blue}1} & {\color{blue}0}  \\
    {\color{red}1} & {\color{red}0} & {\color{red}0} & {\color{red}0} & {\color{red}1} & {\color{red}1} \\
        {\color{blue} 0} & {\color{blue}0} & {\color{blue}0} & {\color{blue}0} & {\color{blue}1} & {\color{blue}1}  \\
    {\color{red}1} & {\color{red}1} & {\color{red}0} & {\color{red}0} & {\color{red}0} & {\color{red}1} \\
        {\color{blue} 1} & {\color{blue}0} & {\color{blue}0} & {\color{blue}0} & {\color{blue}0} & {\color{blue}1}  \\
  \end{bmatrix}.\label{mask2}
\end{equation}

According to $\tilde{\mathbf{M}}$ three partial gradients are assigned to six workers, whose rows are shown in  {\color{red}red}, while  two partial gradients are assigned to the remaining six workers, whose rows are shown in {\color{blue}blue}. In terms of encoding/decoding process these two are equivalent. Therefore, sending an additional coded partial gradient corresponds to adding a  ``virtual'' worker, i.e., the rows in {\color{red}red} correspond to the real workers, while the rows in {\color{blue}blue} to the virtual ones. We note that given $\tilde{\mathbf{M}}$, degree of $h(x)$ and the non-straggling threshold will be 7 and 8, respectively, since  there are exactly seven zeros in each column.\\
 \indent We remind that, in the original GC scheme, the PS waits for $4$ workers to recover the full gradient, while in the proposed scheme each worker can send a coded partial gradient as a virtual worker, after only two computations, and the full gradient can be recovered from any 8 coded partial gradients, including those from the virtual workers. Assume, for example, that three of the workers are non-stragglers, and each of them sends 2 coded partial gradients, while two workers are non-persistent stragglers, and each of them sends only one coded gradient, while the last worker is a persistent straggler. In this case the full gradient can be obtained by the proposed approach but not with the original GC scheme. Hence, the proposed approach improves the per-iteration completion time.\\
\indent In general, if each worker is allowed to send $l$  messages per iteration, we can introduce $K (l-1)$ ``virtual'' workers, resulting in a total of $Kl$ workers. Then, we design a GC scheme for the mask matrix of $Kl$ workers. 
Uncorrelated code design for GC with MMC is defined by the order vector $[m_{0},\ldots,m_{i}]$, where $m_{0}$ is the order of the coded partial gradient sent by the real worker, while $m_{i}$ denotes the order of the coded partial gradient sent by the $i$th virtual worker. For a particular MMC strategy with order vector $\mathbf{m} = [m_{0},\ldots,m_{i}]$, the number of zeros in any column of the mask matrix $M$ is given by $Kl-(\sum^{l}_{i=0}m_{i})$; and thus,  $Kl-(\sum^{l}_{i=0}m_{i})+1$  coded partial gradients are required to recover the full gradient. 

We note that, while the use of coded partial gradients with lower orders increases the recovery threshold, they can be obtained faster, as they allow the PS to exploit the computations carried out by non-persistent stragglers. We leave the optimization of the partial gradient orders depending on system parameters and requirements as future work.

Another important issue regarding MMC is the \textit{communication load}, which denotes the average number of messages received by the PS at each iteration. The communication load increases with the number of virtual workers; therefore, the optimal MMC strategy  depends critically on the communication architecture of the network and the protocol used to transmit messages from the workers to the PS as well as the computation speeds of the workers.\\

\subsection{Clustering}

Next, we introduce {\em clustering}, which can further speed up the computation time. We divide the workers into $P$ equal-size disjoint clusters, where the set of workers in cluster $p$ is denoted by $\mathcal{W}_{p} \subset \mathcal{W}$, $p\in[P]$. Dataset and the corresponding set of partial gradients  $\mathcal{G}=\{g_{1},\ldots,g_{K}\}$ are also divided into $P$ equal-size disjoint subsets, and the set of partial gradients assigned to the $p$th cluster is denoted by $\mathcal{G}_{p}$. In the clustering approach, the workers in the $p$th cluster are responsible for computing
\begin{equation}
\frac{1}{\vert\mathcal{G}_{p}\vert}\sum_{k \in \mathcal{G}_{p}}g_{k},
\end{equation}
and the GC scheme is applied to each cluster independently.\\
\indent At this point, we remark that the fractional repetition scheme in \cite{UCCT.1} is a special case of the proposed clustering approach, where the size of a cluster is equal to the computation load, $r$. As an example consider $K=40$, $r=10$ and $P=4$, where the workers are divided into $P=4$ clusters, while the mini-batches are divided into $4$ subsets, and each cluster is responsible for a different subset. In the fractional repetition scheme the PS waits until at least one worker from each cluster completes and sends its partial gradient. One can observe that if at least $K-r+1=31$ workers  complete and send their computations to the PS, there must be at least one worker from each cluster; hence, the non-straggling threshold is  $K-r+1$. However, the non-straggling threshold represents a worst case scenario. Notice that, even 4 workers, each from a different cluster can be sufficient to obtain the full gradient. On the other hand, the cyclic repetition scheme in \cite{UCCT.1}, which has a circulant mask matrix as in (\ref{mask1}), always has to wait until receiving coded messages from at least $K-r+1$ workers. Although both GC schemes achieve the same optimal non-straggling threshold, their average performance may differ substantially. 

\begin{figure*}
    \centering       
    \includegraphics[scale=0.65]{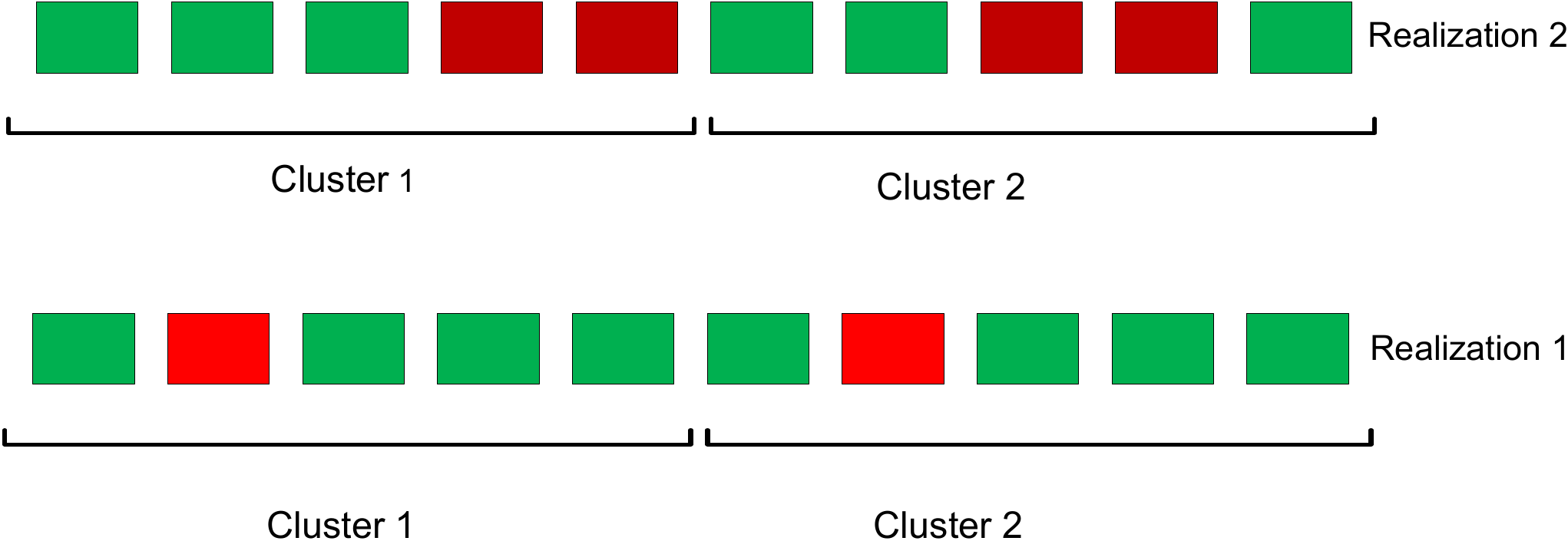}				
    \caption{ Two possible straggler realizations where {\color{green}green}  and  {\color{red}{red}} blocks illustrate the straggler and non-straggler workers respectively.}
    \label{clusters}
\end{figure*}

While the fractional repetition scheme  requires $K$ to be an integer multiple of $r$, the clustering approach outlined above can be applied to any $(K,r)$ pair. When GC is applied with clustering, it is possible to tolerate $r-1$ stragglers in each cluster; thus, for a particular straggler realization, if the full gradient can be obtained with GC (without clustering) then it can also be obtained with clustering, while the converse is not true.\\
\indent  To illustrate this, we consider the case with $K=10$ and $r=3$. When  GC is applied, $8$ non-straggler workers are required to recover the full gradient. Alternatively, clustering the workers into $P=2$ clusters, 3 non-straggler workers from each cluster are required for full gradient recovery. Any straggler realization that is ``good'' for GC (i.e., not more than 2 stragglers) is also good for clustering; however there are certain realizations that are good for the latter, but not for the former. To illustrate this, in Fig. \ref{clusters}, we depict two different straggler realizations. One can observe that Realization 1 is good for both schemes, while in Realization 2, the full gradient recovery can be achieved by only the clustering strategy. We want to emphasize that, although  6 non-straggling workers in Realization 1 are sufficient for full gradient recovery, this does not mean that any 6 non-straggling workers would be sufficient. However, the set of straggler realizations where the full gradient recovery is possible using GC is a subset of the one for clustering. Consequently, while the non-straggling threshold is the same for GC and GC with clustering, this threshold only represents the worst case scenario and as exemplified above, the probability of reaching recovery condition is higher when clustering is employed; and hence  the average computation time can be reduced.

However, when MMC is allowed, clustering may also be disadvantageous. On one hand, more straggling workers can be tolerated on average, on the other hand  GC is applied to each cluster independently; hence, a coded partial gradient from a particular cluster cannot be utilized for another cluster. Consequently, the optimal clustering strategy with MMC depends on the computation statistics of the workers.
\subsection{Hybrid implementation}
The optimal DGD strategy depends critically on the computation time statistics of the workers. In particular, when the computation speeds of the workers are similar, MMC is expected to have a better performance as it can exploit all the computations carried out across the workers; however, when one of the workers is much faster compared to the others, fractional repetition can be preferred. To illustrate this trade-off, consider the case $K=5$ and $r=5$. With the fractional repetition scheme, the PS waits for the fastest worker to finish all the assigned computations; however, with GC with MMC for given order vector $\mathbf{m}= [5,3]$, the PS waits for  3 coded messages sent from 2 workers; hence the overall speed will depend on the second, or even the third fastest worker.

Accordingly, we can propose a hybrid scheme, in which the workers initially behave as dictated by the GC-MM scheme, but if a worker is fast enough to complete all its computations, then it switches to fractional repetition scheme, and sends the average gradient instead of a coded partial gradient. 

\section{Uncoded Computation with MMC}\label{section:uncoded}
 
In uncoded computation, dataset $\mathcal{D}$ is divided into $K$ non-overlapping equal-size subsets $D_{1},\ldots,D_{K}$, where $g_{k}$ denotes the  partial gradient corresponding to dataset $\mathcal{D}_{K},k\in[K]$. To tolerate straggling workers more than one partial gradient is assigned to each worker according to a certain order. Hence, uncoded computation is defined by a partial gradient assignment and order of computation. Let $\mathbf{M}$ be the assignment matrix for the partial gradients to workers, where $\mathbf{M}(j,i)=k$  means that the $k$th partial gradient $g_{k}$ is  computed by the $i$th worker in the $j$th order. This assignment can be random \cite{UCUT.2}, or according to a certain structure \cite{UCUT.4,CC.13}. In this paper, we consider the circular shifted assignment strategy, similar to the one used for GC:
\begin{equation}
\mathbf{M}(j,:)=\text{circshift }([1:N],-(j-1)).
\end{equation}
For instance, for $K=10$ and $r=4$, we have:
\[
\mathbf{M}=
  \begin{bmatrix}
    1 & 2 & 3 & 4 & 5 & 6 & 7 & 8 & 9 & 10\\
    2 & 3 & 4 & 5 & 6 & 7 & 8 & 9 & 10 & 1 \\
    3 & 4 & 5 & 6 & 7 & 8 & 9 & 10 & 1 & 2  \\
    4 & 5 & 6 & 7 & 8 & 9 & 10 & 1 & 2 & 3\\
  \end{bmatrix}.
\]
We highlight that, uncoded computation is actually a special case of the GC with MMC scheme, with message order $m=1$. We remark that the necessary condition to obtain the full gradient, with GC and its multi-message variations, is that each partial gradient is computed by at least one worker. It is easy to see that, uncoded computation will always outperform  GC if we only consider the computation time. Therefore, the main advantage of the GC scheme is to reduce the communication overhead.\\ 
\indent Although  we limit our focus to full gradient recovery in this paper, a partial gradient can be also used to update the parameter vector at each iteration \cite{UCUT.1}. We will show in Section \ref{s:numerical_results} that significant gains can be obtained in both computation time and communication load by ignoring only 5\% of the partial gradients. Lastly, we note that, under the assumption of independent and identically distributed (i.i.d) delays over time and over workers, the obtained partial gradient will be an {\em unbiased estimate} of the full gradient as in the stochastic gradient descent (SGD) approach.
\section{Per Iteration Completion Time Statistics}\label{sec:model}
In this section, we analyze the statistics of  per iteration completion time $T$ for the DGD schemes introduced above. For the analysis we consider a setup with $K$ workers, and  assume that the dataset is also divided into $K$ subsets. For the straggling behavior, we adopt the model in \cite{CC.1} and \cite{CC.2}, and assume that the probability of completing $s$ computations at any server, performing $s$ identical matrix-vector multiplications, by time $t$ is given by
\begin{equation}\label{dist}
F_{s}(t)\defeq
    \begin{cases}
     1-e^{-\mu(\frac{t}{s}-\alpha)}, &  \text{if } t\geq s\alpha, \\
      0,   &  \text{otherwise}. 
    \end{cases}
\end{equation}
The statistical model considered above is a shifted exponential distribution, such that the duration of a computation cannot be less than $\alpha$. We also note that, although the overall computation time at a particular worker has an exponential distribution, the duration of  each computation is assumed to be identical. Let $P_{s}(t)$ denote the probability of completing exactly $s$  computations by time $t$. We have
\begin{equation}
F_{s}(t)=\sum_{s^{\prime}=s}^{r}P_{s^{\prime}}(t),\label{corr}
\end{equation}
where $P_{r}(t)=F_{r}(t)$, since there are a total of $r$ computations assigned to each worker. One can observe from (\ref{corr}) that  $P_{s}(t)=F_{s}(t)-F_{s+1}(t)$, and it can be written as follows:
\begin{equation}
P_{s}(t)= 
    \begin{cases}
     0, &  \text{if } t<s\alpha, \\
     1-  e^{-\mu(\frac{t}{s}-\alpha)} ,  & s\alpha \leq t <(s+1) \alpha,\\
           e^{-\mu(\frac{t}{s+1}-\alpha)}-e^{-\mu(\frac{t}{s}-\alpha)},   &(s+1)\alpha<t.
     \end{cases}
\end{equation} 

We divide the workers into $r+1$ groups according to the number of computations completed by time $t$. Let $N_{s}(t)$ be the number of workers that have completed  exactly  $s$  computations by time $t$, $s = 0, \ldots, r$, and define $\mathbf{N}(t) \triangleq (N_{0}(t),\ldots,N_{r}(t))$, where $\sum_{s=0}^{r}N_{s}(t)=K$. The  probability of a particular realization is given by
\begin{equation}
\mathrm{Pr}(\mathbf{N}(t))=\prod_{s=0}^{r} P_{s}(t)^{N_{s}}{K-\sum_{j<s}N_{j}\choose N_{s}}.
\end{equation}
At this point, we introduce $M(t)$, which denotes the total number of computations completed by all the workers by time $t$, i.e., $M(t)\defeq\sum_{s=1}^{r}s \times N_{s}(t)$, and let $M_{th}$  denote the threshold for obtaining the full gradient. Hence, the probability of recovering the full gradient at PS
by time $t$, $\mathrm{Pr}(T \leq t)$, is given by  $\mathrm{Pr}(M(t) \geq M_{th})$. Consequently, we have
\begin{equation}\label{stat1}
\mathrm{Pr}(T \leq t)=\sum_{\mathbf{N}(t):M(t)\geq M_{th}} \mathrm{Pr}(\mathbf{N}(t)),
\end{equation}
and
\begin{align}
E[T] & = \int_0^\infty \mathrm{Pr}(T > t) dt\\ 
& =\int_0^\infty \left[1 - \sum_{\mathbf{N}(t):M(t)\geq M_{th}} \mathrm{Pr}(\mathbf{N}(t)) \right] dt.
\end{align}

Per iteration completion time statistics of non-straggler threshold based schemes can be derived similarly. For a given non-straggler threshold $K_{th}$, and per server computation load $r$, we can have
\begin{equation}\label{stat2}
\mathrm{Pr}(T \leq t)=\sum_{k=K_{th}}^{K} {K \choose k}(1-e^{-\mu(\frac{t}{r}-\alpha)})^{k}(e^{-\mu(\frac{t}{r}-\alpha)})^{K-k},
\end{equation}
when $t\geq r \alpha$, and $0$ otherwise.
\begin{figure*}
    \centering
         \begin{subfigure}[b]{0.47\textwidth}
        \includegraphics[scale=0.5]{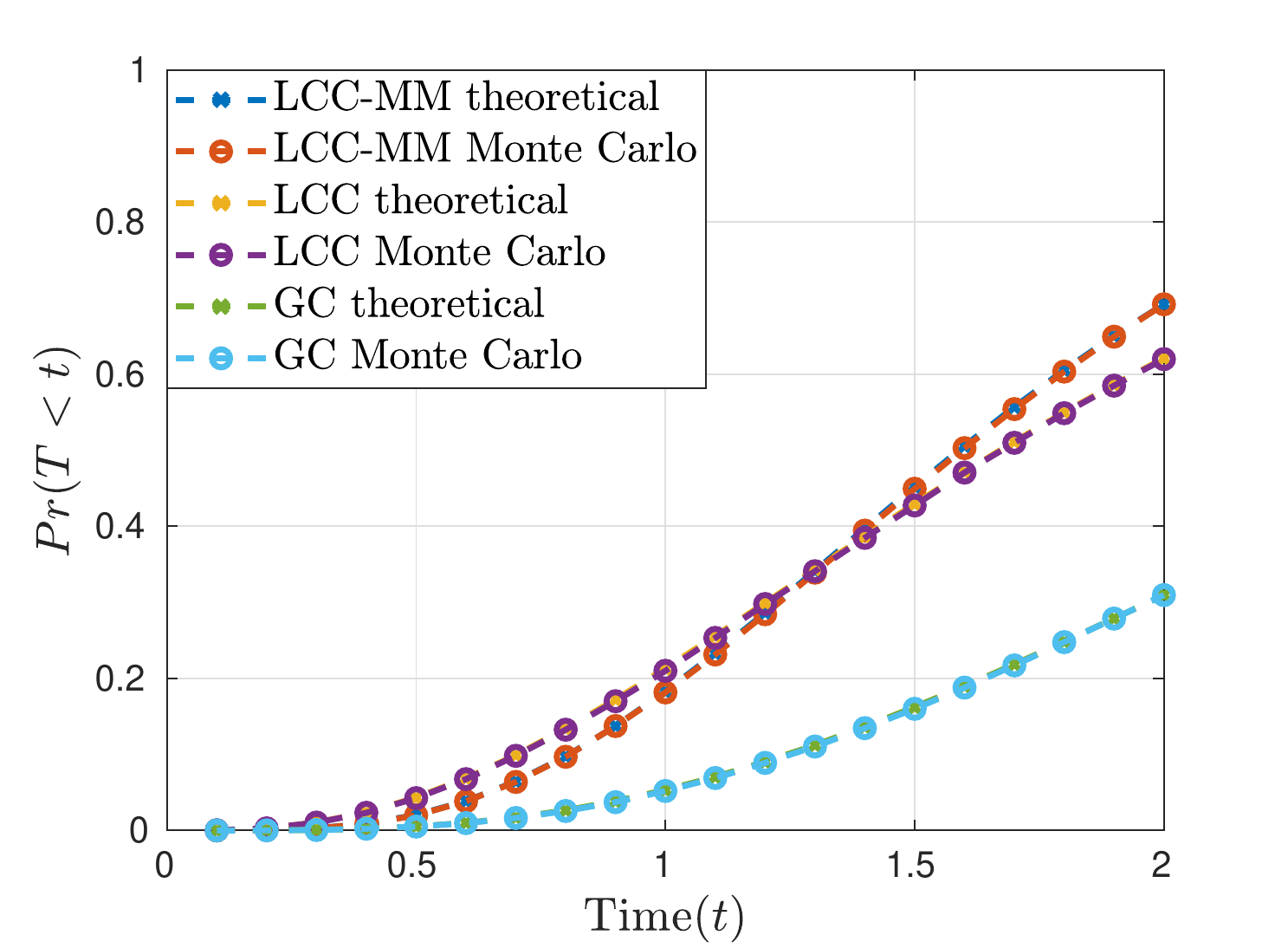}
        \caption{$K=6$, $r=3$}
				\label{N6}
    \end{subfigure}
    \begin{subfigure}[b]{0.47\textwidth}
        \includegraphics[scale=0.5]{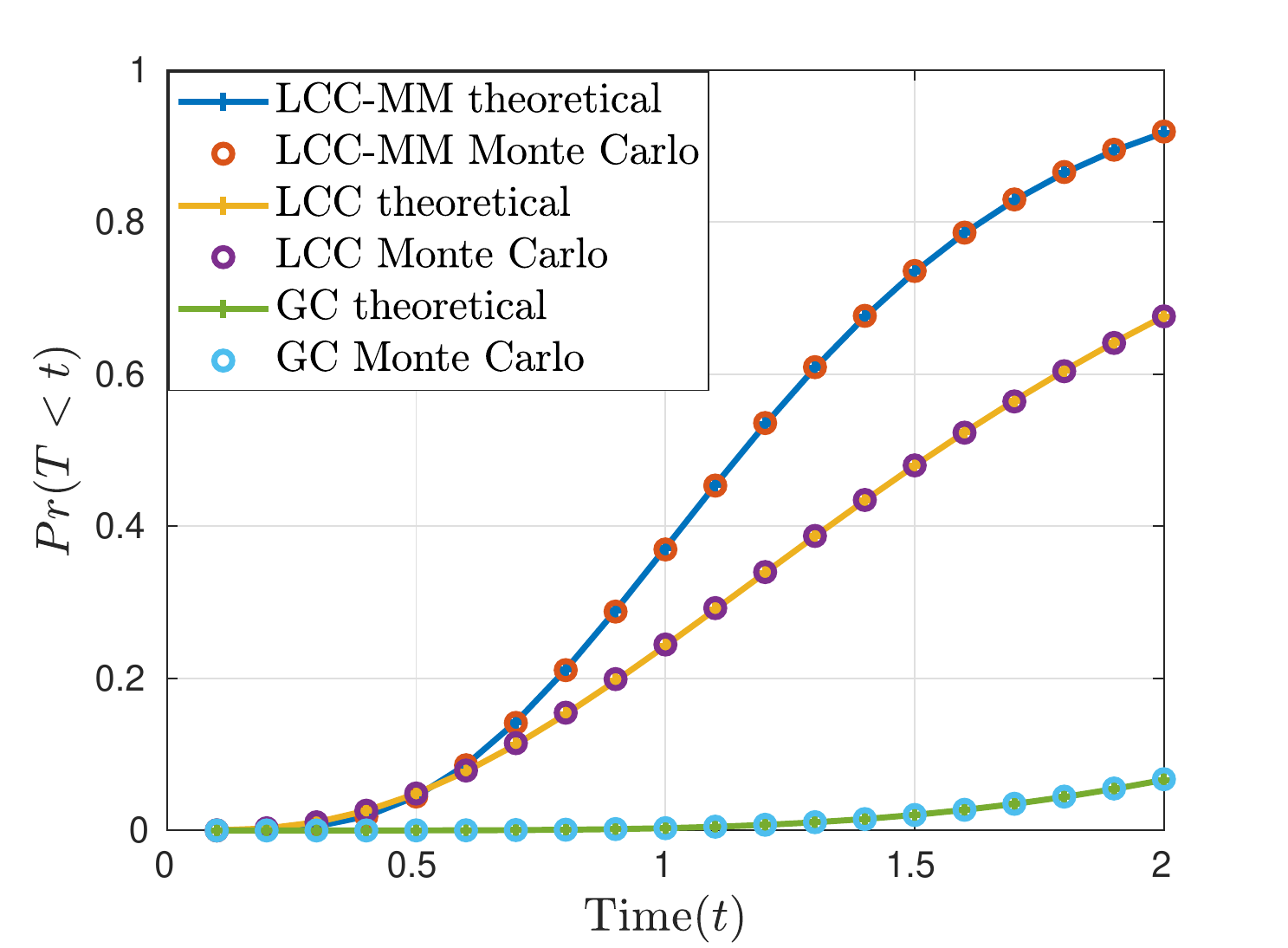}
        \caption{$N=K$, $r=5$}
				\label{N10}
        \end{subfigure}
				\caption{Per iteration completion time statistics. }
		\label{res1}
\end{figure*}
\section{Numerical Results and Discussions}\label{s:numerical_results}
For the numerical results, we consider three different simulation setups, namely model based, data driven and real time implementation. In the first setup, we use the shifted exponential distribution model for the computation time statistics to analyze the average completion time. For the second setup, we initialize 21 Amazon EC2 instances (where the first instance is considered as the PS), then for each EC2 instance we measure the computation time of a certain job as well as communication time with the parameter server, over different times of the day, to form a dataset to analyze the average completion time statistics. Finally, in the third set of simulations, we conduct a real time experiment via implementing a linear regression problem on Amazon EC2 instances through 1000 iterations to monitor the average completion time statistics.
\subsection{Model Based Analysis}
We first verify the correctness of the expressions provided for the per iteration completion time statistics in (\ref{stat1}) and  (\ref{stat2}) through Monte Carlo simulations  generating 100000 independent realizations. Then, we will show that the MMC approach can reduce the average per-iteration completion time, $E[T]$, significantly. In particular, we analyze the per iteration completion time of three different DGD schemes,  GC, LCC,  and LCC with MMC (LCC-MM). For the simulations we consider two different settings, $K=6$, $r=3$ and $K=10$, $r=5$, respectively, and  use the cumulative density function (CDF) in (\ref{dist}) with parameters $\mu=10$ and $\alpha=0.01$ for the completion time statistics.\\
\indent  In Fig. \ref{res1} we plot the CDF of the per iteration completion time $T$ for GC, LCC,  and LCC-MM schemes according to the closed form expressions derived in Section \ref{sec:model} and Monte Carlo simulations. We observe from Fig. \ref{res1} that the two match perfectly.  We also observe that, although the LCC-MM and LCC schemes perform closely in the first scenario (Fig. \ref{res1}-(a)), LCC-MM outperforms the LCC scheme in the second scenario (Fig. \ref{res1}-(b)). This is because, as the computation load $r$ increases, it takes more time for even the fast workers to complete all the assigned computations, which results in a higher number of non-persistent stragglers. Hence, the performance gap between LCC-MM and LCC increases with $r$. Similarly, as expected, since the non-straggling threshold of GC does not scale with $K$, we observe that GC performs better for small $r$ when the $K/r$ ratio is preserved.\\ 
\begin{figure*}
         \begin{subfigure}{0.5\textwidth}
        \includegraphics[scale=0.5]{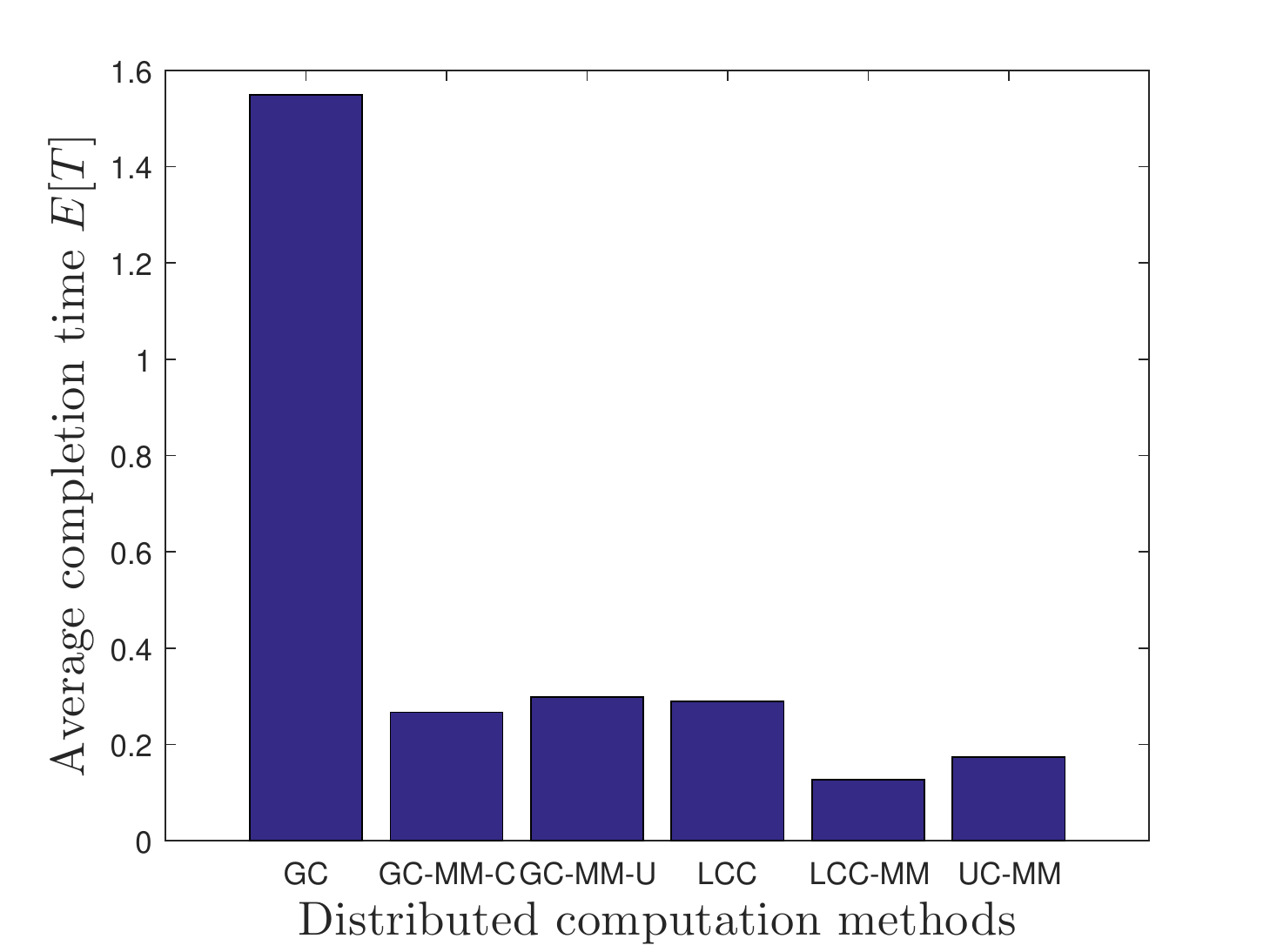}
        \caption{Average completion time performance}
				\label{comp_time}
    \end{subfigure}
    \begin{subfigure}{0.5\textwidth}
        \includegraphics[scale=0.5]{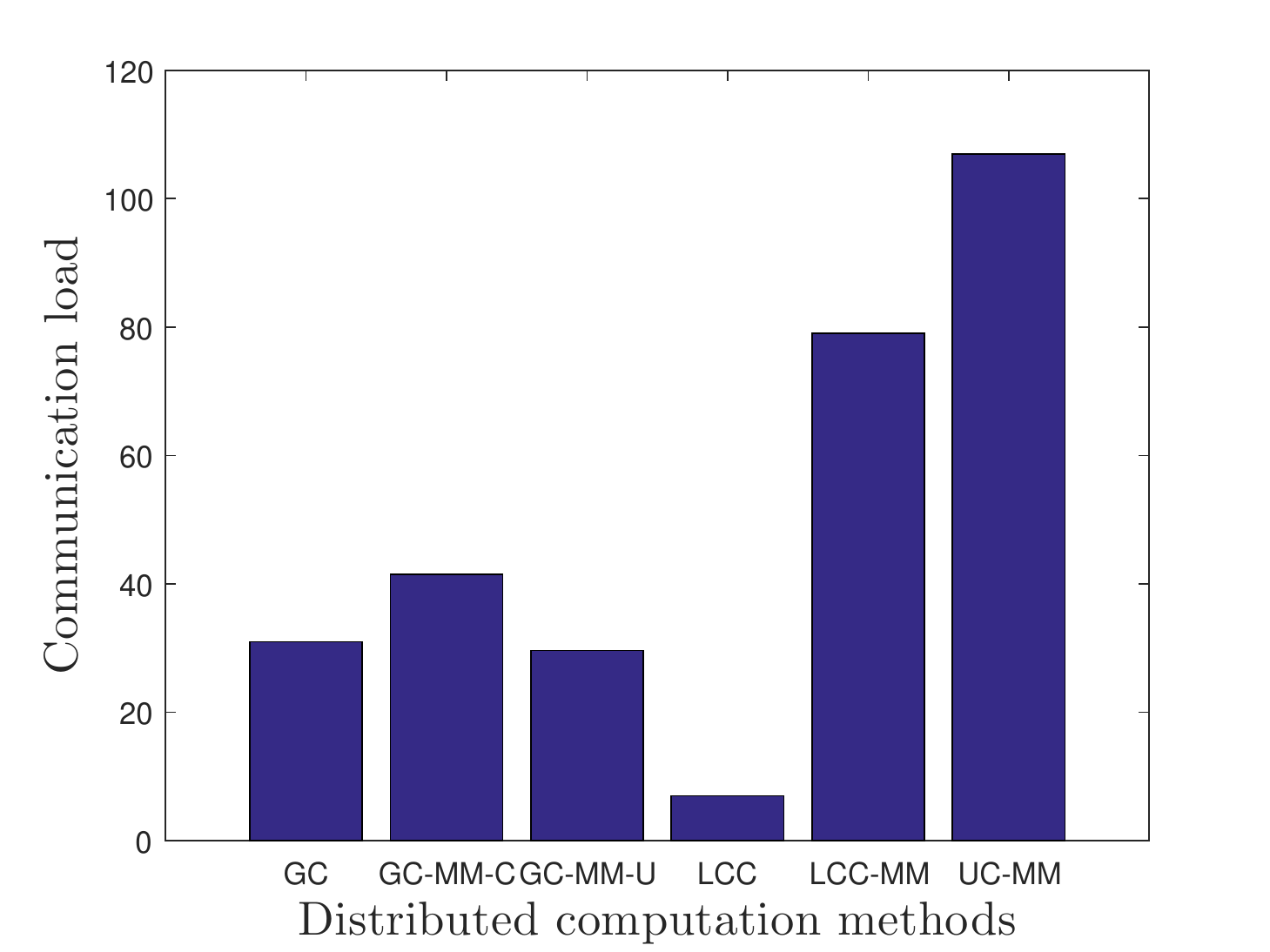}
        \caption{Communication load performance}
				\label{comm_time}
        \end{subfigure}
				\caption{Per iteration completion time and communication load statistics.}
		\label{sim1}
\end{figure*}
\begin{figure*}
         \begin{subfigure}{0.5\textwidth}
        \includegraphics[scale=0.5]{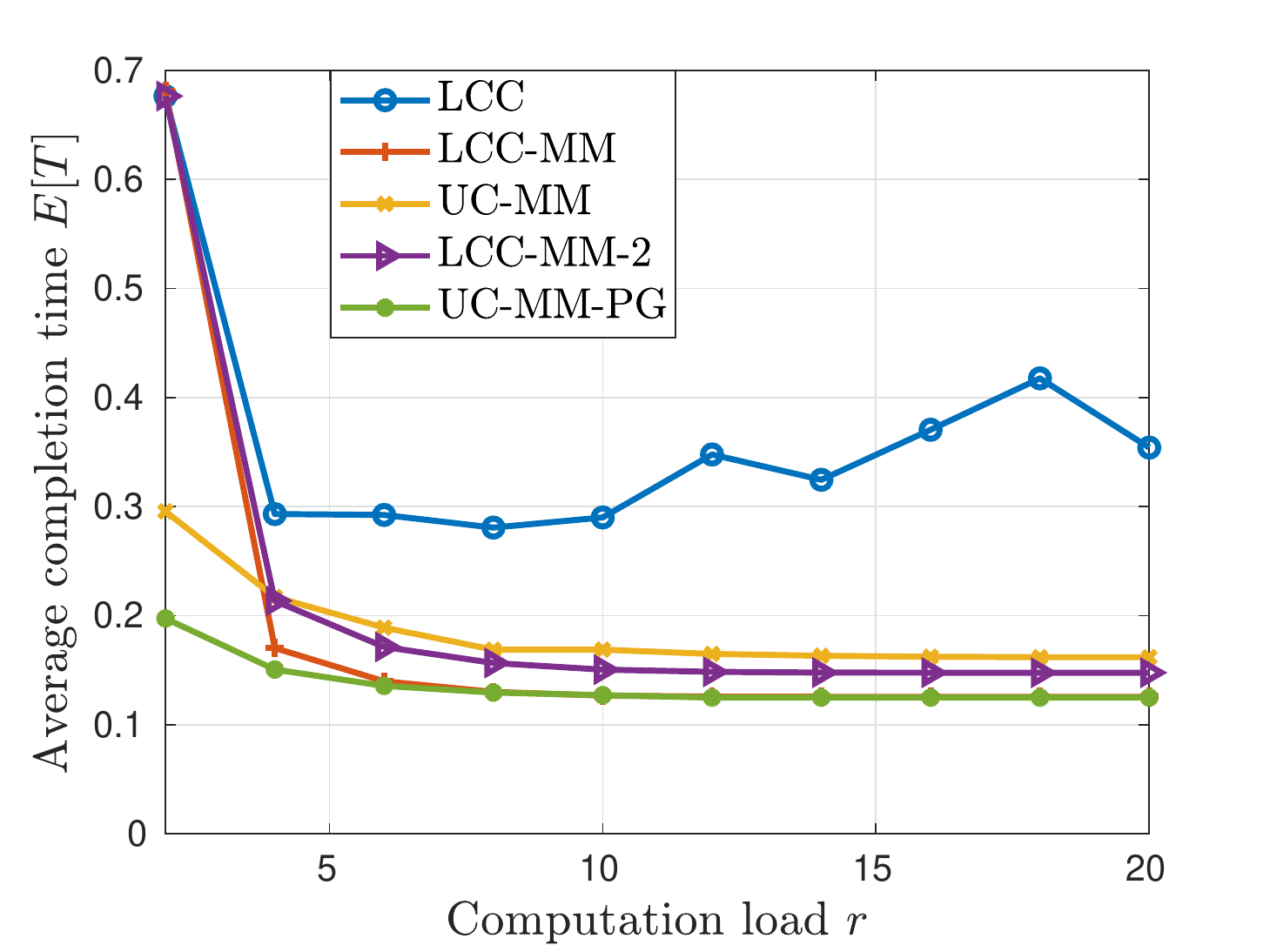}
        \caption{Average completion vs. computation load}
				\label{comp}
    \end{subfigure}
    \begin{subfigure}{0.5\textwidth}
        \includegraphics[scale=0.5]{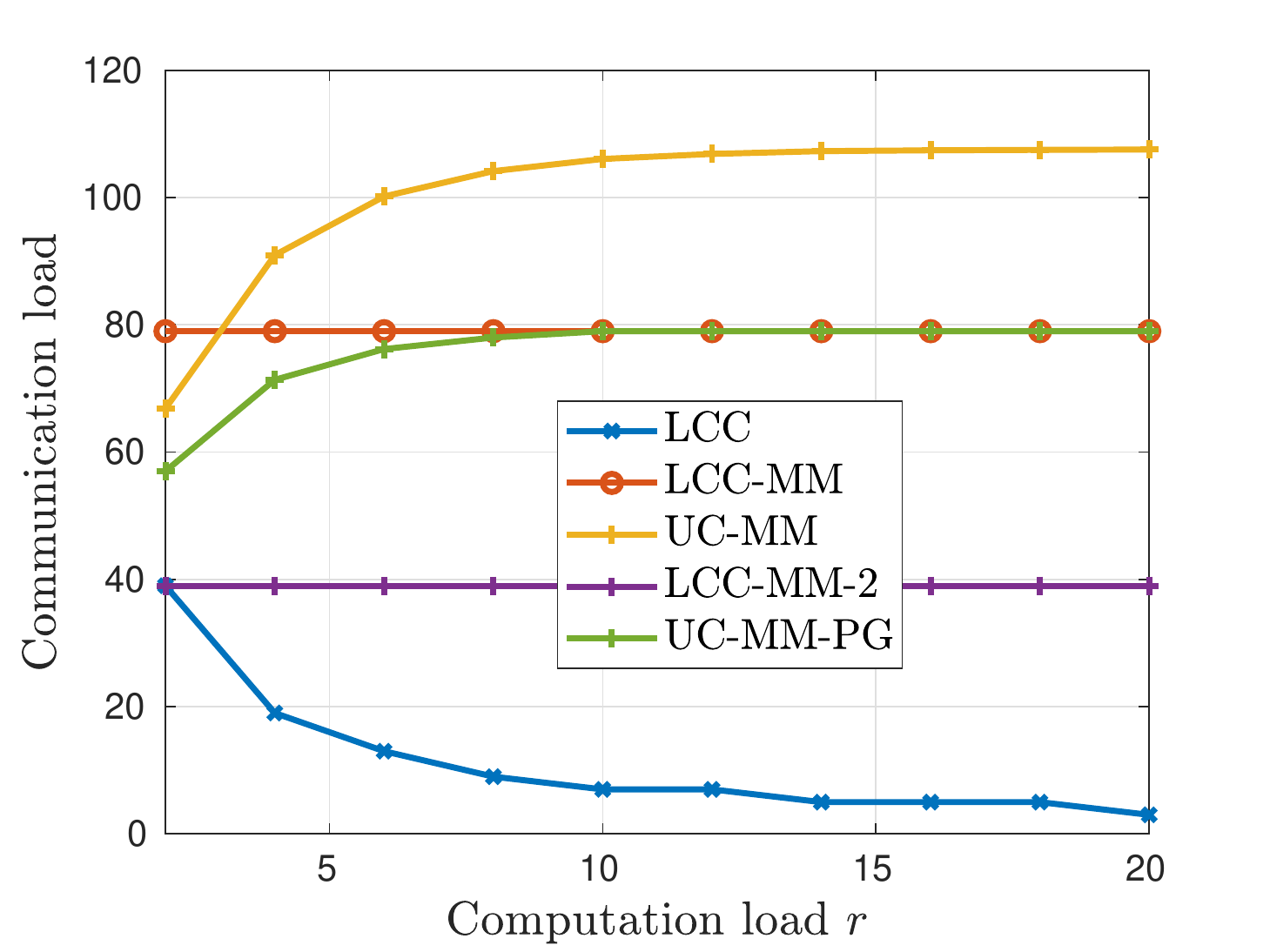}
        \caption{Communication load vs. computation load.}
				\label{comm}
        \end{subfigure}
				\caption{Per iteration completion time and communication load statistics.}
		\label{sim2}
\end{figure*}
\indent Next, we consider the setup from \cite{CC.4}, where $K=40$ workers are employed for DGD with a computation load of $r=10$, and analyze the performance of six different DGD schemes, namely, GC, GC with MMC and uncorrelated design (GC-MM-U), GC with MMC and correlated design (GC-MM-C), LCC, LCC-MM and uncoded computation with MMC (UC-MM). For the design of GC-MM-U, we divide the workers into four equal-size clusters, and we use  uncorrelated code structure with order vector $\mathbf{m}=[6, 8, 10]$, so that each worker can send up to 3 coded partial gradients. Similarly, for the design of GC-MM-C, we again divide the workers into four equal-size clusters and use the correlated code structure with order $m=6$, so that each worker can send up to 5 coded partial gradients. For the computation time statistics, we use the distribution in (\ref{dist}) with parameters $\mu=10$ and $\alpha=0.01$. In the performance analysis, we consider both  the average per iteration completion time $E[T]$ and the communication load, measured by the average total number of transmissions from the workers to the PS, and the results obtained from $100000$ Monte Carlo realizations are illustrated in Fig. \ref{sim1}.  We observe that  LCC-MM approach can provide approximately $50\%$ reduction in the average completion time compared to  LCC, and more than $90\%$ reduction compared to GC. A more interesting result is that the UC-MM scheme outperforms both LCC and GC. This result is especially important since UC-MM has no decoding complexity at the PS. Hence, when the decoding time of PS is also included in the average per iteration completion time this improvement will be even more significant. We also observe that LCC-MM scheme achieves the minimum average completion time. However, Fig. \ref{sim1}(b) highlights that the MMC schemes, particularly LCC-MM and UC-MM, induce much higher communication load compared to the conventional single message schemes. The results illustrated in Fig. \ref{sim1} also show that the multi-message variations of GC can perform as well as LCC in terms of the average per-iteration completion time, while inducing much lower communication overhead compared to the LCC-MM and UC-MM schemes.\\
\indent Finally, on the same setup, we analyze the performance of the DGD schemes with respect to the computation load $r$, and compute both the average per iteration completion time $E[T]$ and the communication load for ten different $r$ values, i.e., $r=2,4,\ldots,20$. In Section \ref{sec:introduction}, we identified two main drawbacks of the single-message coded computation schemes; namely, over-computation and under-utilization. In Fig. \ref{sim2}(a), these drawbacks are explicitly demonstrated. One can observe that after a certain point, the average completion time of LCC starts to increase with  $r$, which reflects over-computation. The gap between the LCC and LCC-MM highlights under-utilization of the computation resources.\\
\indent From Fig. \ref{sim2}(a), we observe that the UC-MM scheme consistently outperforms LCC for all the computation load values. More interestingly, UC-MM performs very close to LCC-MM, and for a  small $r$, such as $r=2$, it can even outperform LCC-MM. Hence, in terms of the computation time UC-MM can be considered as a better option compared to LCC especially when $r$ is low.\\
\indent On the other hand, in Fig. \ref{sim2}(b) we observe that, in terms of the communication load the best scheme is LCC, while the UC-MM introduces the highest communication load. We also  observe that the communication load of  LCC-MM remains constant with $r$, whereas that of the LCC (UC-MM) scheme monotonically decreases (increases) with $r$. Accordingly,  the communication load of the LCC and UC-MM schemes are closest at $r=2$. Hence, from both  
Fig. \ref{sim2}(a) and Fig. \ref{sim2}(b) we can conclude that when $r$ is low, UC-MM might be a better option compared to  LCC taking into account the computation time, the communication load and decoding complexity together. We also want to underline the fact that although LCC-MM achieves a lower average completion time, MMC increases the communication load as well as the decoding complexity. 
\begin{remark}
An important aspect of the average per-iteration completion time that is ignored here, and by other works in the literature, is the decoding complexity at the PS. Among these three schemes, UC-MM has the lowest decoding complexity, while LCC-MM has the highest. However, as discussed in Section \ref{sec:codedcomp}, the number of transmissions as well as the decoding complexity  can be reduced via increasing the number of polynomials used in the decoding process. To illustrate this, we consider a different implementation of the LCC-MM scheme, where two polynomials are used, denoted by LCC-MM-2. In this scheme, for given $r$, coded inputs correspond to  evaluation of two polynomials, each of degree $N-2$, at $r/2$ different points. Each worker sends a partial result to the PS after execution of two computations, which correspond to the evaluation of these two polynomials at the same point. Since two polynomials are used, the number of transmissions is reduced by approximately half compared to LCC-MM as illustrated in Fig. \ref{sim2}(b). A noticeable improvement is achieved in the communication load, at the expense of a relatively small increase in the average per iteration completion time as illustrated in Fig. \ref{sim2}(a).
\end{remark}
\indent Another important advantage of the UC-MM scheme is its applicability to partial gradient scenario. The objective of  all the straggler avoidance schemes explained in this paper is to recover the full gradient at the PS. Accordingly, with UC-MM, the PS waits until it receives all $K$ partial gradients to terminate the iteration. However, to reduce the computation time PS may terminate an iteration after receiving only $\tilde{K}<K$ partial gradients out of $K$ \cite{UCUT.1}. We refer to this variation of UC-MM scheme as UC-MM-PG. For the UC-MM-PG scheme, the key design parameter is the tolerance rate $\frac{K-\tilde{K}}{K}$ and for our analysis we set the tolerance rate to $5\%$. The results in Fig. \ref{sim2}(a) show that when $r$ is  small, UC-MMC-PG  can reduce the average completion time up to $70\%$ compared to LCC, and up to $33\%$ compared to UC-MMC; while only 2 out of 40 gradient values are missing at each iteration. In addition to an improvement in the average completion time, the UC-MMC-PG scheme can also reduce the communication load as shown in Fig. \ref{sim2}(b). We remark that, in partial gradient approach the estimated gradient, due to missing partial gradients, is not the original gradient but an estimate of it. Although, each update is less accurate compared to  full-gradient updates, since the parameter vector is updated over many iterations, partial gradient approach may converge to the optimal value faster than the full-gradient approach. Indeed, stochastic gradient descent is an extreme case of this partial gradient approach, and is commonly used in practise. Moreover, tolerance rate can be dynamically updated through iterations to achieve better convergence results.

\subsection{Data Driven Simulations}
In this setup, we initialized 21 Amazon EC2 t2.micro instances, where the first one is labeled as the parameter server. We use the MPI protocol, particularly mpi4py \cite{mpi}, to establish connections between instances. For the computation, we consider a matrix-vector multiplication with sizes $3000\times 3000$ and $3000\times1$, respectively, which is the core computation task for GD in a linear regression problem  assuming that the whole dataset is divided into 20 subsets each containing $3000$ data points and each data point is a vector of $3000$ parameters. We measure the computation time using time.time() command before and after each computation. For  message passing we use non-blocking communication with  Issend and Irecv commands for message sending and receiving, respectively. Further, we use wait() command to verify the time instant when the message is successfully received and again we use the time.time() command to measure the time.\\
\indent For  data collection, we do {point-to-point} analysis such that in each simulation we use only one instance and the parameter server.  The chosen instance performs the computation (the assigned matrix-vector multiplication) and sends the result to the PS, which is repeated after receiving  a new vector from the PS. In total, we form a measurement set of size $3000$ for both computation and communication latency for each node. These measurement sets are then used for our average per-iteration time analysis. We want to note that, in practise one of the predominant factors affecting the average completion time is the congestion at the PS due to the MPI protocol; however, this is very much dependent on the particular protocol used, and can be reduced or eliminated with more efficient communication protocol. For example, by employing a hierarchical framework with multiple PSs congestion issue can be resolved in large scale implementations. Hence, we  first analyze the average completion time ignoring the effects of  congestion. We refer to these simulations as {\em data driven}, which are based on the assumptions that the communication channels from workers to PS are orthogonal.\\
\begin{figure*}
    \begin{subfigure}{0.5\linewidth}
        \includegraphics[scale=0.5]{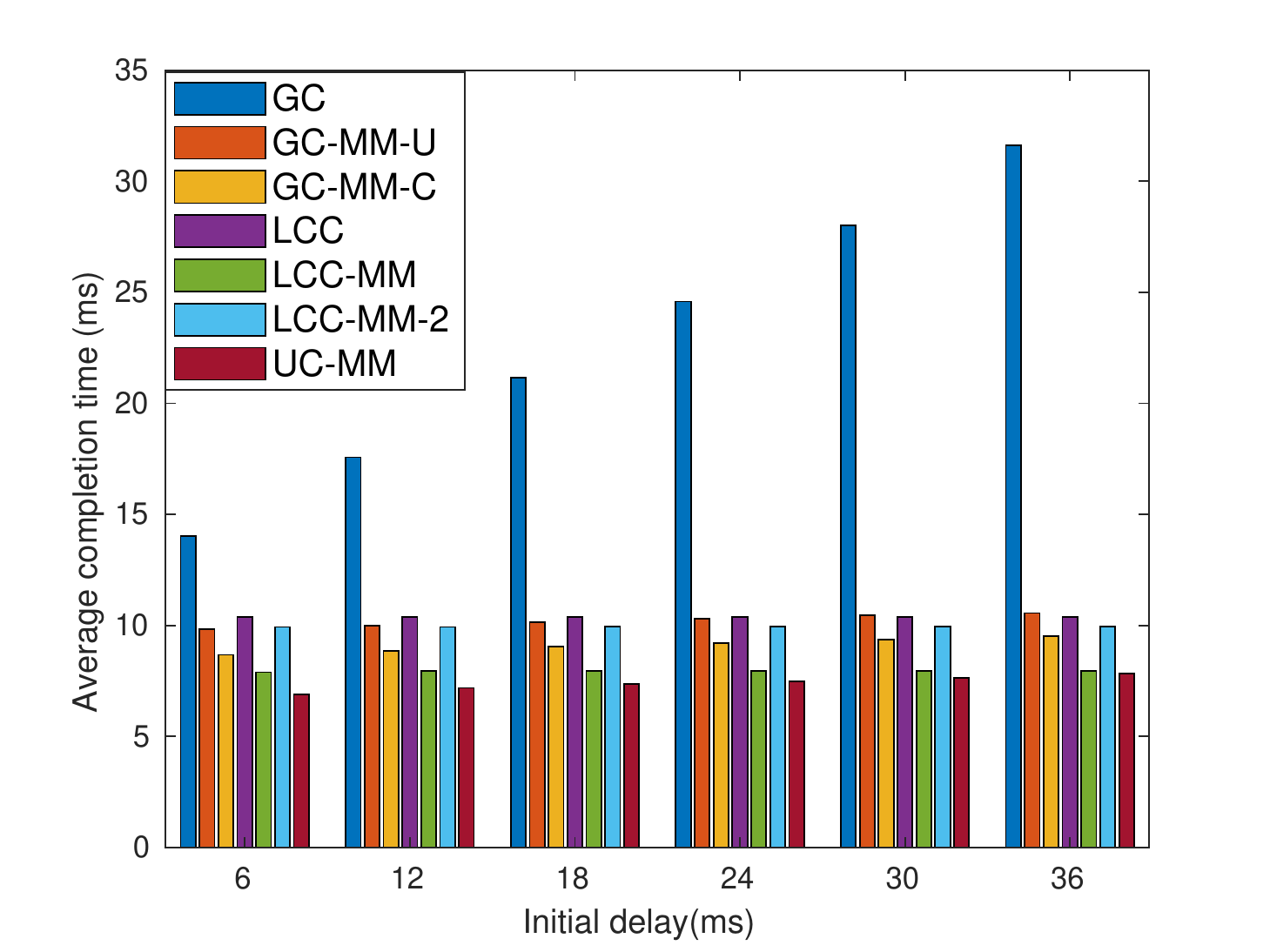}
            \subcaption{r=4 and p=0.5}
				\label{sc11}
    \end{subfigure}
    \smallskip
    \begin{subfigure}{0.5\linewidth}
        \includegraphics[scale=0.5]{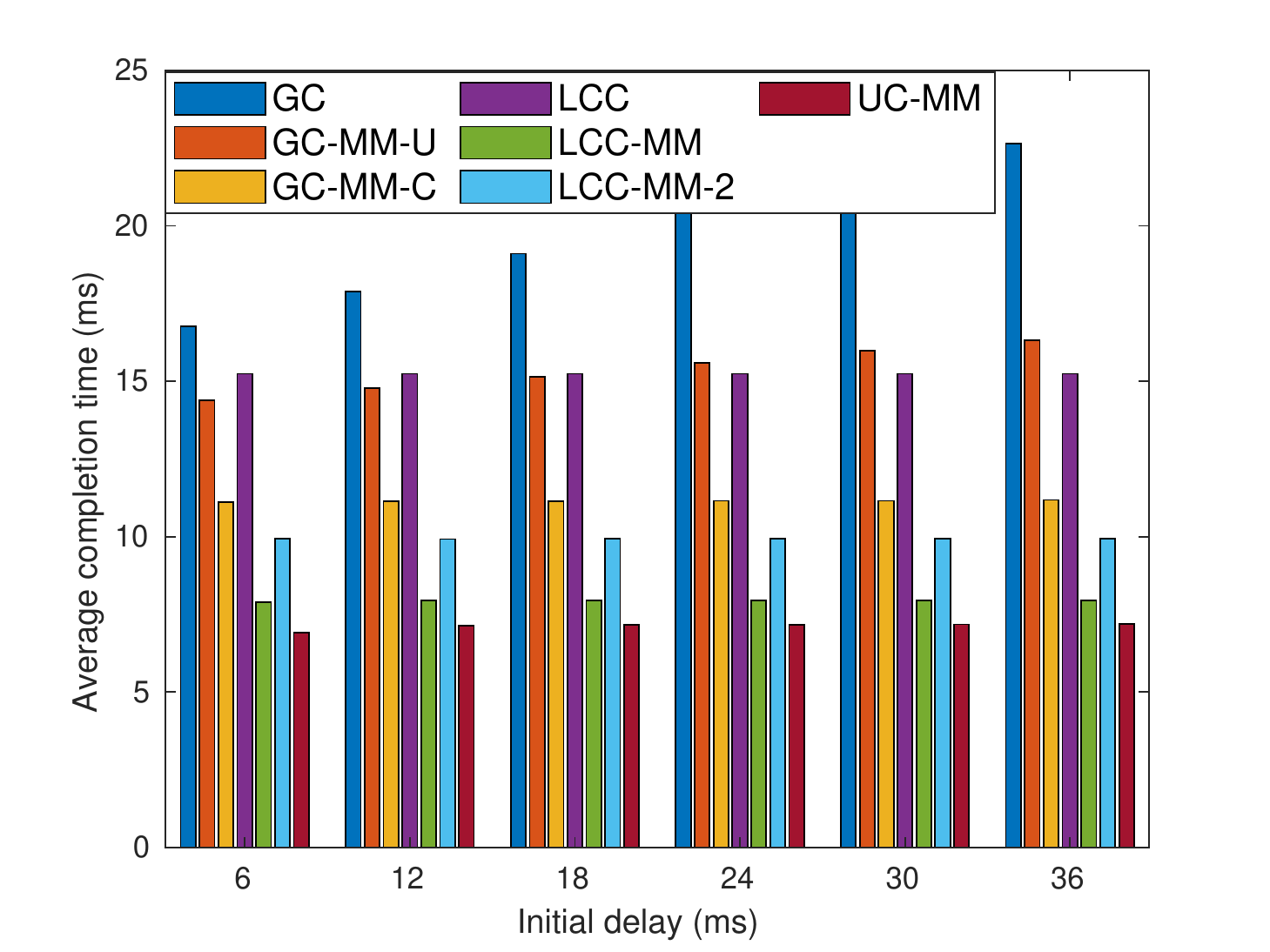}
            \subcaption{r=6 and p=0.2}
				\label{sc12}
    \end{subfigure}
    \smallskip
    \begin{subfigure}{0.5\linewidth}
        \includegraphics[scale=0.5]{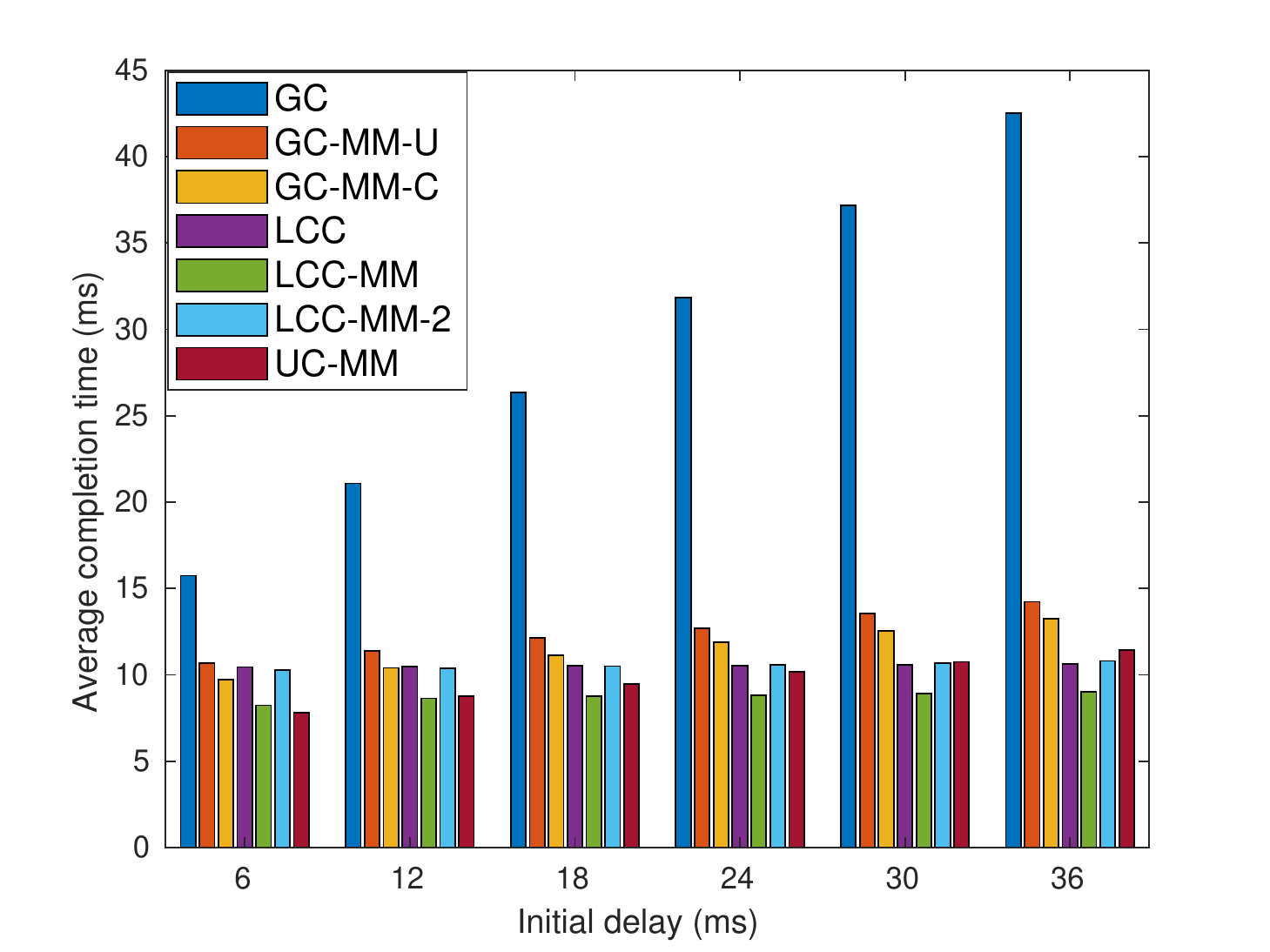}
            \subcaption{r=4 and p=0.3}
				\label{sc13}
    \end{subfigure}
    \smallskip
    \begin{subfigure}{0.5\linewidth}
        \includegraphics[scale=0.5]{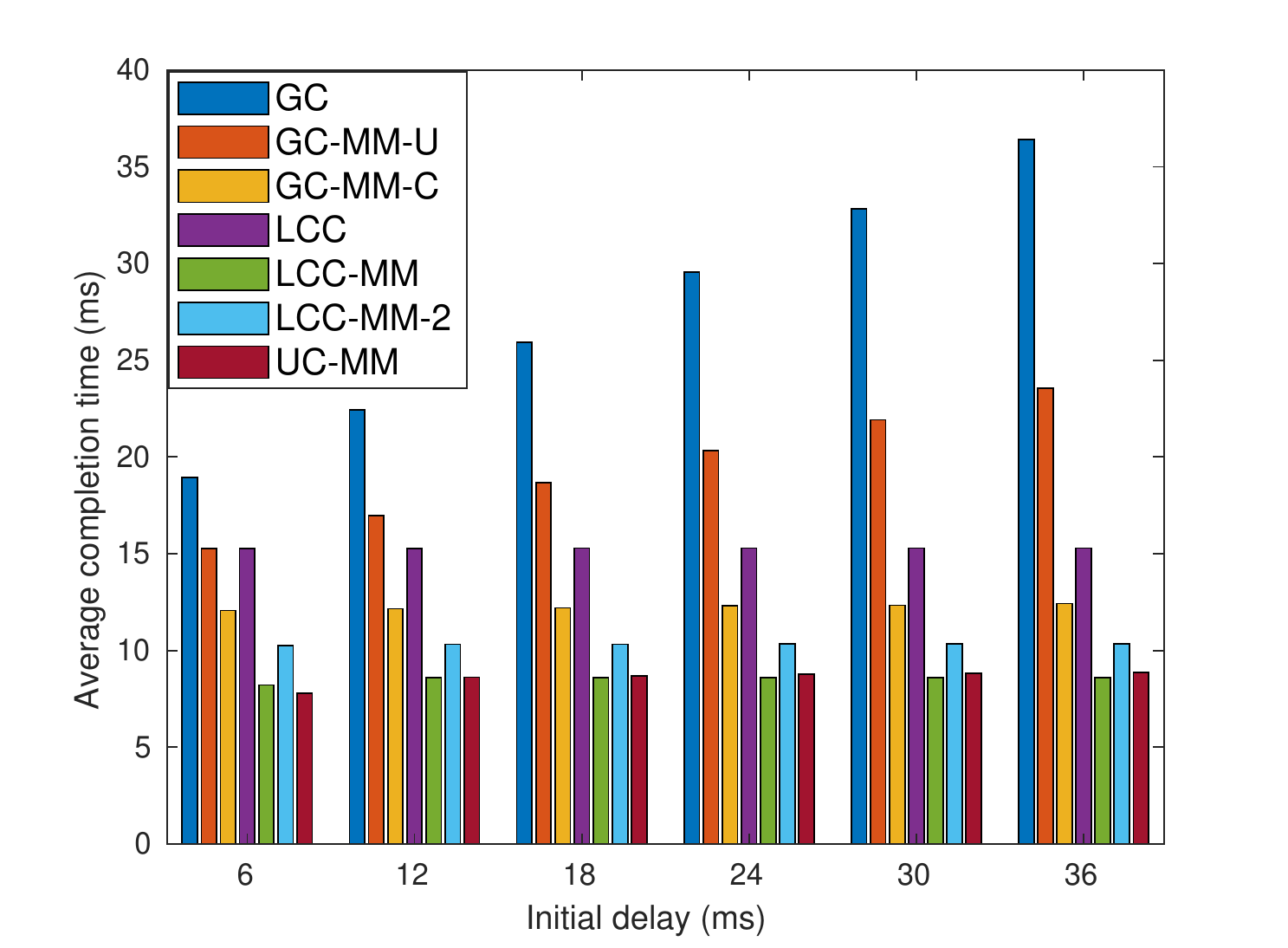}
            \subcaption{r=6 and p=0.3}
				\label{sc14}
    \end{subfigure}
    \smallskip
    \begin{subfigure}{0.5\linewidth}
        \includegraphics[scale=0.5]{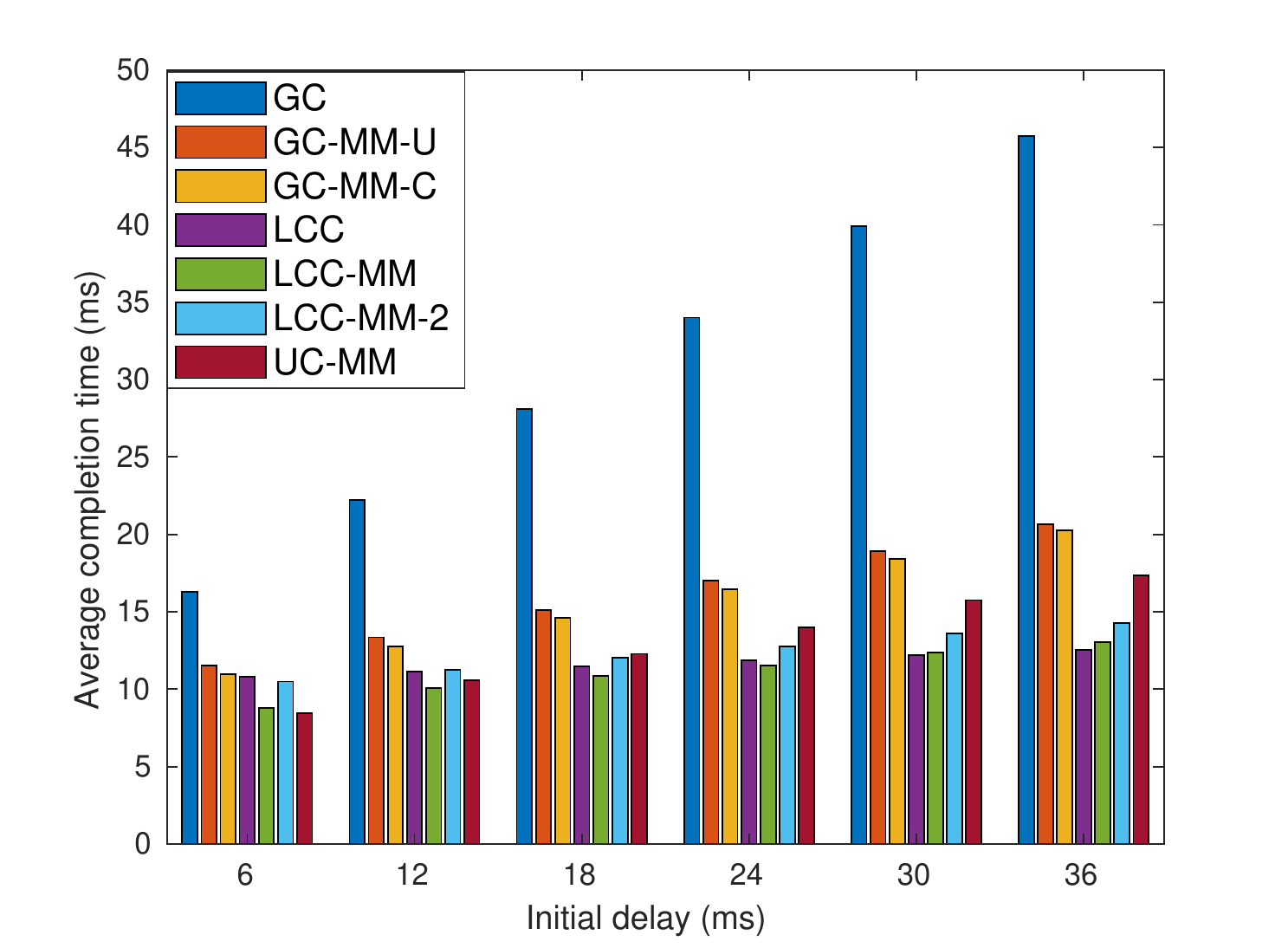}
            \subcaption{r=4 and p=0.4}
				\label{sc15}
    \end{subfigure}
    \begin{subfigure}{0.5\linewidth}
        \includegraphics[scale=0.5]{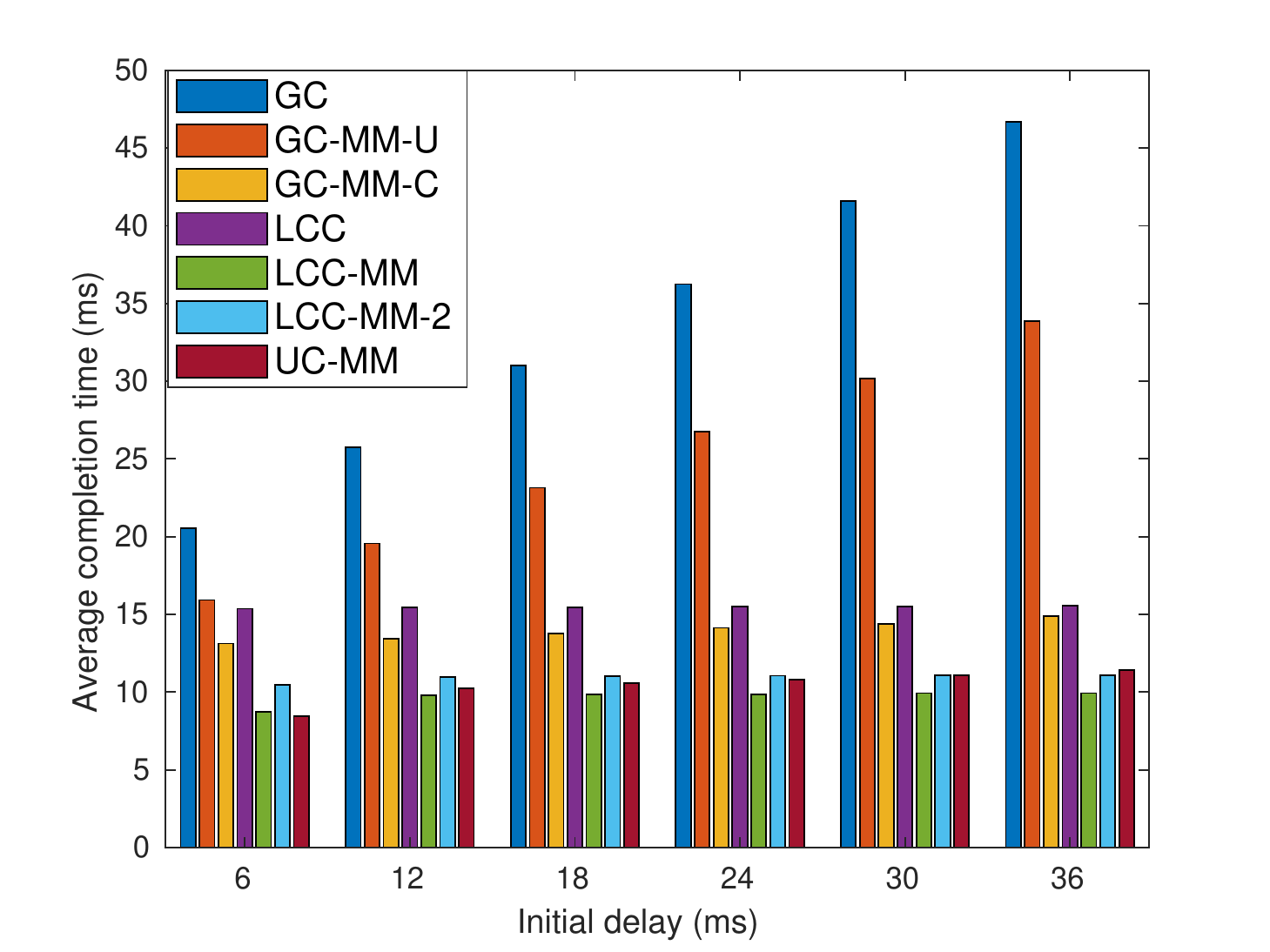}
            \subcaption{r=6 and p=0.4}
				\label{sc16}
				
    \end{subfigure}
    \caption{Average completion time analysis for GC, GC-MM-U, GC-MM-C, LCC LCC-MM, LCC-2 and UC-MM schemes with random fixed initial delay.}
    \label{sc1}
\end{figure*}
\indent We consider two different scenarios. In the first scenario we randomly delay the computation time of the instances for a fixed duration. In the second simulation, in addition to computational delay, we add exponentially distributed delay to the communication latency. 
\begin{figure*}[t]
         \begin{subfigure}{0.5\linewidth}
        \includegraphics[scale=0.5]{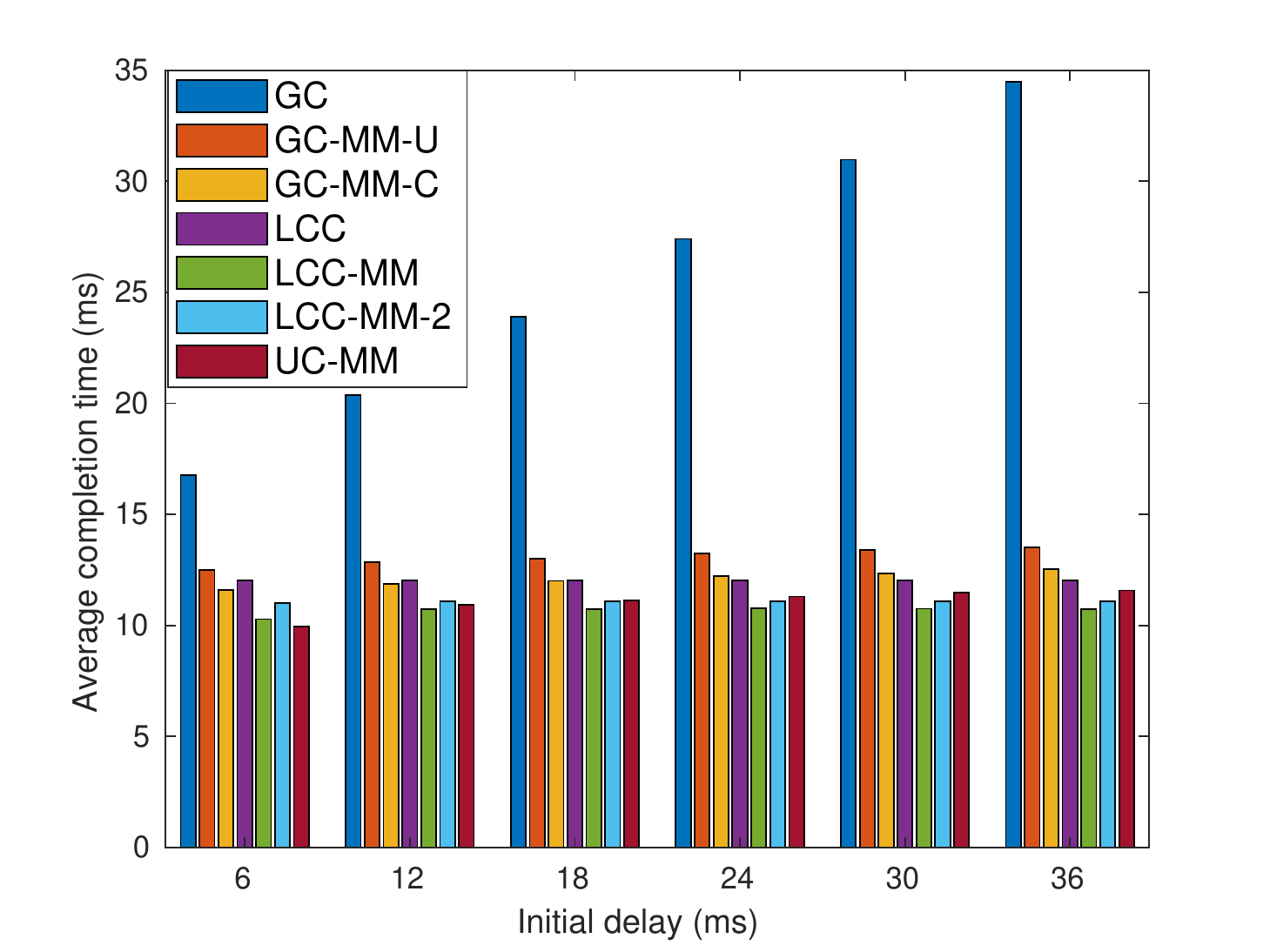}
        \caption{r=4, p=0.2 $\mu=2$}
				\label{sc21}
    \end{subfigure}
    \smallskip
    \begin{subfigure}{0.5\linewidth}
        \includegraphics[scale=0.5]{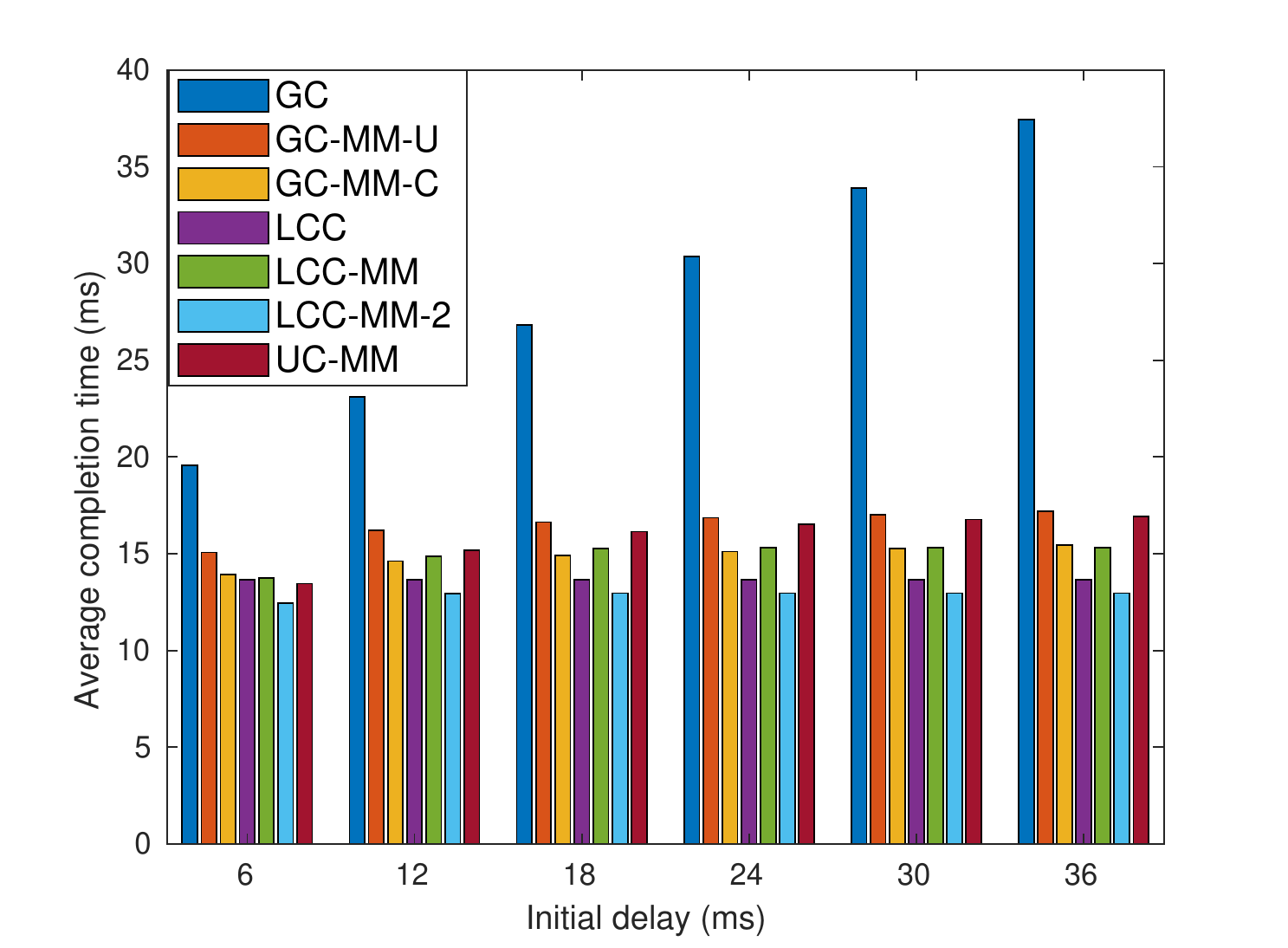}
        \caption{r=4, p=0.2 $\mu=4$}
				\label{sc22}
    \end{subfigure}
    \smallskip
    \begin{subfigure}{0.5\linewidth}
        \includegraphics[scale=0.5]{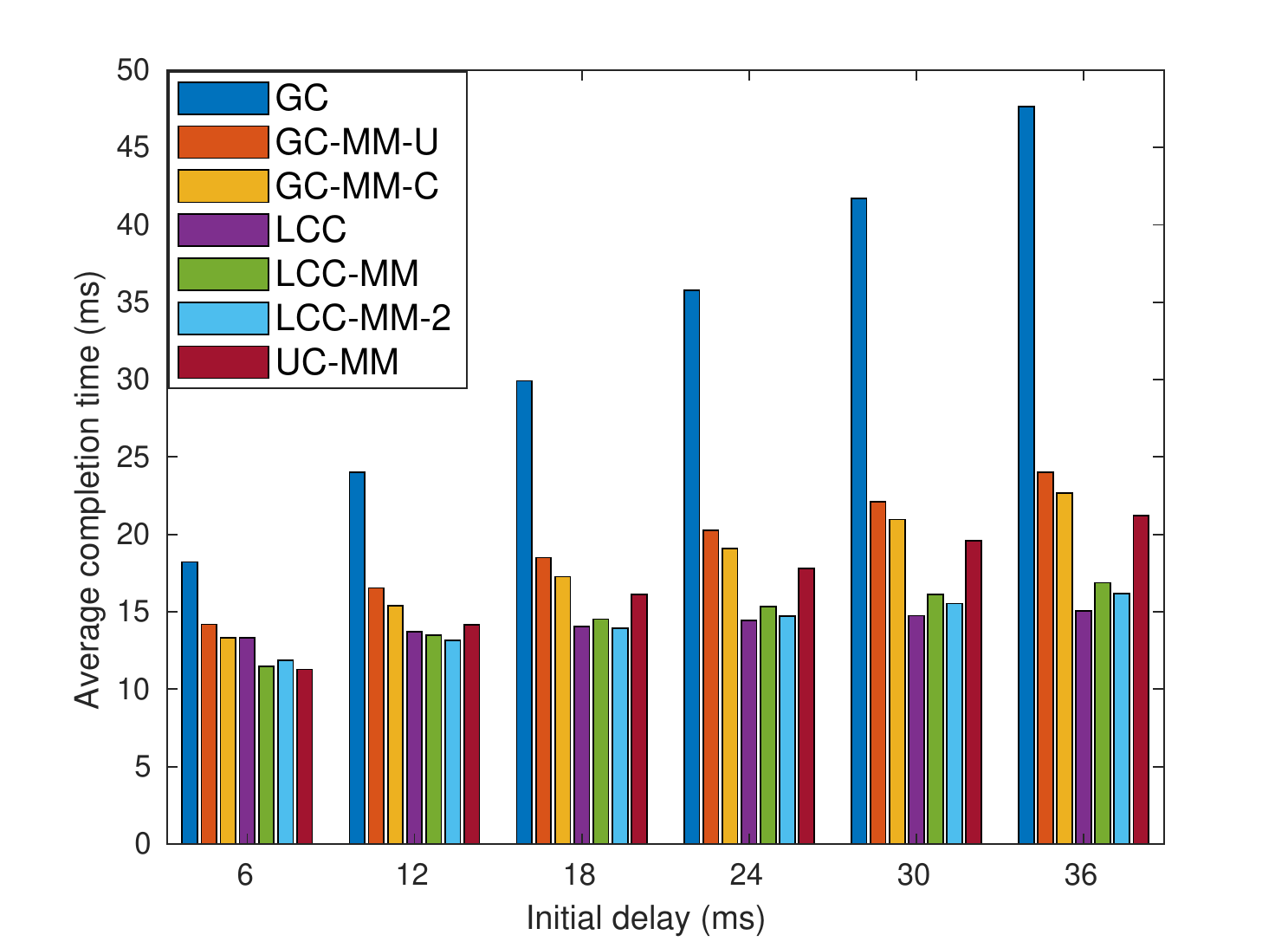}
        \caption{r=4, p=0.4 $\mu=2$}
				\label{sc23}
    \end{subfigure}
    \smallskip
    \begin{subfigure}{0.5\linewidth}
        \includegraphics[scale=0.5]{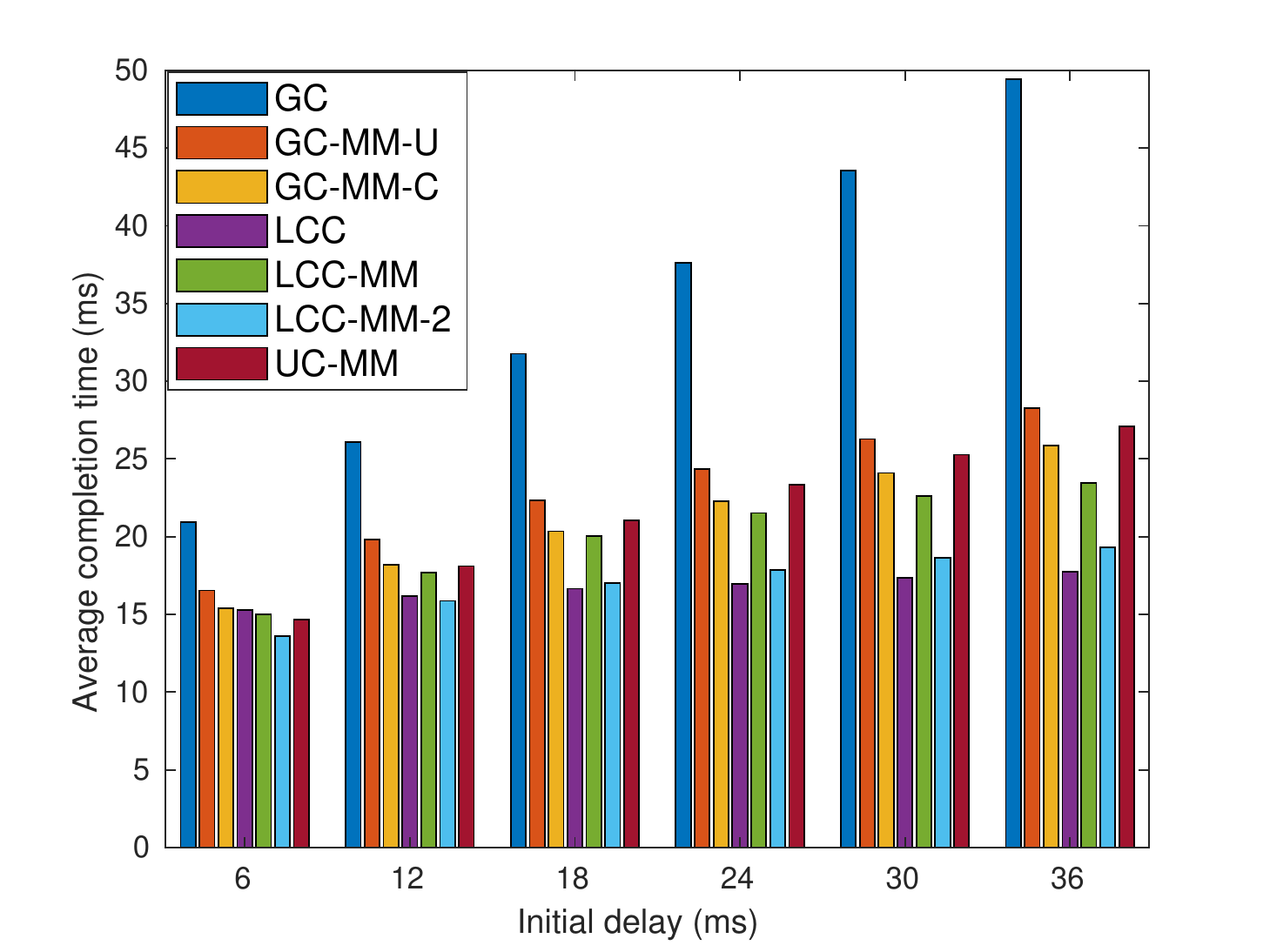}
        \caption{r=4, p=0.4 $\mu=4$}
				\label{sc24}
	 \end{subfigure}
	 			\caption{Average completion time analysis for GC, GC-MM-U, GC-MM-C, LCC LCC-MM, LCC-2 and UC-MM schemes with random fixed initial delay and exponential communication delay when $r=4$.}
		\label{sc2}
\end{figure*}
\subsubsection{Scenario 1}
We introduce the term {\em delay probability}, denoted by $p$, to refer to the probability of a machine to be delayed. This delay can be due to the computation process, as mostly argued in the literature, a possible access failure (connection lost), or the queuing delay due to congestion of computation tasks. For the simulations, we consider a fixed additional delay that comprise all aforementioned delays, which we refer as the {\em initial delay}. Fixed initial delay approach have been also used for simulations in \cite{CC.4,UCCT.1}.\\
\indent In our simulations, we consider failure probabilities 0.2, 0.3 and 0.4, and computational loads of $r=4$ and $r=6$. We use the GC-MM-C scheme with message order 3 (with cluster size of 5) and 4 (with cluster size of 10) when $r=4$ and $r=6$, respectively. Similarly, we use the GC-MM-U scheme with  message order vector $\boldsymbol{m}=[3,4]$ (with cluster size of 5) and $\boldsymbol{m}=[4,5,6]$ (with cluster size of 10) when $r=4$ and $r=6$, respectively. We refer to each $(r,p)$ pair as a sub-scenario and consider six of them in total. For each sub-scenario we vary the initial delay in the range of 6 to 36 miliseconds (ms), and the results are shown in Fig. \ref{sc1}.\\
\indent From the results, an immediate observation is that multi message schemes perform better than their single message counterparts when the computation load $r$ is high. We note that, although a higher computation load reduces the non-straggler threshold, it also increases the computation time of the non-straggler workers. Hence, when the ratio of non-straggler threshold to the number of workers is less than $1-p$; that is, when the delay probability is over-estimated, we observe the limitation due to over-computation, and single message schemes performs poorly as clearly illustrated in Fig. \ref{sc12}. On the other hand, MMC has flexibility of either collecting fewer computations from a large set of workers, e.g., when $p$ is low, or collecting more  computations from fewer workers, e.g., when $p$ is high. This flexibility makes MMC schemes, especially LCC-MM and UC-MM, better options compared to their single message counterparts.\\
\indent Simulation results also point out that although LCC is superior to the GC scheme, proposed variations of GC, particularly GC-MM-C, can outperform LCC in certain cases. Besides, we observe that the correlated GC design, GC-MM-C, performs better compared to the uncorrelated design, GC-MM-C, especially when $r$ is large. \\
\indent Finally, the simulation results, especially those with $r=4$, show that, as $p$ increases, i.e., as $1-p$ gets close to the ratio of non-straggler threshold to the number of workers, comparative performance of the LCC scheme improves and even outperforms LCC-MM and UC-MM schemes. This observation highlights the fact that when the PS is limited to receive computations from the same subset of workers, which is the case when $p$ is large, LCC may perform better. 
\subsubsection{Scenario 2}\label{subsubsec:sc2}
\begin{figure}
    \centering       
    \includegraphics[scale=0.35]{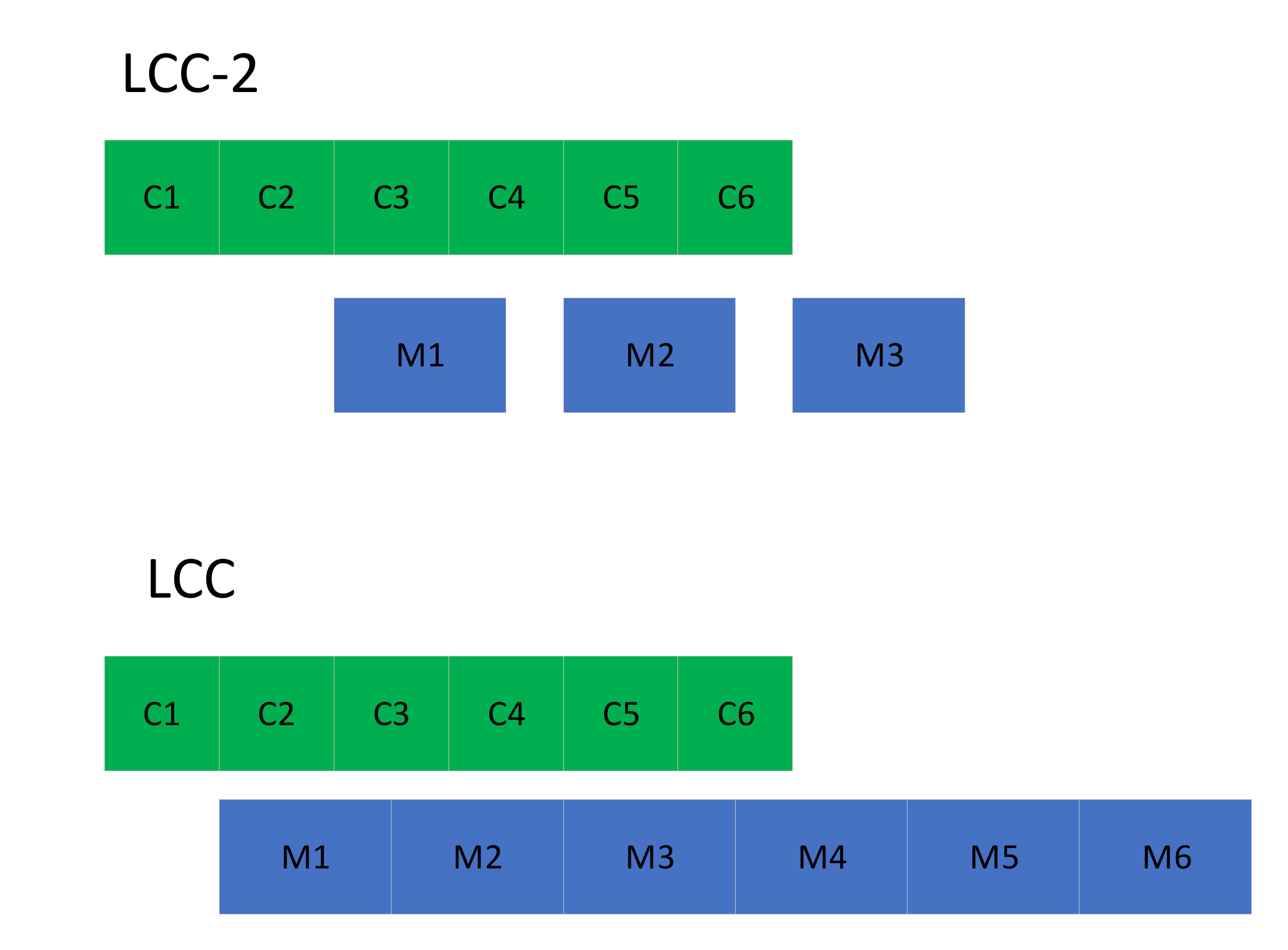}				
    \caption{ Overlapping communication and computation, where {\color{green}green}  and  {\color{blue}{blue}} blocks illustrate the computation  and the communication steps, respectively.}
    \label{overlap}
\end{figure}

 In the previous simulations, we focus on  worker based delays by using an initial delay parameter. We remark that with  non-blocking communication approach communication and computation can be executed in parallel, however each worker can send a message when the corresponding computation is completed and the previous message is successfully received by the PS as illustrated in Fig. \ref{overlap}. Hence, under certain scenarios where the  communication latency is higher than the computation latency  MMC strategy might be inefficient. In other words, the success of the MMC strategy depends on the ratio between the average computation and communication latency. To this end, we extend our previous analysis by adding additional exponentially distributed delays with parameter $\mu$ to the communication latency to demonstrate the impact of the communication latency on the MMC schemes.\\
\indent We first set $r=4$, and consider 4 sub-scenarios each corresponding to a different $p,\mu$ pair, where $p$ takes values 0.2 and 0.4, and $\mu$ takes values 2 and 4. For each sub-scenario we again change the initial delay in the range of 6 to 36 ms, and the results are illustrated in Fig. \ref{sc2}.\\    
\begin{figure*}[t]
    \begin{subfigure}{0.5\linewidth}
        \includegraphics[scale=0.5]{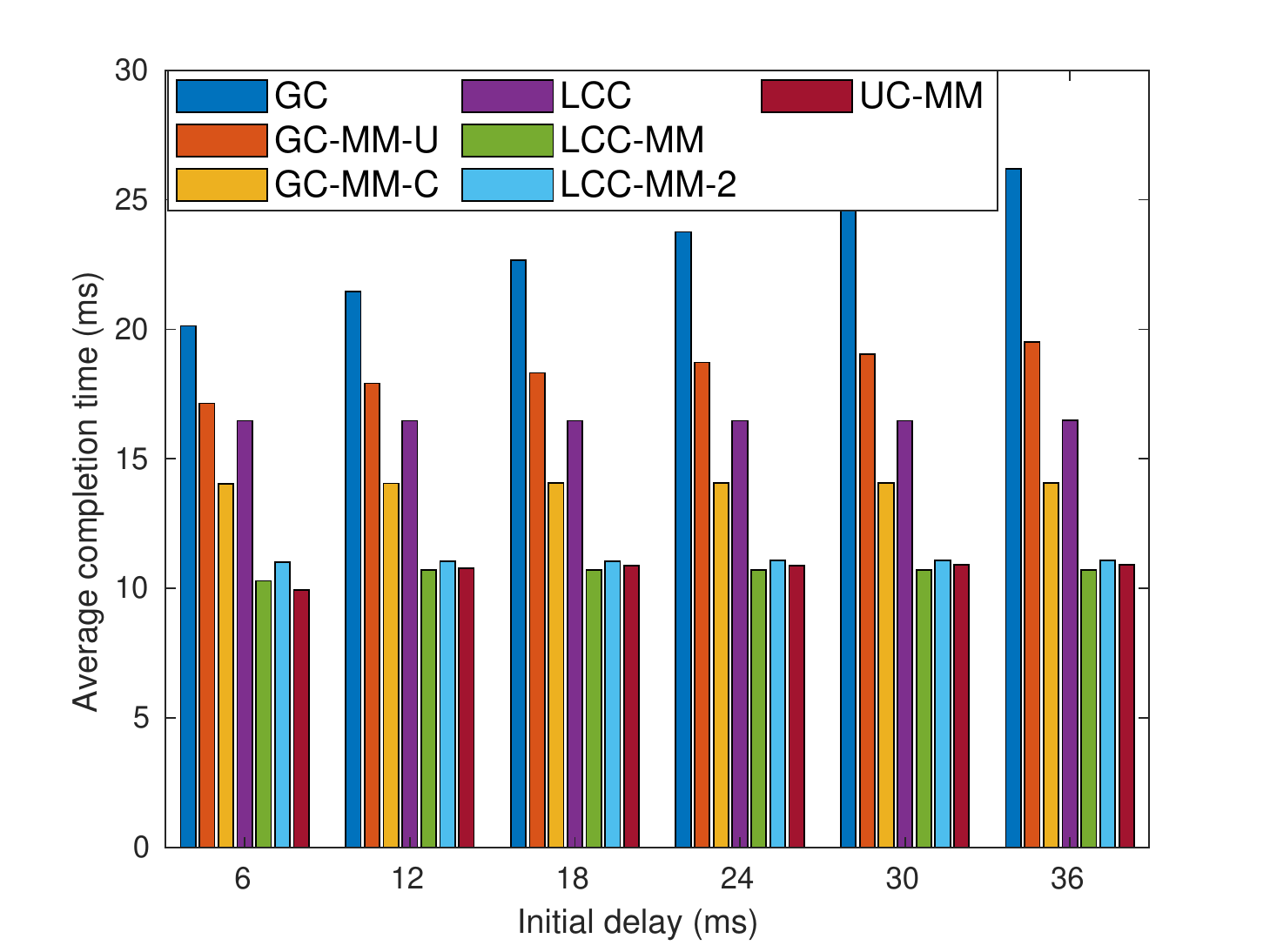}
        \caption{r=6, p=0.2 and $\mu=2$}
				\label{sc31}
    \end{subfigure}
    \smallskip
    \begin{subfigure}{0.5\linewidth}
        \includegraphics[scale=0.5]{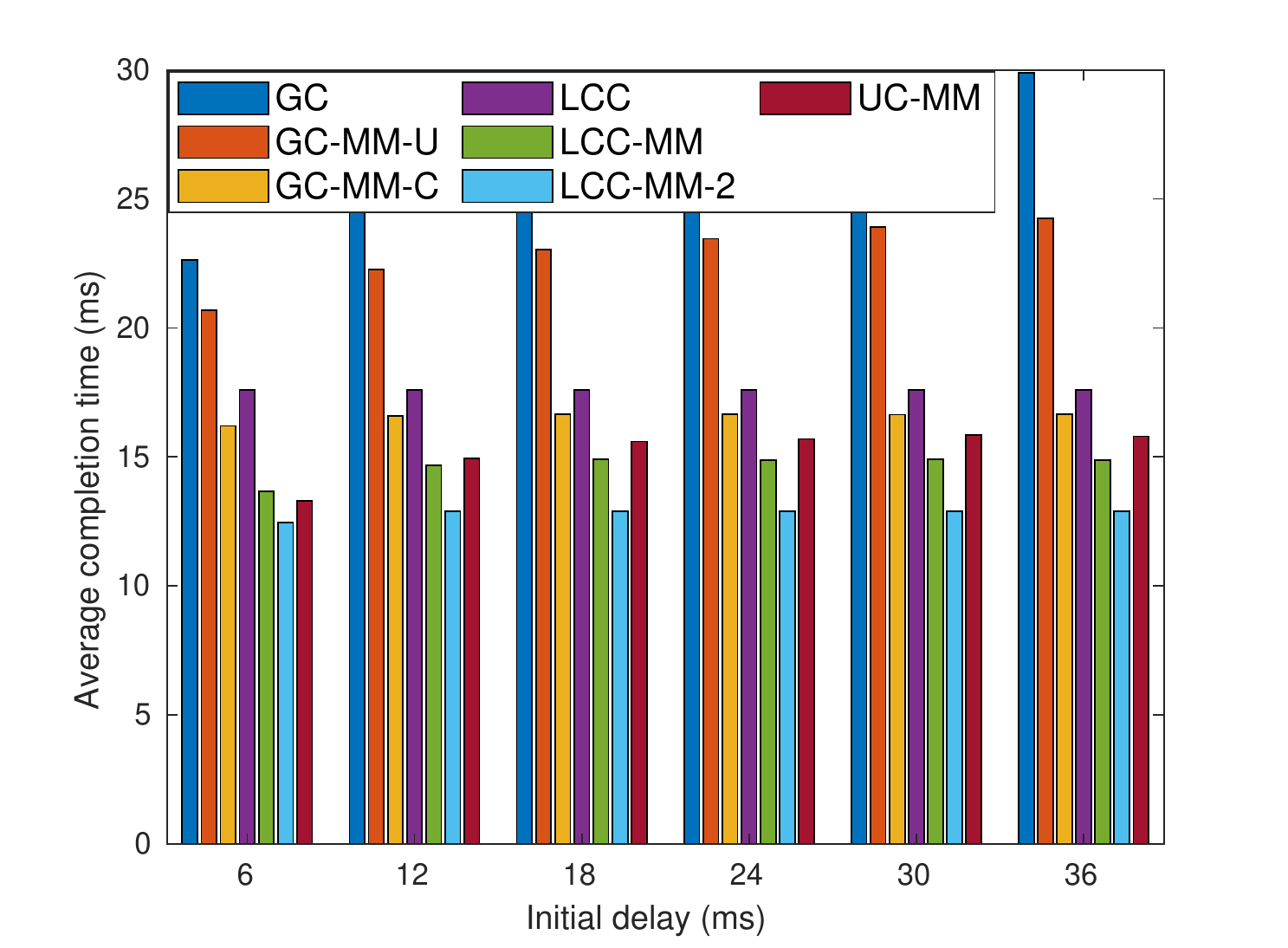}
        \caption{r=6, p=0.2 and $\mu=4$}
				\label{sc32}
    \end{subfigure}
    \smallskip
    \begin{subfigure}{0.5\textwidth}
        \includegraphics[scale=0.5]{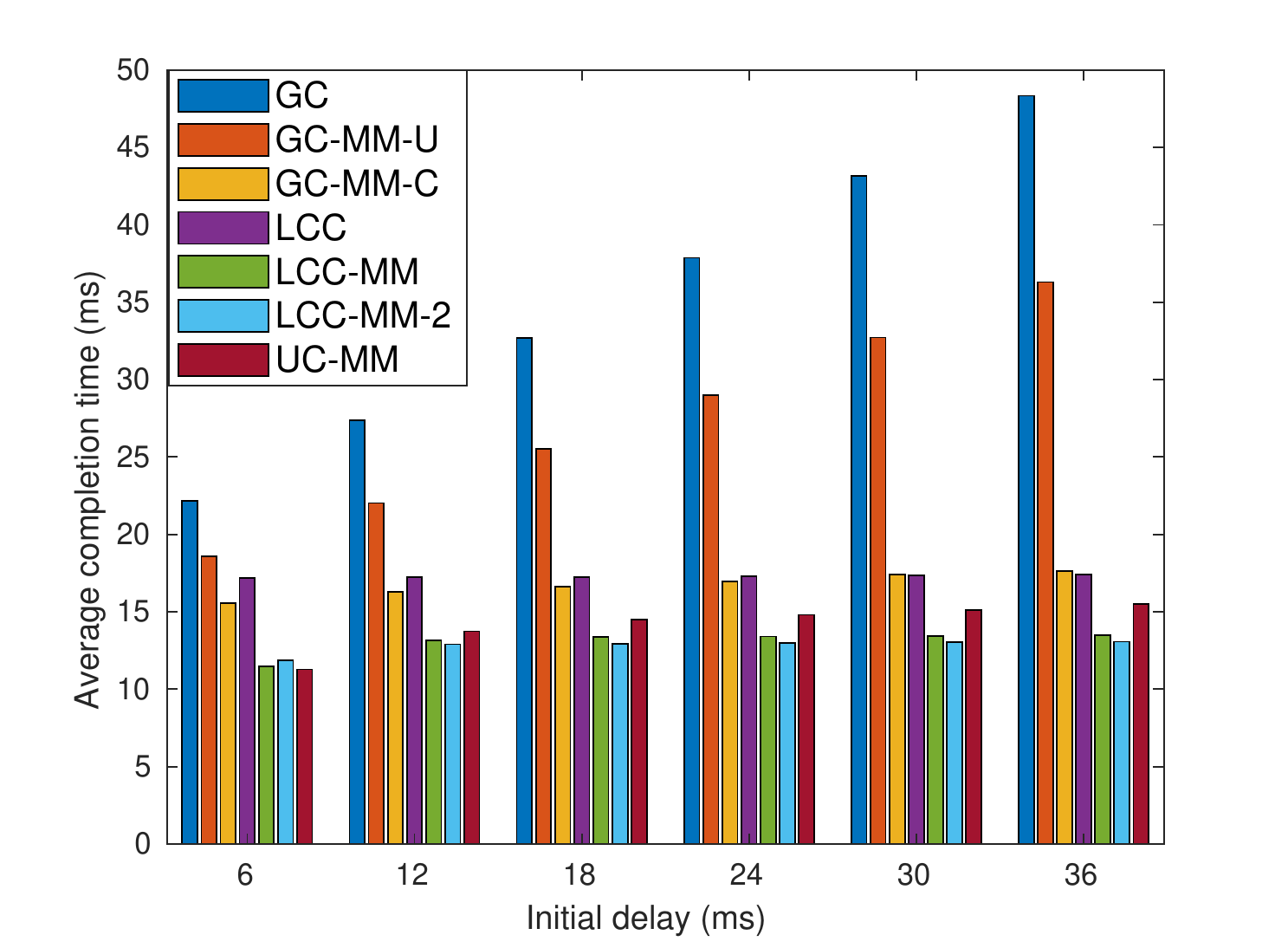}
        \caption{r=6, p=0.4 and $\mu=2$}
				\label{sc33}
    \end{subfigure}
    \smallskip
    \begin{subfigure}{0.5\linewidth}
        \includegraphics[scale=0.5]{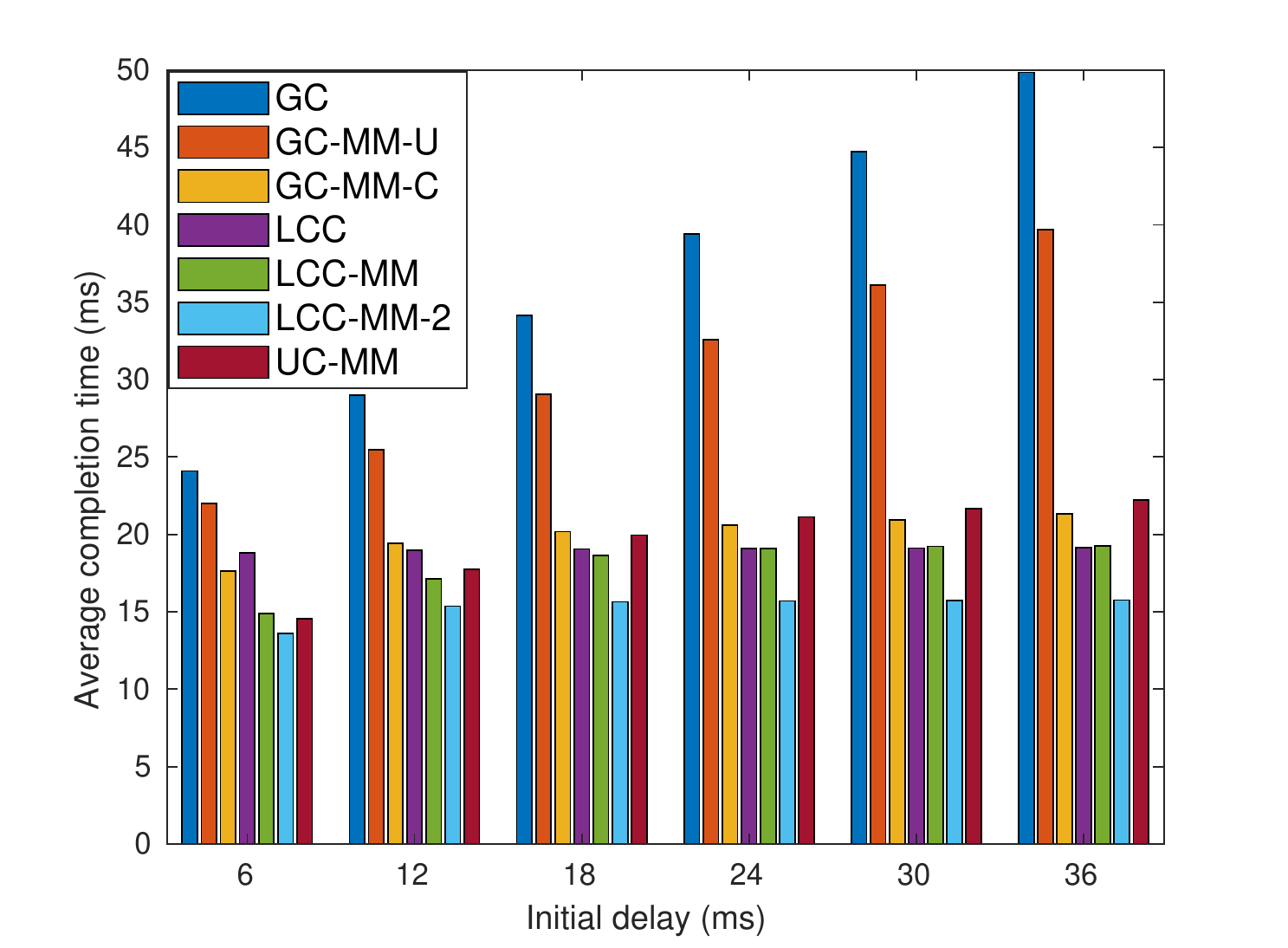}
        \caption{r=6, p=0.4 and $\mu=4$}
				\label{sc34}
    \end{subfigure}
				\caption{Average completion time analysis for GC, GC-MM-U, GC-MM-C, LCC LCC-MM, LCC-2 and UC-MM schemes with random fixed initial delay and exponential communication delay when $r=6$}
		\label{sc3}
\end{figure*}
One can easily observe, comparing Fig. \ref{sc22} and Fig. \ref{sc11}, that, even for small $p$, MMC schemes, especially UC-MM, can lose their advantage over single-message schemes when the communication latency is considerably high. Indeed, LCC and its multi-message variations, LCC-MM and LCC-MM-2, outperform  UC-MM in all four sub-scenarios except the first one, in which UC-MM performs slightly better than LCC. Another interesting observation is that when $\mu=4$, GC-MM-C out performs UC-MM especially when $p$ is low. Hence, when the communication latency in the network is large, GC with MMC can be preferred against UC-MM.\\  
 \begin{figure*}[t]
         \begin{subfigure}{0.5\textwidth}
        \includegraphics[scale=0.5]{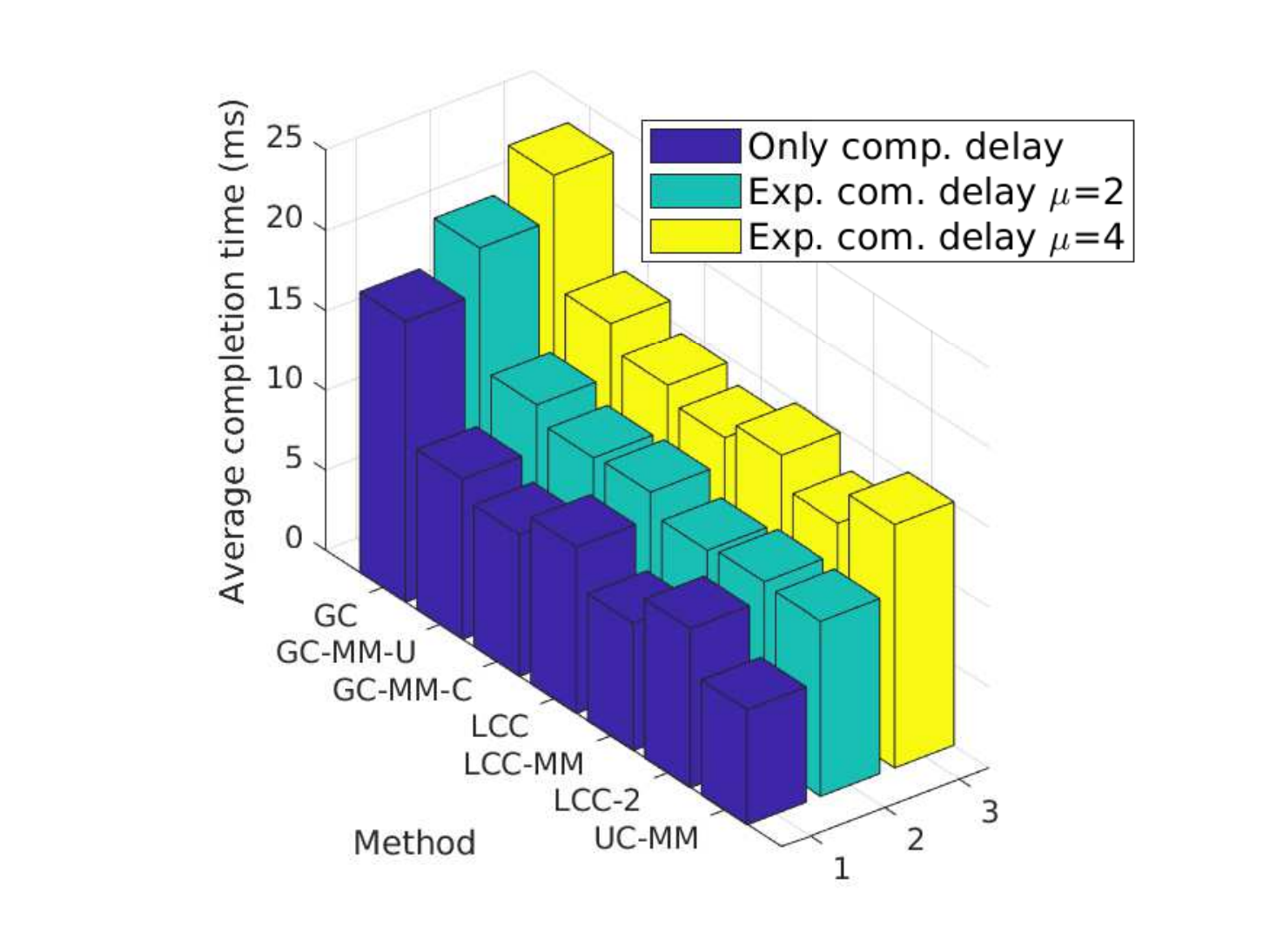}
        \caption{failure probability p=0.2, initial delay 12 ms}
				\label{cmpr1}
    \end{subfigure}
    \smallskip
    \begin{subfigure}{0.5\textwidth}
        \includegraphics[scale=0.5]{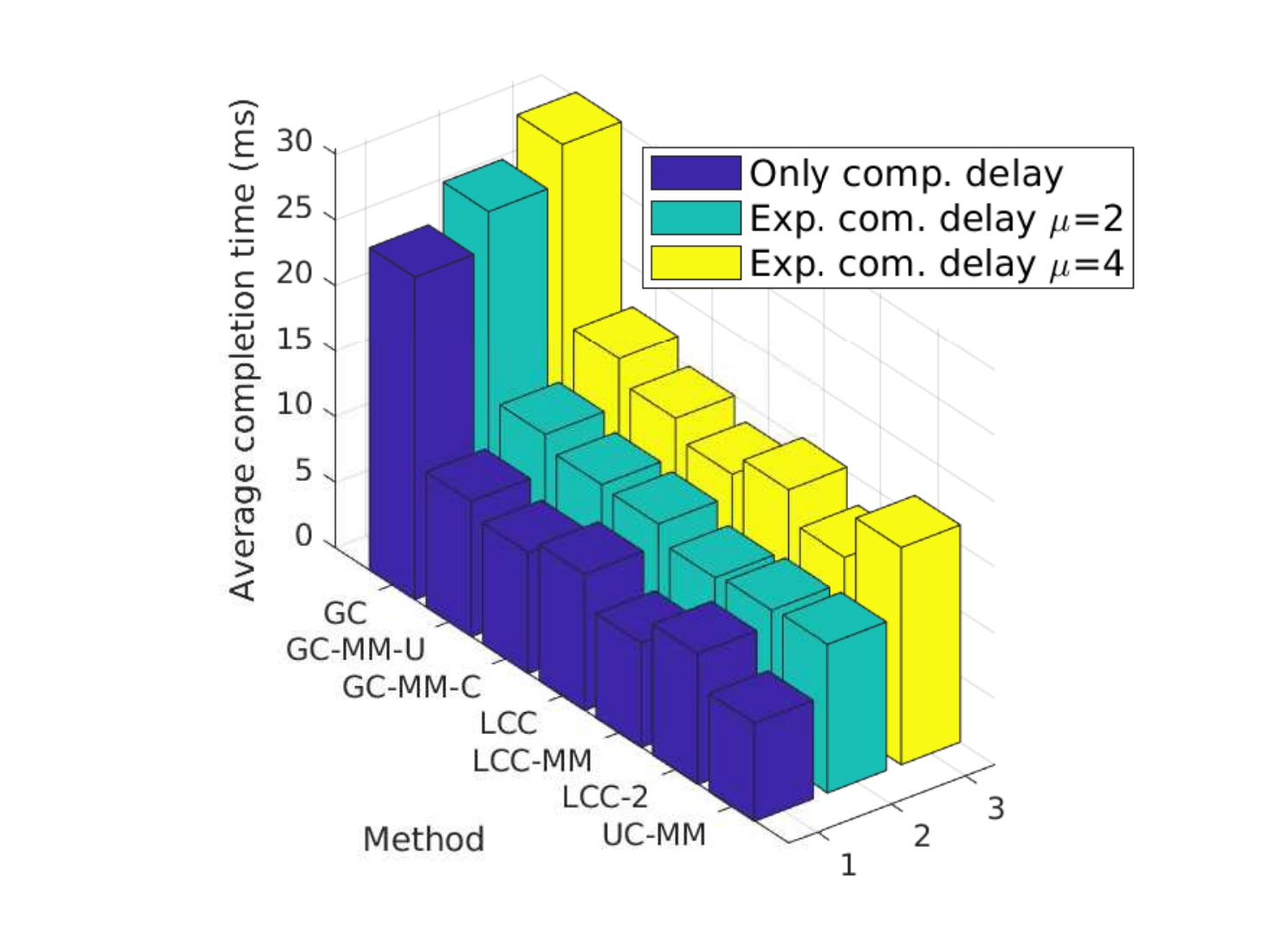}
        \caption{failure probability p=0.2, initial delay 24 ms}
				\label{cmpr2}
        \end{subfigure}
        \smallskip
                 \begin{subfigure}{0.5\textwidth}
        \includegraphics[scale=0.5]{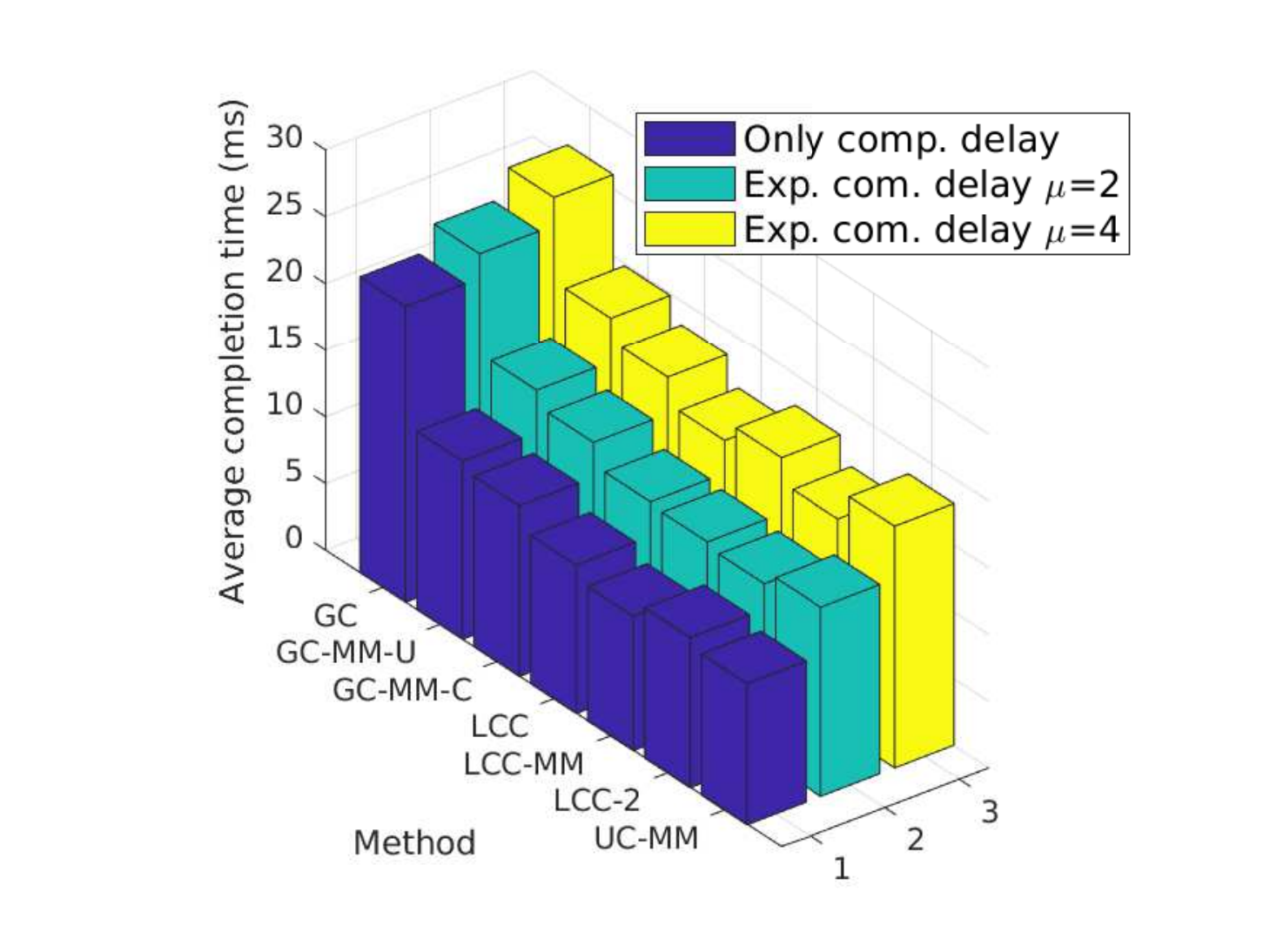}
        \caption{delay probability p=0.4, initial delay 12 ms}
				\label{cmpr3}
    \end{subfigure}
             \begin{subfigure}{0.5\textwidth}
        \includegraphics[scale=0.5]{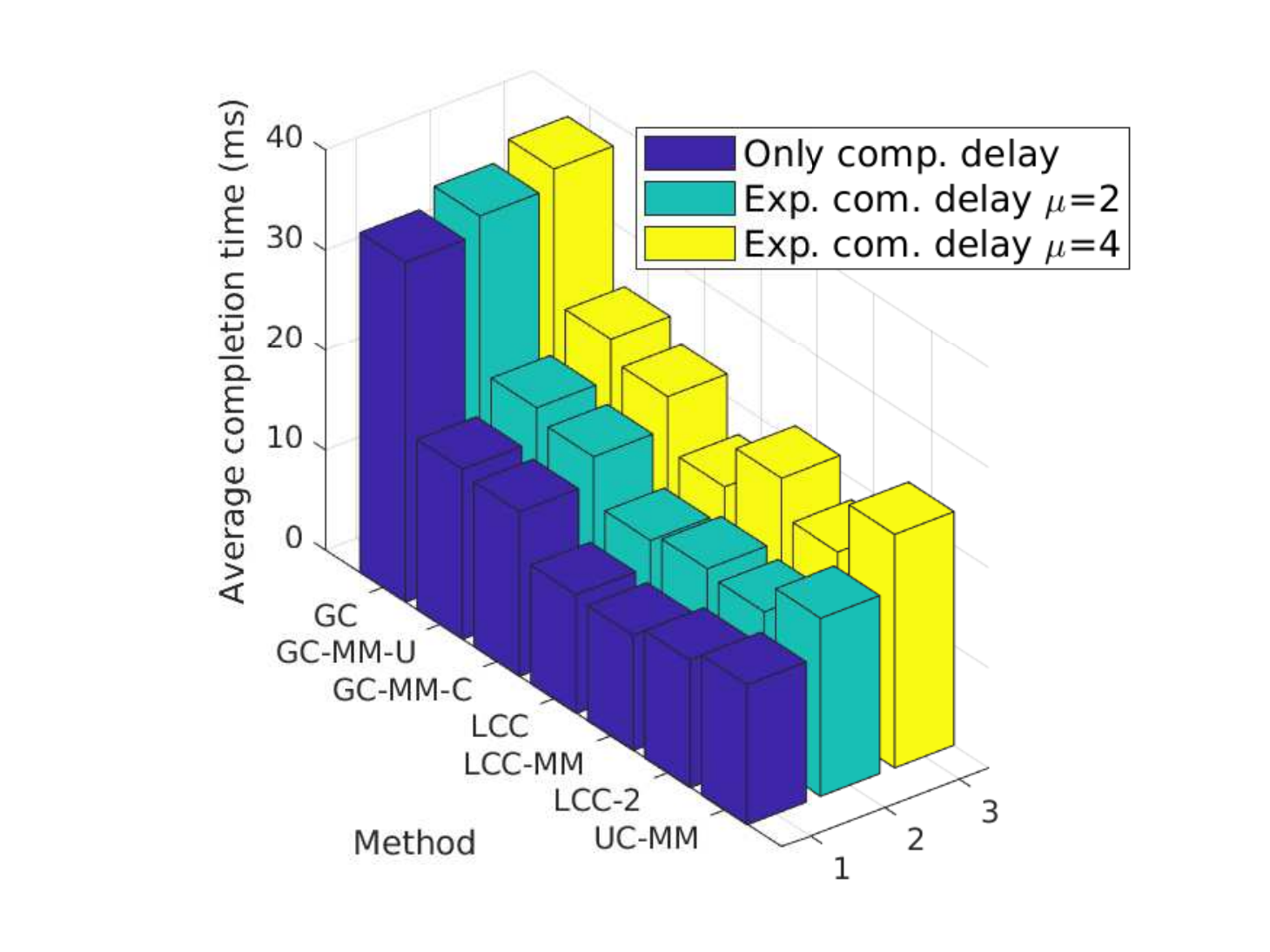}
        \caption{delay probability p=0.4, initial delay 24 ms}
				\label{cmpr4}
    \end{subfigure}
    \smallskip
				\caption{Per-iteration completion time for different schemes with $r=4$. }
		\label{cmpr}
\end{figure*}
\indent We repeat the simulations for the same four sub-scenarios with communication load $r=6$, and the results are illustrated in Fig. \ref{sc3}. Although the results show similarities with the previous one, we can identify some variations. First of all, as we expected, the relative performance of LCC deteriorates, due to the over-computation, especially when $p=0.2$. When $p=0.2$, compared to the case of $r=4$, LCC loses its advantage against UC-MM. We also observe that in none of these four sub-scenarios LCC is the best one. \\
\indent Fig. \ref{sc2} and Fig. \ref{sc3} point out that LCC-MM-2 can be a better alternative compared to  LCC, LCC-MM and UC-MMC when both the communication latency and communication load $r$ are high. This is because LCC-MM-2,  improves upon the LCC-MM and UC-MM schemes via reducing the number of messages sent at each iteration as well as increasing the time between two communication rounds which better overlaps the communication and computation processes  as illustrated in Fig. \ref{overlap}. Thanks to overlapped communication time, LCC-MM-2 scheme is more robust to communication latency compared to LCC-MM and UC-MM.\\
\indent Finally, to monitor the marginal effect of the communication latency, we pick four particular cases; $p=0.2$ with $12$ ms initial delay, $p=0.2$ with $24$ ms initial delay, $p=0.4$ with $12$ ms initial delay, and $p=0.4$ with $24$ ms initial delay; and plot the performance of all the schemes with respect to $\mu$ in  Fig. \ref{cmpr}.  We observe that the average completion time increases with respect $\mu$; however, while GC and LCC  exhibit a gradual increase, LCC-MM and UC-MM experience a step increase with $\mu$. Particularly, when $p=0.2$, it is clear how UC-MM and LCC-MM schemes lose their advantages  with increasing communication latency.
\subsection{Real Time Simulations}
\begin{figure*}[t]
         \begin{subfigure}{0.5\textwidth}
        \includegraphics[scale=0.5]{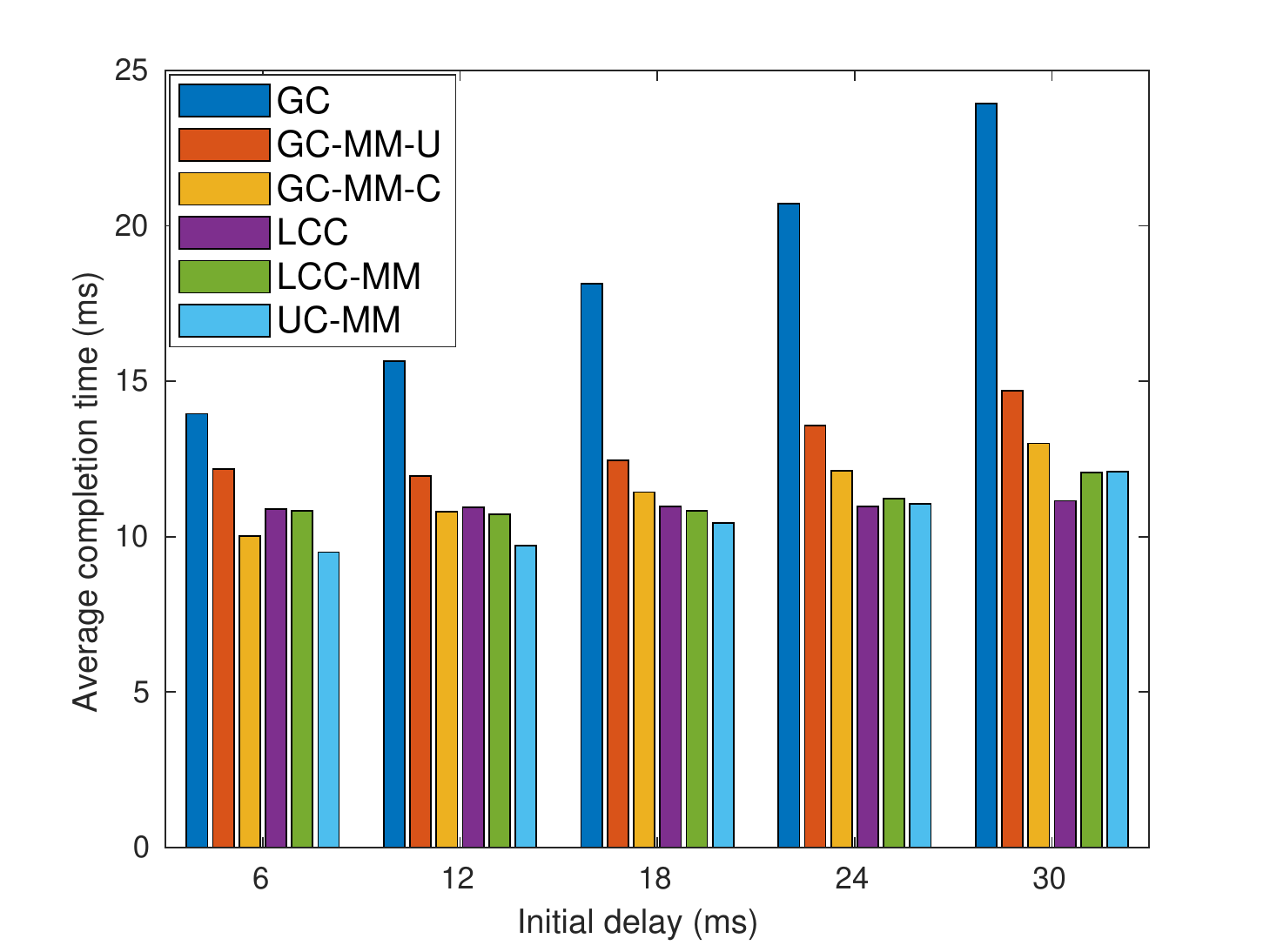}
        \caption{p=0.1}
				\label{real3p1}
    \end{subfigure}
    \smallskip
    \begin{subfigure}{0.5\textwidth}
        \includegraphics[scale=0.5]{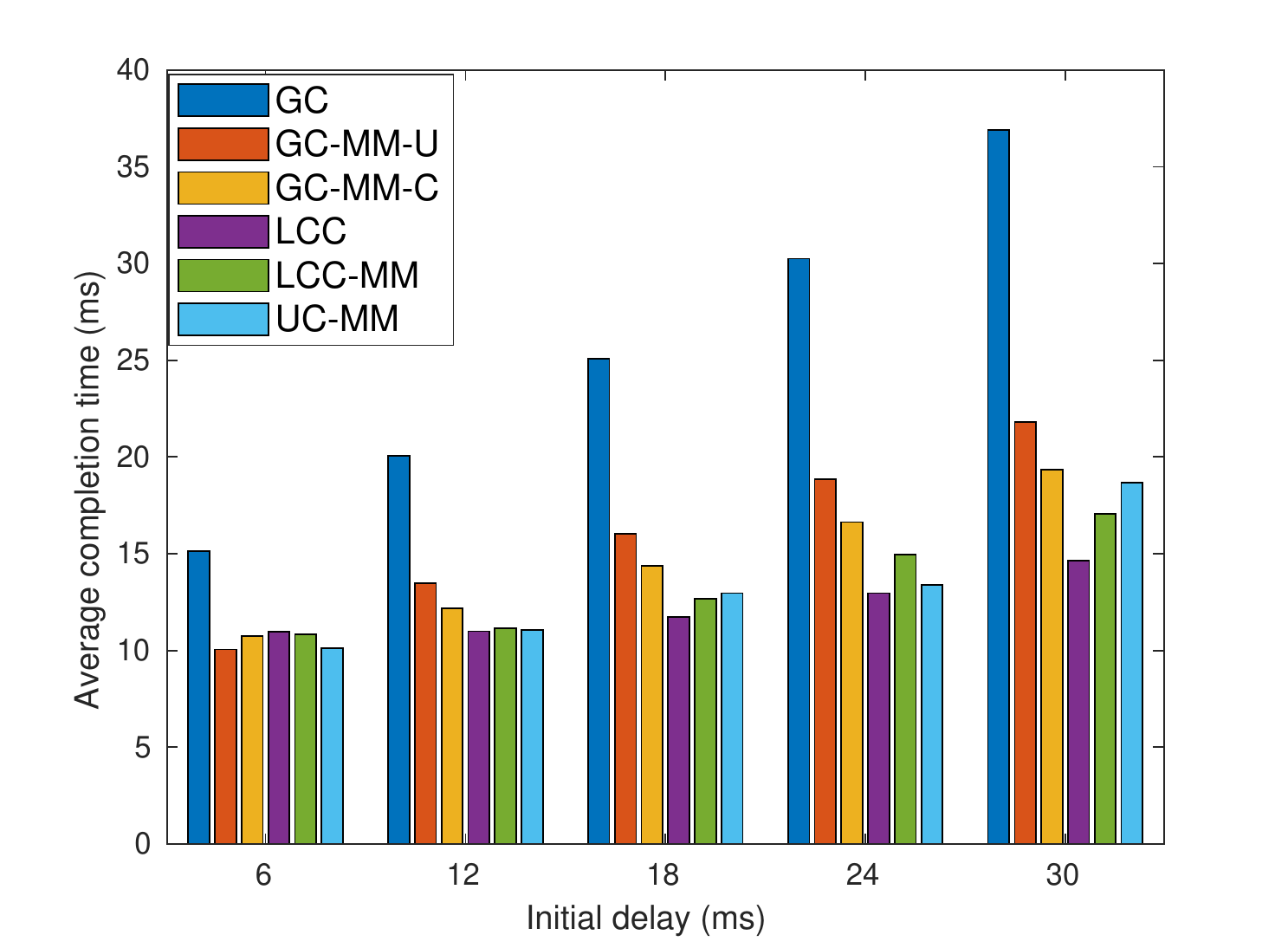}
        \caption{p=0.2}
				\label{real3p2}
        \end{subfigure}
        \smallskip
                 \begin{subfigure}{0.5\textwidth}
        \includegraphics[scale=0.5]{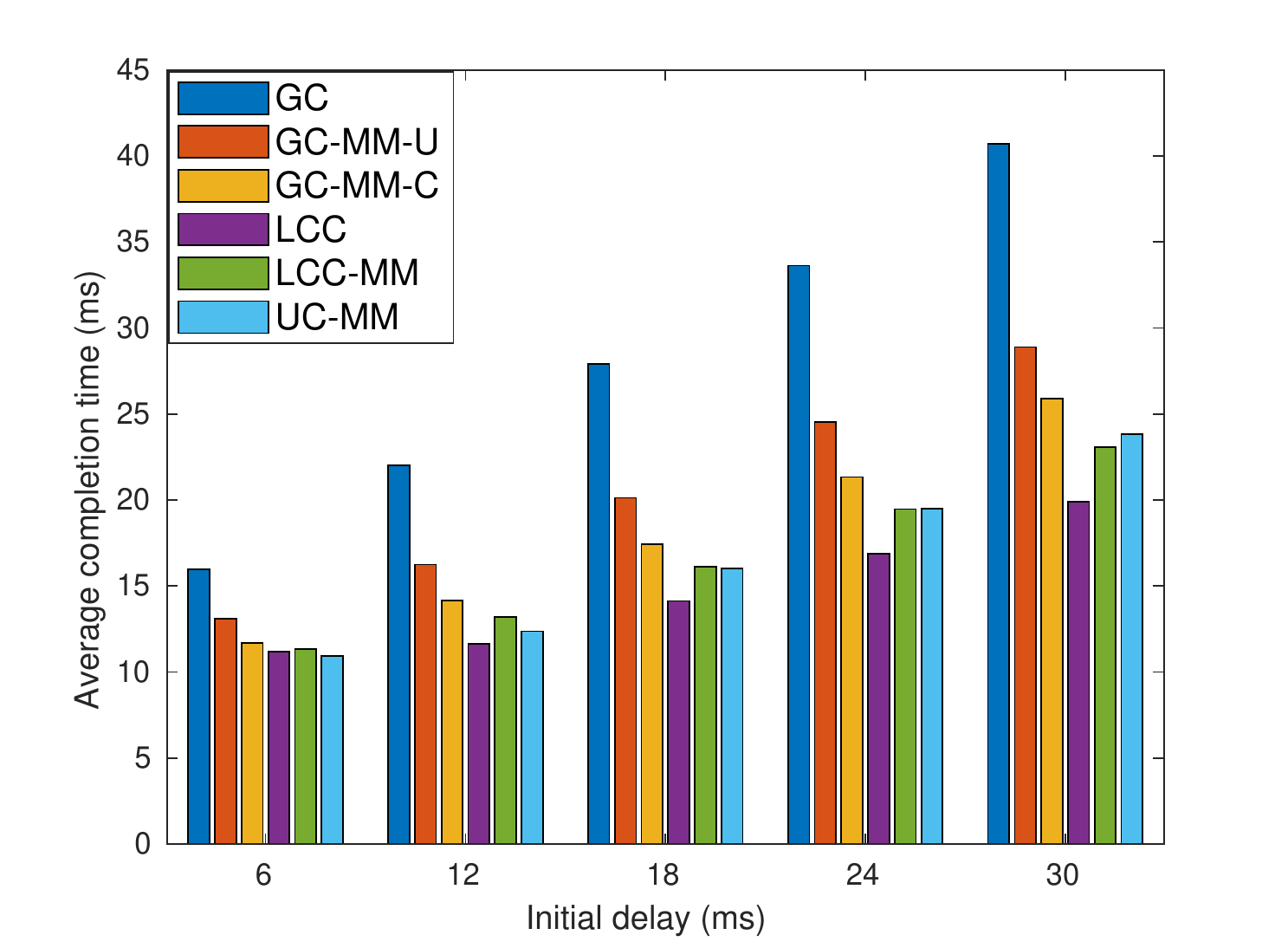}
        \caption{p=0.3}
				\label{real3p3}
    \end{subfigure}
    \smallskip
             \begin{subfigure}{0.5\textwidth}
        \includegraphics[scale=0.5]{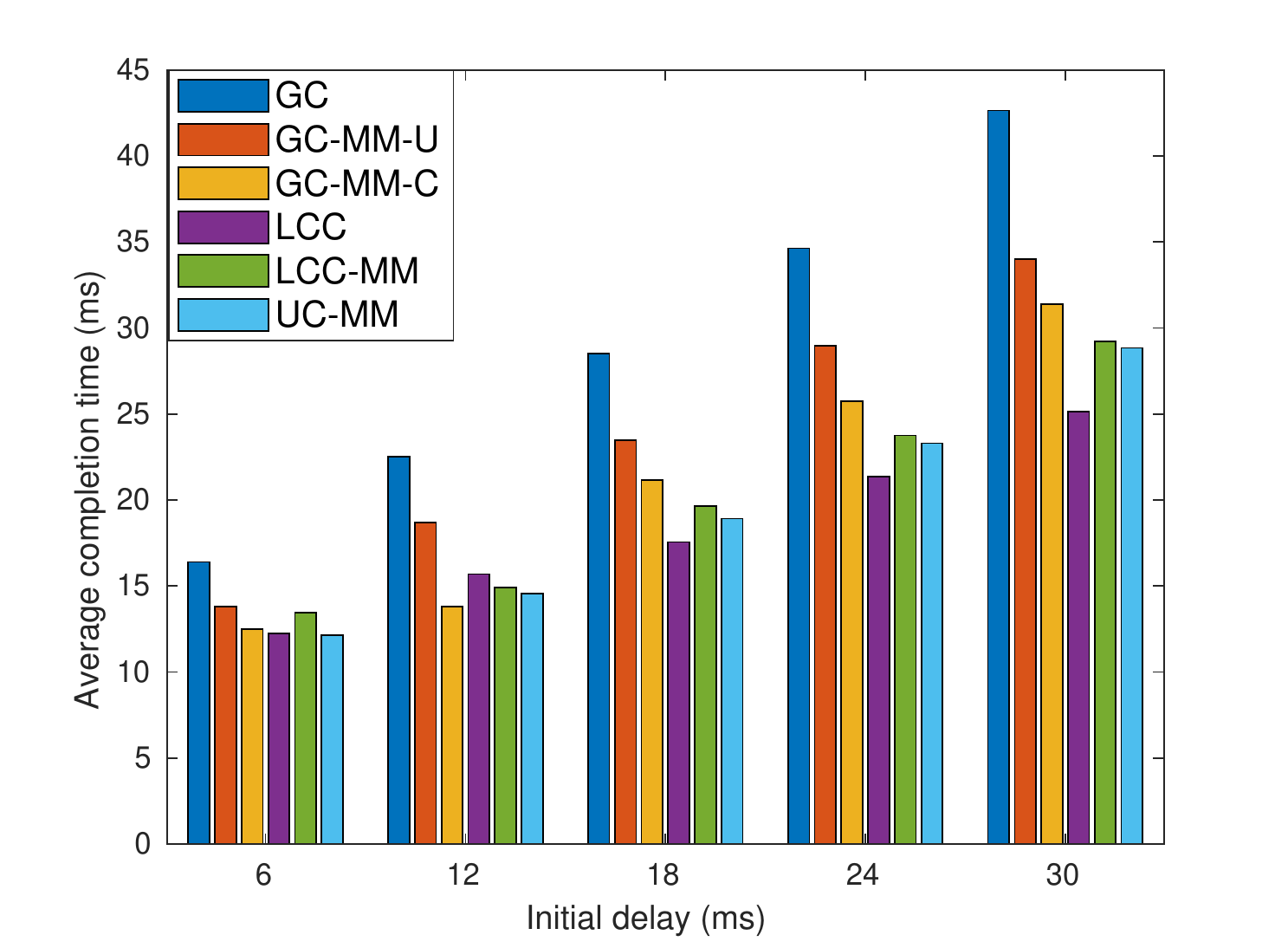}
        \caption{p=0.4}
				\label{real3p4}
    \end{subfigure}
    \smallskip
				\caption{Per iteration completion time for different schemes with $r=3$. }
		\label{real3}
\end{figure*}

\begin{figure}
         \begin{subfigure}{0.5\textwidth}
        \includegraphics[scale=0.5]{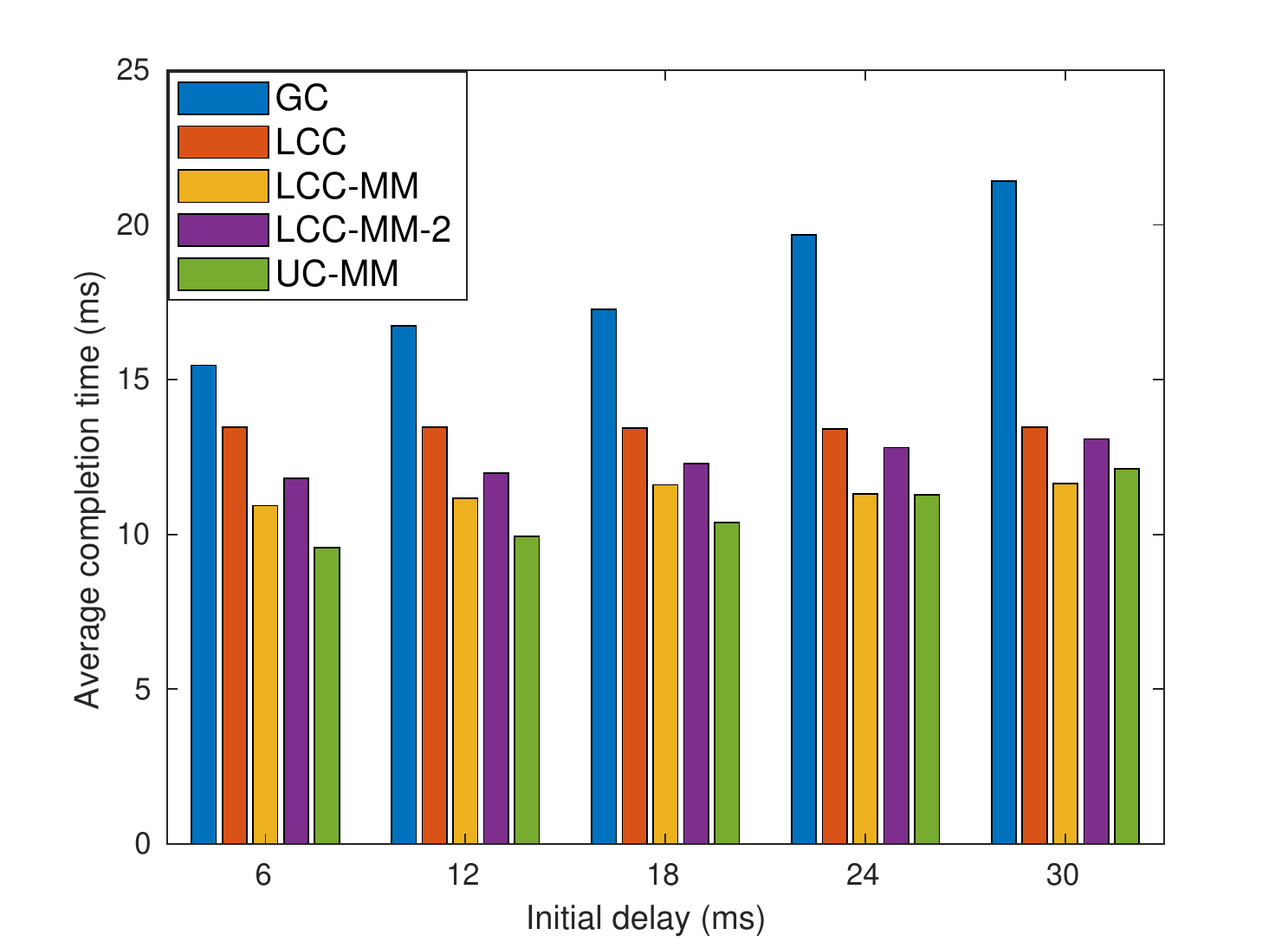}
        \caption{delay probability p=0.1}
				\label{realr4p1}
    \end{subfigure}
    \smallskip
    \begin{subfigure}{0.5\textwidth}
        \includegraphics[scale=0.5]{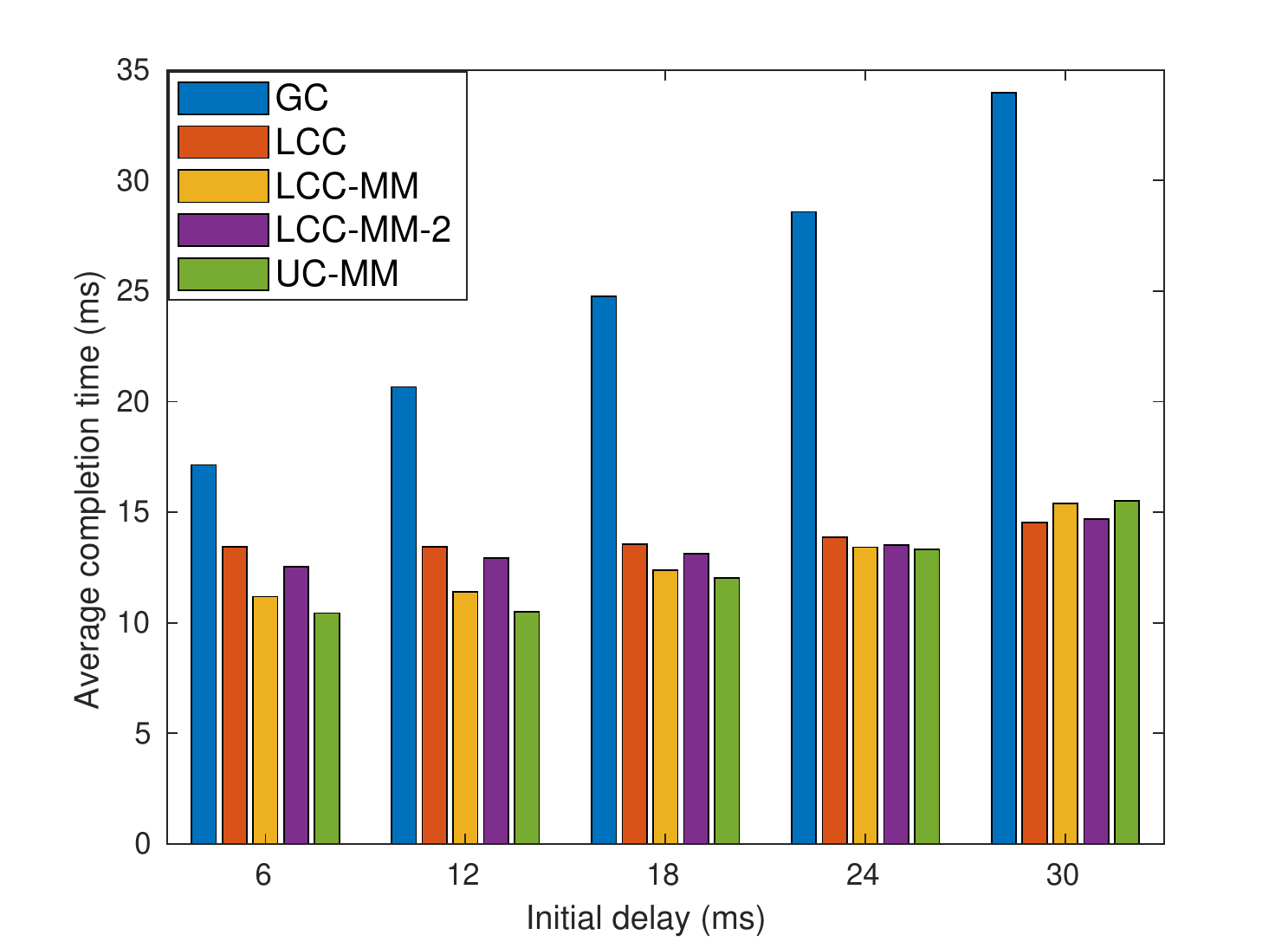}
        \caption{delay probability p=0.2}
				\label{realr4p2}
        \end{subfigure}
        \smallskip
                 \begin{subfigure}{0.5\textwidth}
        \includegraphics[scale=0.5]{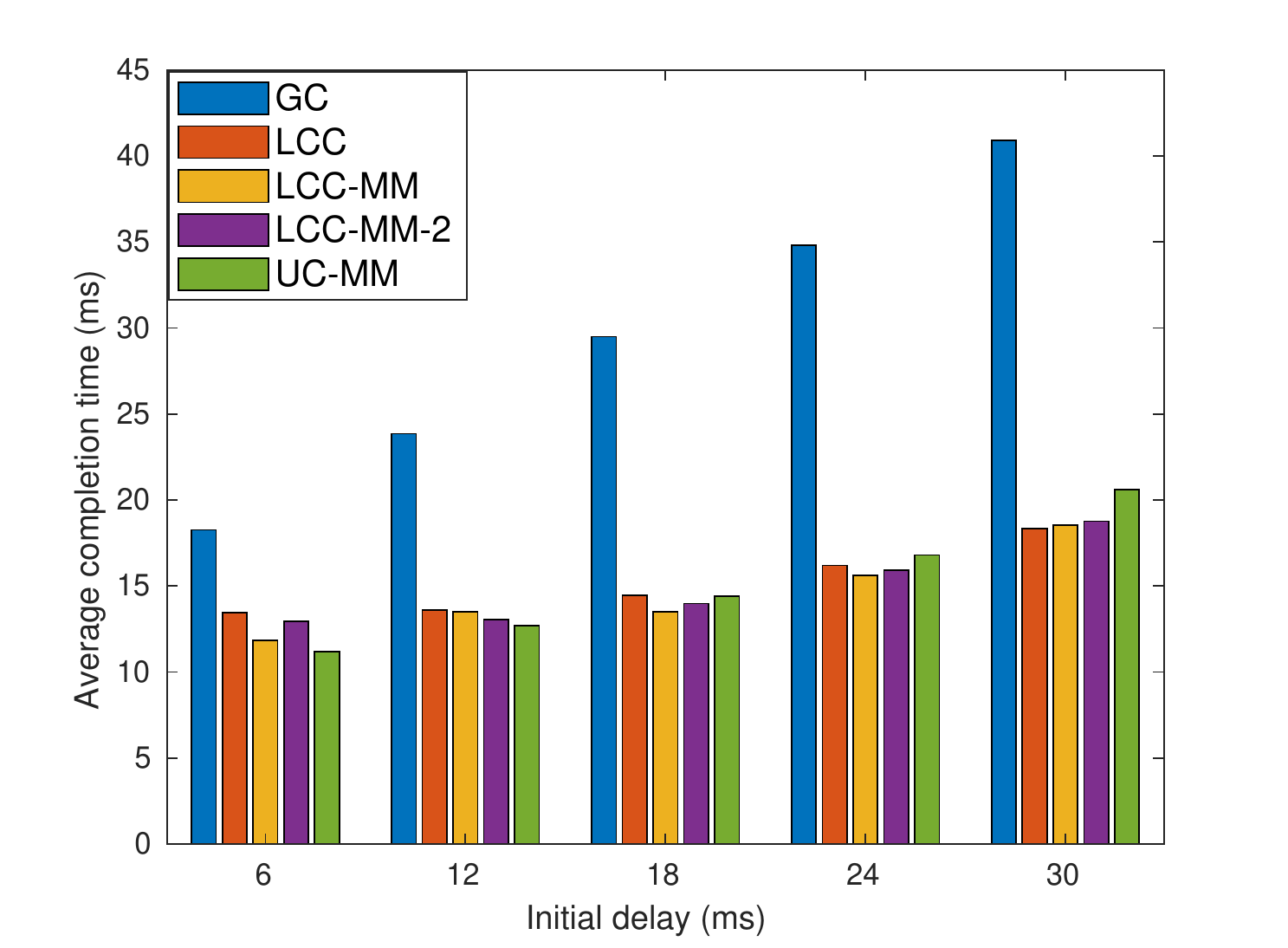}
        \caption{delay probability p=0.3}
				\label{realr4p3}
    \end{subfigure}
    \smallskip
             \begin{subfigure}{0.5\textwidth}
        \includegraphics[scale=0.5]{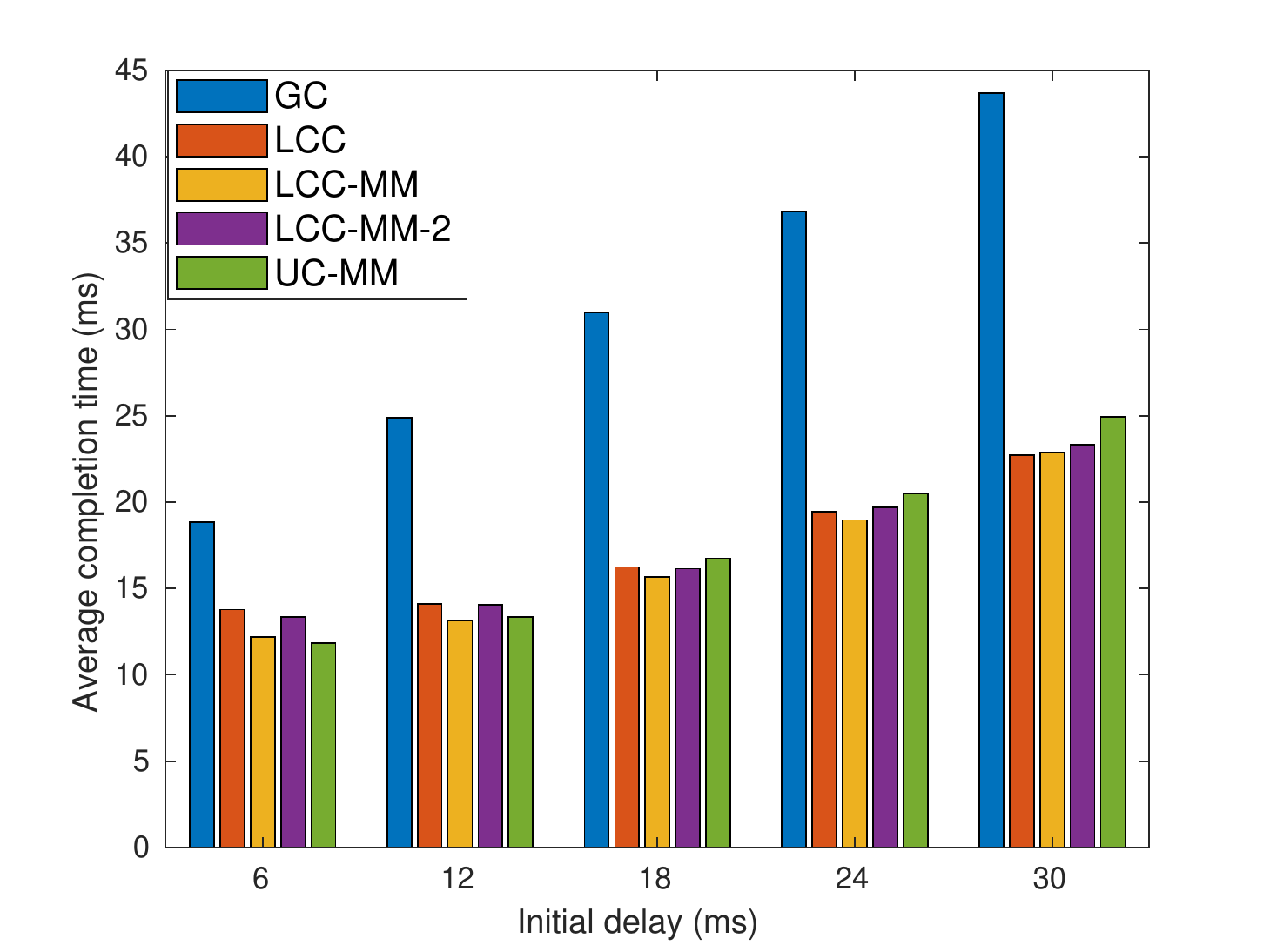}
        \caption{delay probability p=0.4}
				\label{realr4p4}
    \end{subfigure}
    \smallskip
				\caption{Per iteration completion time for different schemes with $r=4$.}
		\label{realr4}
\end{figure}

\begin{figure*}
         \begin{subfigure}{0.5\textwidth}
        \includegraphics[scale=0.5]{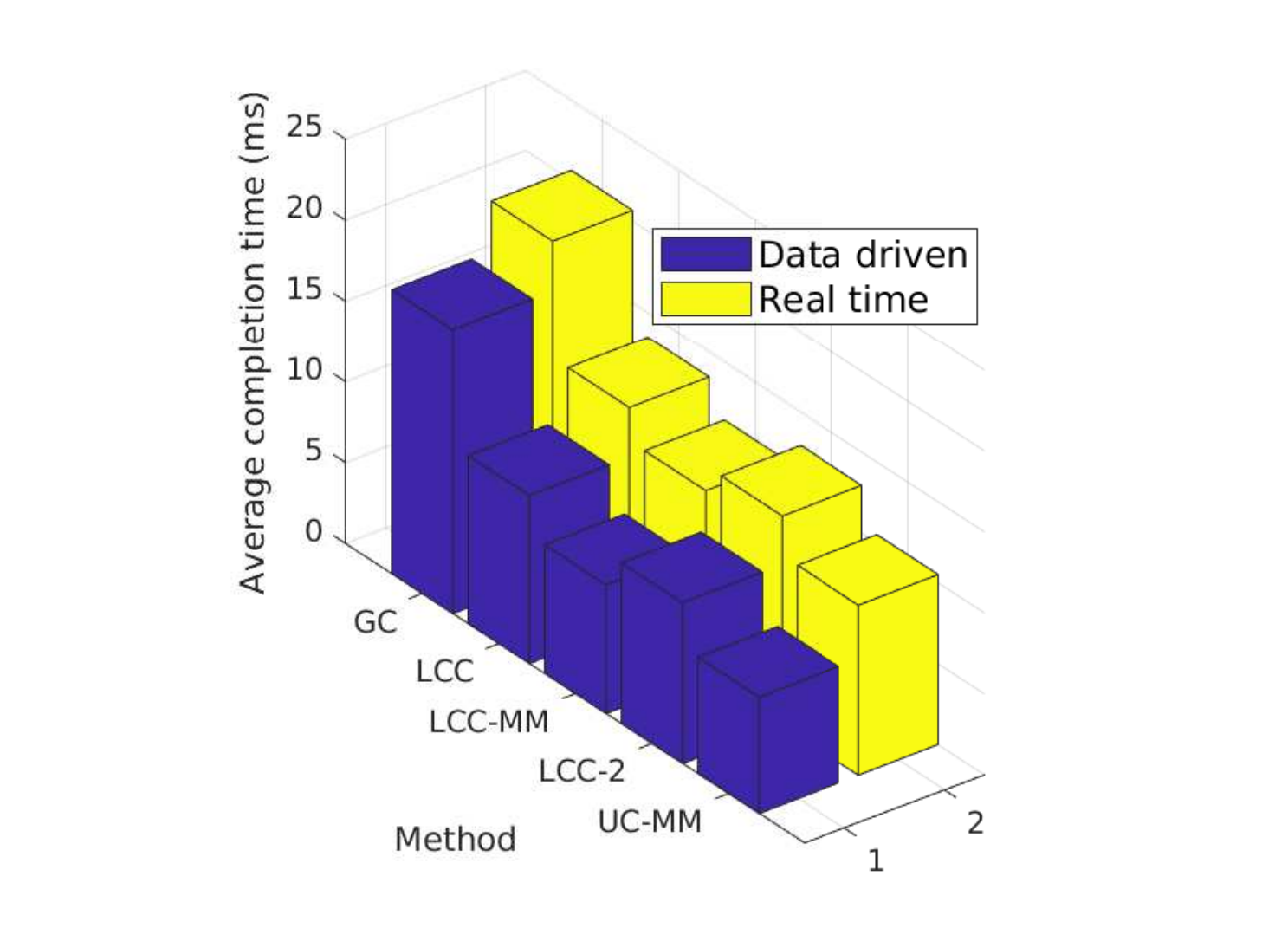}
        \caption{ p=0.2 and initial delay is 12 ms}
				\label{cmpr21}
    \end{subfigure}
    \begin{subfigure}{0.5\textwidth}
        \includegraphics[scale=0.5]{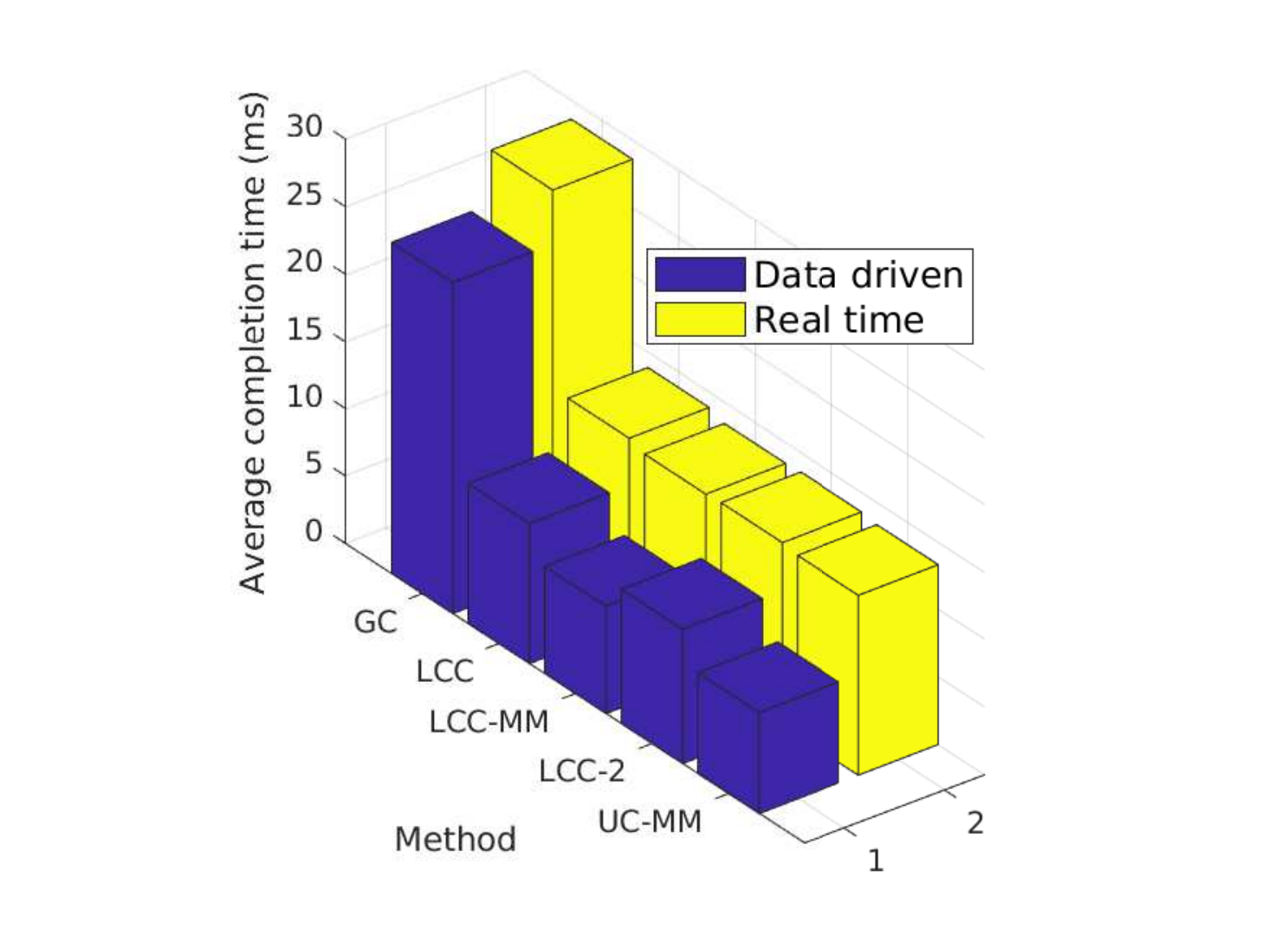}
        \caption{p=0.2 and initial delay is 24 ms}
				\label{cmpr22}
        \end{subfigure}
				\caption{Comparison of data driven simulation results with real time simulation results.}
		\label{cmpr2}
\end{figure*}

\begin{figure}
    \centering        \includegraphics[scale=0.6]{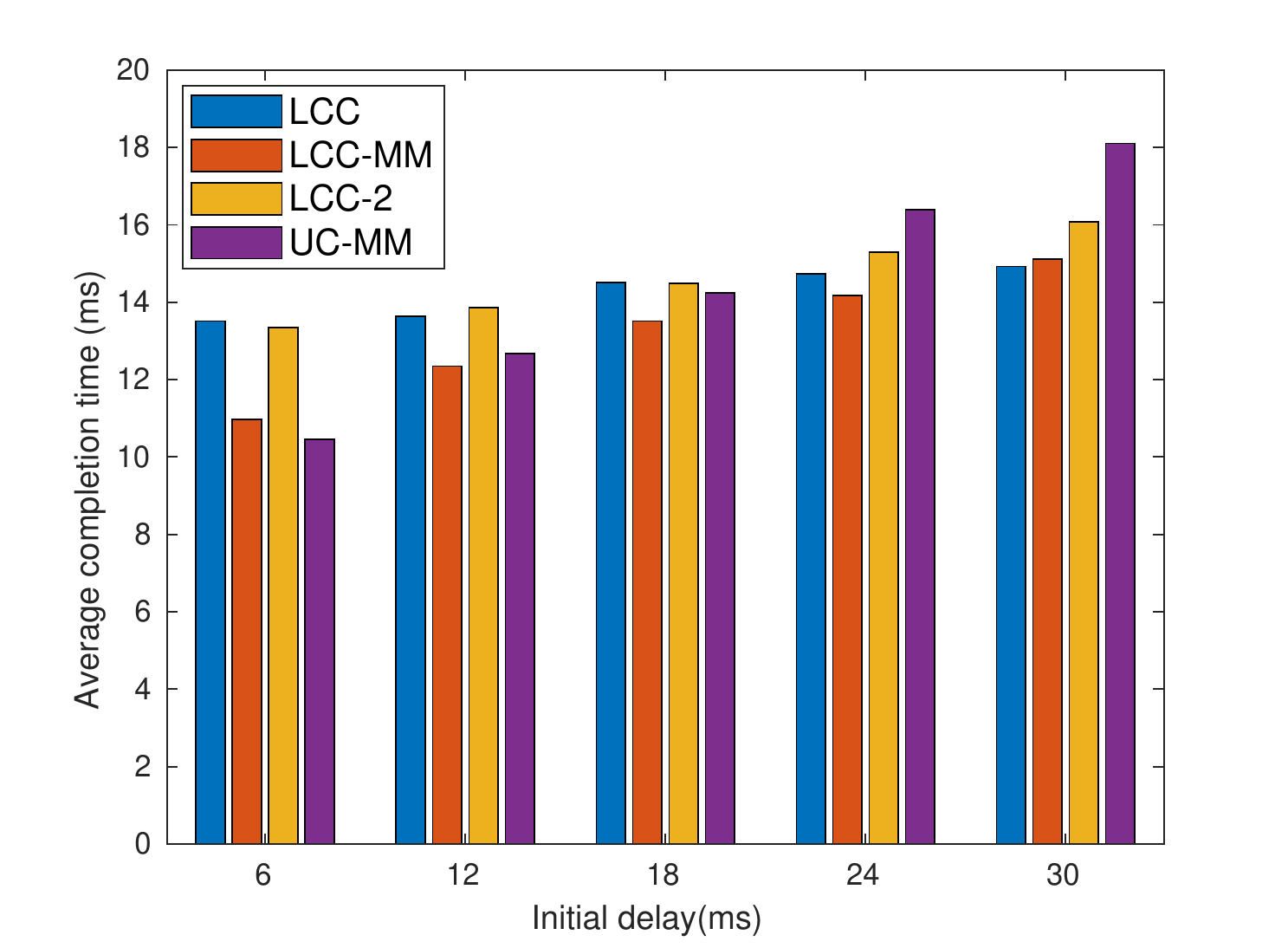}				\caption{ Performance under uncorrelated delay.}
            \label{uncorr}
\end{figure}

Data driven simulations ignore the effect of congestion on the completion time statistics. To remedy this, we perform real time analyses on Amazon EC2 servers. Similarly to the data driven simulations, we initialize 21 Amazon EC2 t2.micro instances, where the first one is labeled as the PS, and we use the MPI protocol to establish connections between the these instances. At the beginning of each iteration, after receiving the model update from the PS, we randomly induce a fixed delay at each instance using time.sleep() command. Then,  the PS  waits until the required condition to complete an iteration, which depends on the scheme employed, is met. We present the average completion time over $1000$ iterations. We first set $r=3$, and consider four different sub-scenarios with p=0.1, 0.2, 0.3, 0.4. In each scenario, we change the initial  delay from 6 ms to  30 ms and the results are illustrated in Fig. \ref{real3}.\\
\indent Although GC-MM-C and UC-MM  outperform  LCC when both initial delay and $p$ are low,
 in general, LCC achives the best performance, especially when $p$ is large. Nevertheless, we also observe that GC-MM-C and UC-MM  perform close to LCC, particularly when $p$ is low; indeed, in all the cases, the performance gap between  LCC and UC-MM is at most $26\%$. Hence, when the decoding complexity of the LCC scheme as well as the initial data encoding process are taken into consideration, UC-MM scheme is still a  strong candidate for distributed computation.\\ 
\indent Next, we set $r=4$ and repeat the simulation as in the previous case, but this time we compare the performance of the GC, LCC, LCC-MM, LCC-MM-2, and UC-MM schemes. We observe that when $p$ is low, i.e., $p=0.1$ and $p=0.2$, MMC schemes LCC-MM and UC-MM outperform  others. Indeed, UC-MM can perform up to 40\%  better compared to LCC. On the other hand, when we consider larger $p$ values, LCC, LCC-MM, LCC-MM-2, and UC-MM schemes have similar performances, although average completion time  of UC-MM scheme is slightly higher when the initial delay is large.\\
\indent We remark that, although the real time simulation results present similar trends with our initial data driven analysis, we observe some differences as well. In particular, when $r=4$ and $p=0.2$, we expect UC-MM and LCC-MM schemes to perform much better based on our data driven analysis illustrated in Fig. \ref{sc11}. However, as we discussed in Section \ref{subsubsec:sc2}, communication latency is also an important factor for the average completion time statistics, and the multi-message schemes are more prone to communication delays. Our interpretation for the results in Fig. \ref{realr4p2} is that the performance of the UC-MM and LCC-MM schemes are limited due to the congestion at the PS. To show the  effect of congestion more explicitly we limit our focus to two cases with initial delay of 12 ms and $p=0.2$, and initial delay of 24 ms and $p=0.2$. For these two cases, we compare the data driven simulation results of GC, LCC, LCC-MM, LCC-MM2 and UC-MM schemes with their real time counterparts in Fig. \ref{cmpr2}.\\
\indent It is clear from Fig. \ref{cmpr21} that all the schemes suffer from the congestion, but its effect is more significant for multi-message schemes. Fig. \ref{cmpr22}, further shows that multi-message schemes, particularly LCC-MM and UC-MM, may lose their advantage due to congestion. We emphasize that these observations are consistent with our data driven simulation results with exponential communication delay.\\
\indent One of the most interesting observations from the real time simulation results is the trend of the LCC schemes, particularly LCC and LCC-MM-2, with respect to initial delay. According to data driven results in Fig. \ref{sc1}, LCC scheme should be robust to the initial delay; hence, we expect the average completion time of the LCC scheme not to increase with initial delay, but the real time simulation results in Fig. \ref{realr4p4} seem to be inconsistent with this intuition. However, this discrepancy results from the way communication delay is introduced in real time scenarios. When we introduce delay using the time.sleep() command in real time simulations, the instance might be still sleeping in the next iteration since average completion time is less than the initial delay in general. In other words, an initial delay at a particular iteration can affect the following iterations, which is not the case in data driven simulations. This impact becomes more visible as $p$ increases. To verify our reasoning we repeat the simulations for $p=0.4$ with different initial delays, but this time we terminate the delay when the iteration is completed, so that the delay in one iteration has no impact on the following iterations. The corresponding simulation results are illustrated in Fig. \ref{uncorr}, which support our interpretation.\\
\indent This observation leads to a new discussion on the modelling of delay at workers. In the literature, the delay is mostly associated with the computation process. In that case, one can argue that, after each iteration uncompleted jobs will be terminated, so that the delay will not affect the following iterations. On the other hand, it is also possible to observe  delays due to access failure or scheduled tasks for other clients. Such delays are not limited to a single iteration, causing correlation among delays over consecutive iterations.\\
\indent Next, we consider both correlated and uncorrelated delays for completeness of our analysis. We set $r=6$, and for both correlated and uncorrelated scenario we analyze two cases with $p=0.1$ and $p=0.3$. In the case of correlated delay with $p=0.1$, compared to LCC, LCC-MM  and UC-MM  can achieve 36-40\% and  42-58\%  reduction in the average completion time, respectively. Similarly, they achieve around 48\% and \%60 reduction, respectively, when the delay is uncorrelated. When the delay is uncorrelated, we observe similar trends for $p=0.3$, such that  both LCC-MM and UC-MM achieve around 40\% reduction in the average completion time compared to LCC. When the delay is correlated, UC-MM still outperforms LCC, but its performance deteriorates with the increase in the  initial delay and when the initial delay is large LCC-MM becomes a better option. We believe that understanding the impact of  correlation in the delay over time is an interesting future research direction.
\begin{figure*}
         \begin{subfigure}{0.5\textwidth}
        \includegraphics[scale=0.5]{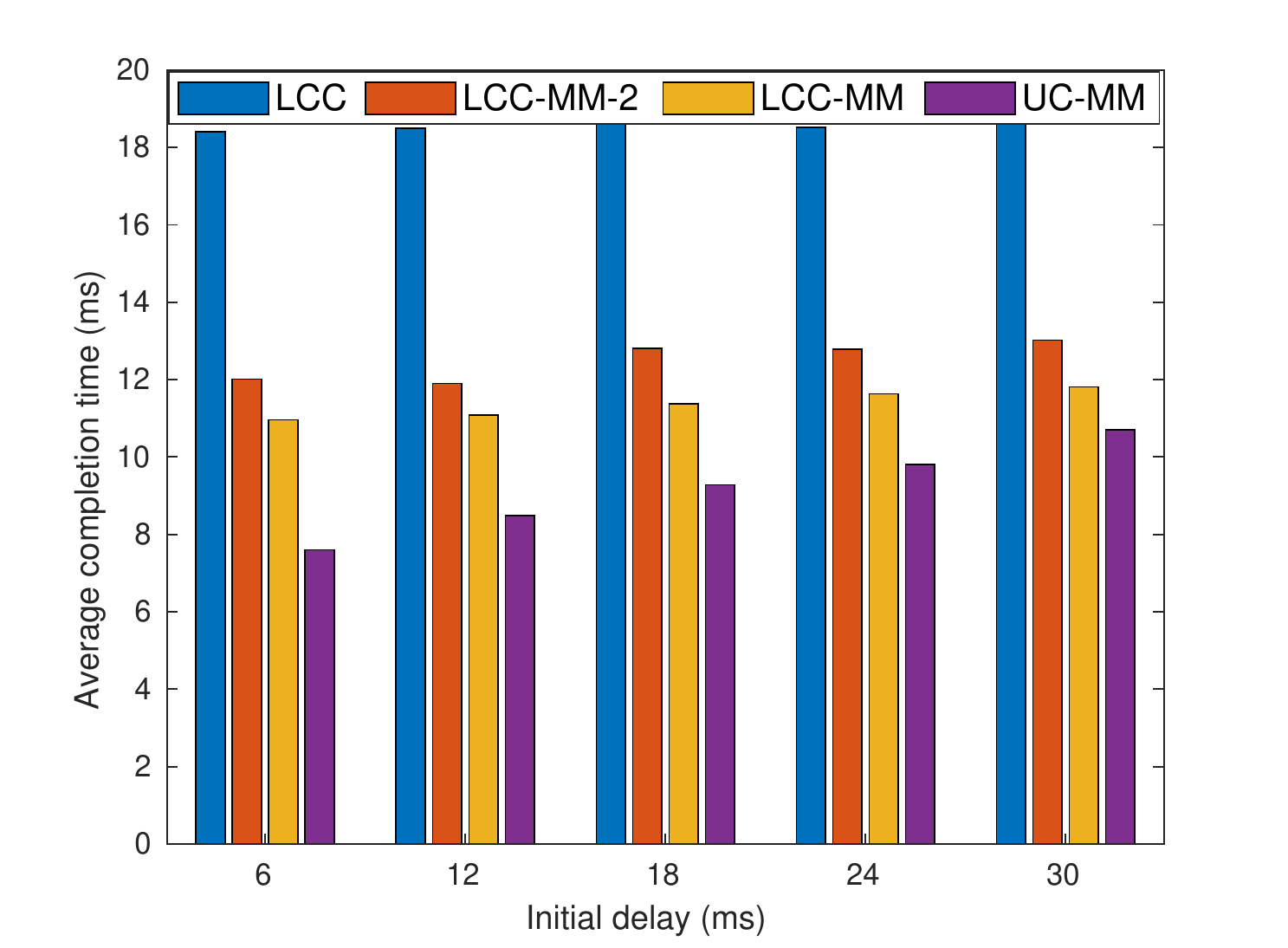}
        \caption{Correlated delay, p=0.1}
				\label{r4p1}
    \end{subfigure}
    \smallskip
    \begin{subfigure}{0.5\textwidth}
        \includegraphics[scale=0.5]{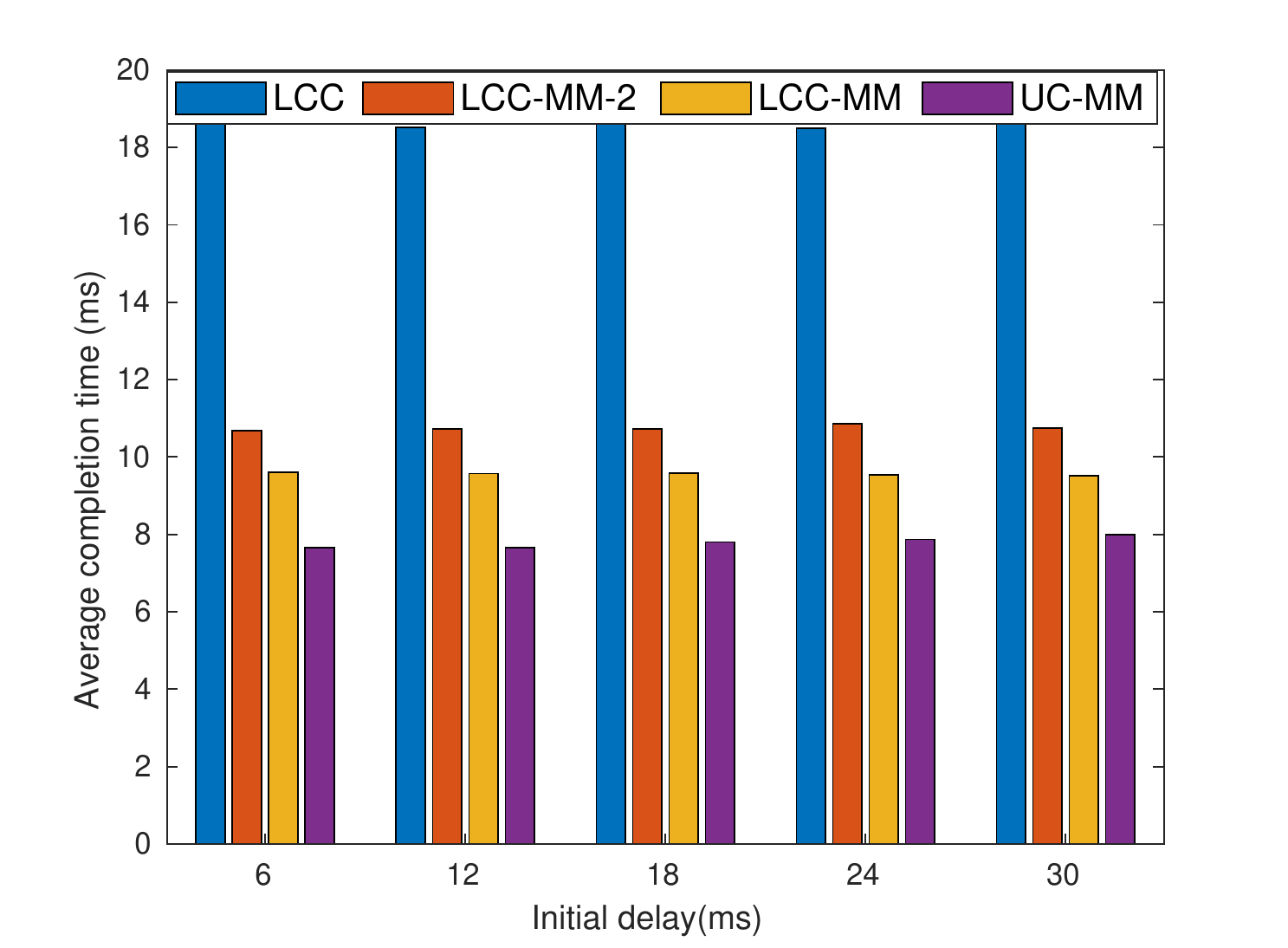}
        \caption{Uncorrelated delay, p=0.1}
				\label{r4p2}
        \end{subfigure}
        \smallskip
                 \begin{subfigure}{0.5\textwidth}
        \includegraphics[scale=0.5]{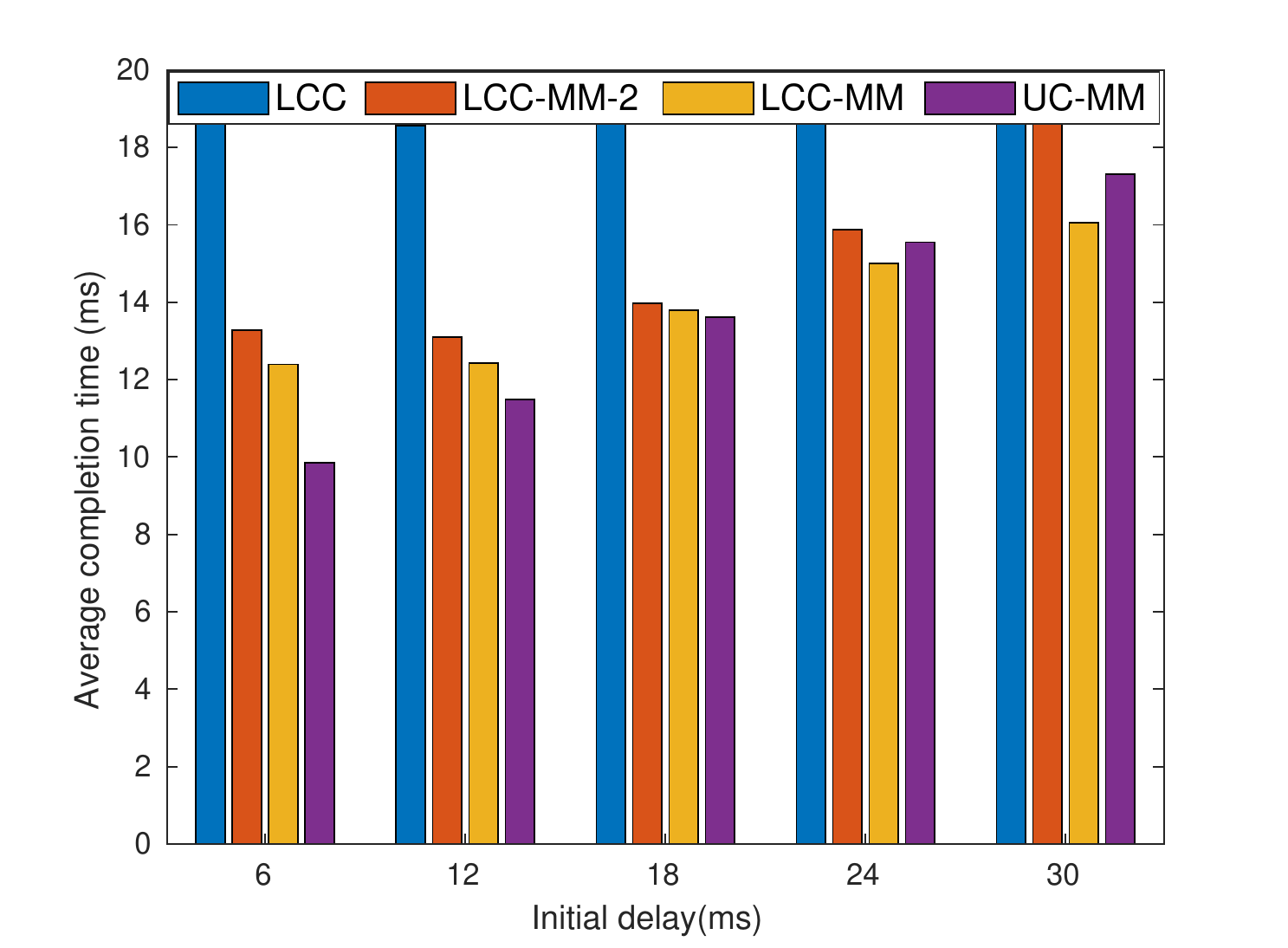}
        \caption{Correlated delay, p=0.3}
				\label{r4p3}
    \end{subfigure}
    \smallskip
             \begin{subfigure}{0.5\textwidth}
        \includegraphics[scale=0.5]{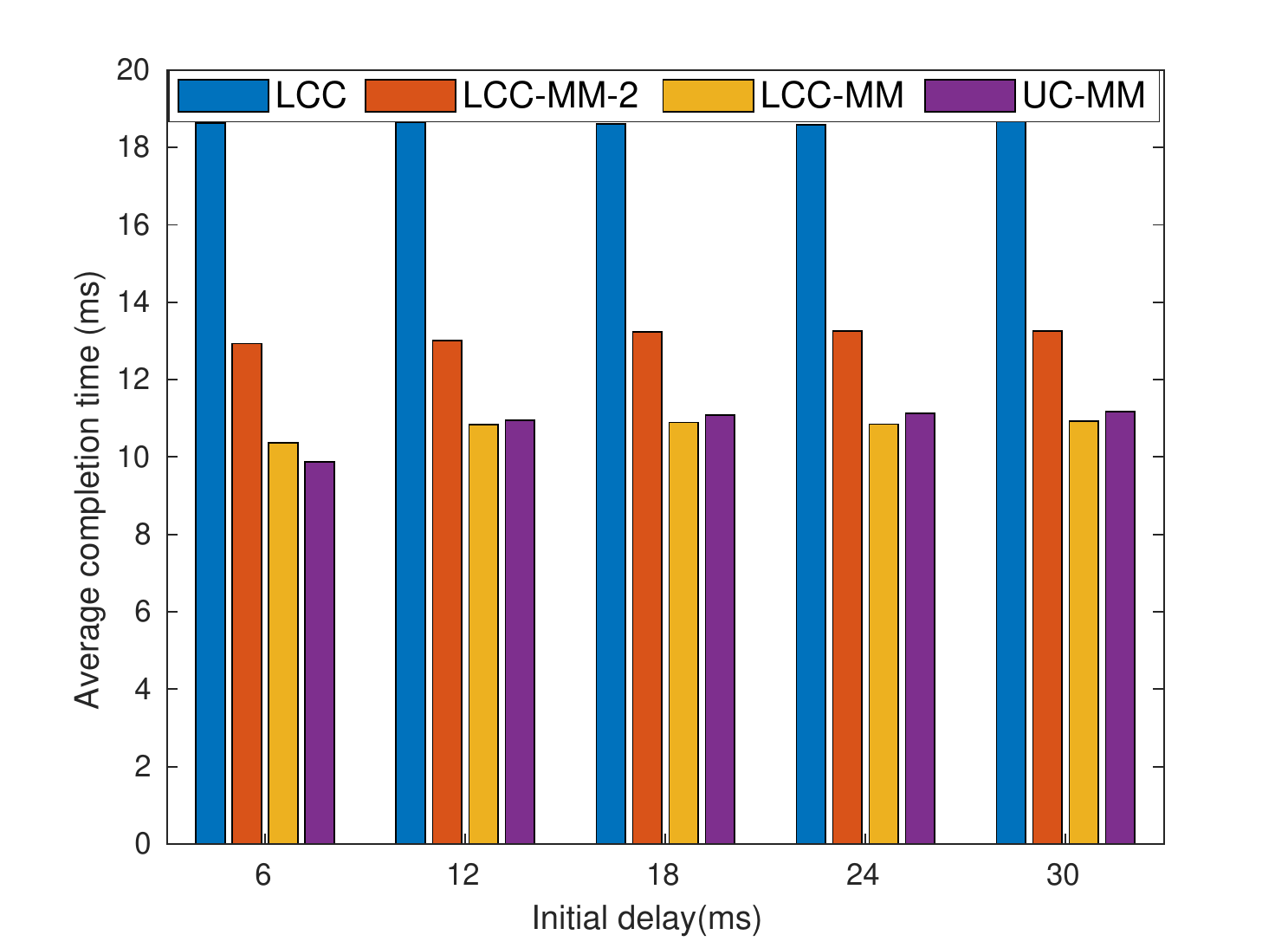}
        \caption{Uncorrelated delay, p=0.3}
				\label{r4p4}
    \end{subfigure}
    \smallskip
				\caption{Per iteration completion time for different schemes with $r=6$. }
		\label{}
\end{figure*}

\subsection{Discussions}
Comparing data driven and real time simulation results, we have shown that network congestion, especially in large scale implementations, might be a predominant issue for the performance. For real time implementations, we expect the communication latency to scale with the number of instances which limits the advantage of the MMC approach.
However, a hierarchical network architecture, where the instances are grouped and multiple PSs are employed, can be used to alleviate the congestion, and thus MMC approach can   still be beneficial.\\
\indent In this paper, we have mostly limited our focus to the exact recovery of the full gradient, however partial gradient recovery, as well as gradient approximation, are both important research directions, which have been recently studied by several works \cite{CC.AG1,CC.AG2,CC.AG3,CCP.1,CCP.2}. We remark that partial gradient or approximate gradient recovery, can reduce the computation time by allowing less accurate updates. Besides, the most popular optimization framework for {\em deep learning} is SGD, which basically uses {\em unbiased estimation} of the gradient using randomly sampled data \cite{SGD}. Therefore, one can argue that the full gradient may not be required for a successful implementation of the GD framework in many machine learning applications. However, as already discussed in \cite{UCCT.1}, missing partial gradients may cause GD algorithm to diverge in some cases, particularly when an acceleration strategy, such as Nesterov’s accelerated gradient, is employed. In addition, even for the SGD implementation, it is shown that the number of required iterations for training can be reduced by increasing the batch size \cite{train0}, which is actually the main motivation behind large scale implementations \cite{train1,train2}. Furthermore, impact of the stragglers on the convergence may also depend on the dataset, its distribution among the workers (such as i.i.d./non-i.i.d. distributions) and the straggler realizations.\\
\indent Finally, we want to note that, in this paper, for the overall latency analysis we take into account the computation time and the communication time, but ignore the latency at PS due to the encoding complexity. As discussed in \cite{UCCT.3}, the implemented code structure also plays an important role in the overall latency. However, in the scope this paper, our main focus has been to introduce a design framework for distributed learning with MMC, and we also note that different code structures can be incorporated with the introduced framework. Hence, we leave the MMC strategy with reduced decoding complexity as a future extension of this work.
\section{Conclusion}
We have introduced novel coded and uncoded DGD schemes when MMC is allowed from each worker at each iteration. First, we have provided a closed-form expression for the per iteration completion time statistics of these schemes under a shifted exponential computation time model, and verified our results with  Monte Carlo simulations. Then, we have compared  these schemes with other DGD schemes in the literature in terms of the average computation and communication loads incurred.\\ 
\indent We have observed that allowing multiple messages to be conveyed from each worker at each GD iteration can reduce the average completion time significantly by expoiting non-straggling workers at the expense of an increase in the average communication load.  We have also observed that UC-MM with simple circular shift can be  more efficient  compared to coded computation approaches when the workers have limited storage capacity. We emphasize that, despite benefits of coded computation in reducing the computation time, their relevance in practical big data problems is questionable due to the need to jointly transform the whole dataset, which may not even be possible to store in a single worker.
In this paper, we have performed comprehensive simulations with different parameters to highlight the fundamental trade-offs in the practical implementation of the distributed computation in the context of gradient descent for machine learning applications.


%


\IEEEpeerreviewmaketitle

\bibliographystyle{IEEEtran}
\bibliography{IEEEabrv,ref}

\end{document}